\documentclass{svproc}
\usepackage{url}
\usepackage{type1cm}        	% activate if the above 3 fonts are not available on your system
\usepackage{makeidx}         	% allows index generation
\usepackage{graphicx}        	% standard LaTeX graphics tool when including figure files
\usepackage{multicol}        	% used for the two-column index
\usepackage[bottom]{footmisc}	% places footnotes at page bottom
\usepackage{newtxtext}       	% selects Times Roman as basic font
\usepackage{newtxmath}       	% selects Times Roman as basic font
\usepackage{subfigure}
\usepackage{color}
\usepackage{lscape}%rotate table
\usepackage{multirow}
\usepackage{eurosym}
\usepackage{comment}

\usepackage{amssymb}  % $\bigcirc$ 用
\usepackage{pifont}   % \ding{55} 用
\usepackage{enumitem} %アルファベットenumarate

\begin{document}
\mainmatter % start of a contribution
\title{Effect of vehicle groups on heterogeneous disordered traffic flow}
%
% also used for the TOC unless
% \toctitle is used
%
\author{Akihito Nagahama\inst{1} \and Nichika Asai\inst{1} \and Claudio Feliciani\inst{2,3} \and Xiaolu Jia\inst{4} \and Katsuhiro Nishinari\inst{2,5}} 
\authorrunning{Akihito Nagahama et al.} % abbreviated author list (for running head)
\tocauthor{}
\institute{Graduate School of Informatics and Engineering, The University of Electro-Communications, 1-5-1 Chofugaoka, Chofu, Tokyo 182-8585, Japan,\\
\email{naga0862@uec.ac.jp}
\and
Department of Aeronautics and Astronautics, School of Engineering, The University of Tokyo, 7-3-1 Hongo, Bunkyo-ku, Tokyo 113-8656, Japan
\and
Waseda Institute for Advanced Study, Waseda University, 1-21-1 Nishiwaseda, Shinjuku-ku, Tokyo, 169-0051, Japan
\and
Beijing Key Laboratory of Traffic Engineering, Beijing University of Technology, 100 Pingleyuan, Chaoyang-district, 100124, Beijing, China
%thexiaolujia@bjut.edu.cn
\and
Research Center for Advanced Science and Technology, The University of Tokyo, 4-6-1 Komaba, Meguro-ku, Tokyo 153-8904, Japan}

\maketitle % typeset the title of the contribution

%（和訳）
% heterogenous disordered traffic（車線規律が緩く多様な車種が混在する交通）では，さまざまな車種が厳密な車線規律なしに走行する中で，自己組織化的な車両グループがしばしば形成される。こうしたグループ形成は認識されているものの，マクロな交通ダイナミクスに与える影響は依然として不明である。本研究は，グループの出現頻度および構成が，heterogeneous disordered traffic における流量–密度関係にどのように影響するかを検討する。実交通映像観測から得た軌跡データを用い，交通の異質性を考慮した流量–密度図を構築するために，3種類のPCU推定手法を適用した。
% 解析の結果，グループ割合（すなわち，グループに属すると分類される車両の割合）が，流量特性に対して非線形かつ交通状態依存的な影響を持つことが示された。具体的には，グループ割合が中程度（30–60\%）のとき，中・高密度条件でより高い流量が観測され，一方で50\%を超えると低密度または高密度の極端側へ分布が偏る傾向が見られた。さらに，台数ベースとPCUベースのグループ割合の比較から，とくにバイクのような小PCU車両が中心のグループでは，正規化手法がグループ動態の解釈に大きく影響することが示された。加えて，自由流条件ではグループ割合が低いほど流量が高まる一方，エントロピー単体では，エントロピーと速度の関連は traffic situation を通じて一貫して示されなかった。
%代表的な傾向と極値的な高流量事例を対比することで，類似した密度条件およびグループ割合条件の下でも，交通が低効率モードと高効率モードを取り得ることが示唆された。
% 全体として，本研究の知見は，グループの出現（割合）がマクロ交通ダイナミクスの形成に重要であることを示す一方で，グループ内部構成と交通性能を結び付けるには，単一のエントロピー指標を超えるより詳細な構造記述子が必要となり得ることを示唆する。
% これらの知見は，インフラ整備や交通規制が限定的な異種混合交通環境に適した，ボトムアップ型の交通制御戦略の検討に有用な示唆を与える。

\begin{abstract}
In heterogeneous disordered traffic, where various vehicle types operate without strict lane discipline, self-organized vehicle groups often emerge. While the formation of such groups has been recognized, their influence on macroscopic traffic dynamics remains unclear. This study investigates how the prevalence and composition of vehicle groups affect flow--density relationships in heterogeneous disordered traffic. Using trajectory data from real-world video observations, we apply three distinct Passenger Car Unit (PCU) estimation methods to construct flow--density diagrams that account for traffic heterogeneity.
The analysis reveals that group proportions, i.e., the proportion of vehicles that are classified as belonging to groups, have a nonlinear and traffic-situation-dependent impact on flow characteristics. 
Specifically, moderate group proportions (30--60\%) were more clearly associated with peak-flow or high-flow cases under denser traffic conditions, rather than consistently producing the highest representative flow in the median-based analysis.
Comparisons between vehicle-count-based and PCU-based group proportions indicate that normalization methods significantly affect the interpretation of group dynamics, particularly when groups consist mainly of small-PCU vehicles such as motorcycles. 
Additionally, lower group proportions were associated with relatively favorable flow under free-flow conditions, while the entropy-based analysis indicates that the association between entropy alone and speed is not consistently observed across traffic situations.
By contrasting representative trends and extreme high-flow cases, the results further suggest that traffic under similar density and group-proportion conditions can exhibit low-efficiency and high-efficiency modes.
Overall, the findings highlight the importance of group prevalence in shaping macroscopic traffic dynamics, while suggesting that linking internal group composition to traffic performance may require richer structural descriptors beyond a single entropy measure.
These results suggest condition-dependent implications for bottom-up traffic control: encouraging favorable vehicle formations in denser traffic and preserving fluid leader--follower relations in free-flow traffic may help improve heterogeneous traffic flow, especially when supported by minimal infrastructure or driver-assistance technologies.

\end{abstract}
\section{Introduction}
\label{sec:intro}
%交通渋滞たいへんだね　heterogeneous disordered trafficって呼ぶよ
The rapid proliferation of automobiles has played a crucial role in driving global economic growth in the 20th century, enhancing daily convenience, and facilitating the movement of people and goods. However, rapid population growth and economic development worldwide have led to increased motorization, exacerbating traffic congestion. Traffic congestion negatively impacts various aspects of society, including the economy, the environment, and public health. For instance, a study~\cite{IDB} estimated that in 2016, the European Union (EU) suffered an economic loss of approximately \euro200 billion, equivalent to 1.4\% of its GDP, due to congestion.
Traffic congestion is not solely a concern for developed countries. 
In developing countries, traffic congestion causes severe delays in urban mobility. According to a study~\cite{akbar2023fast}, cities like Dhaka, Lagos, and Manila rank among the slowest in the world, with average travel times up to three times longer than those in the fastest cities. In Dhaka, for instance, average vehicular speed is less than one-third that of Flint, USA, resulting in substantial time losses for daily commuters and freight movement alike.
Many other examples were also reported in literature~\cite{elmansouri2020urban,samal2021adverse,fattah2022insights}.
In many developing regions, congestion is further exacerbated by the presence of diverse vehicle types, including motorcycles, three-wheelers, cars, and other various types of vehicles which often operate with minimal adherence to lane discipline. Such traffic conditions, characterized by a high degree of heterogeneity and disorder, are referred to in this study as heterogeneous disordered traffic.

%hetero-disordered trafficにおいて、様々な車種が自己組織化する例が報告されています。
%例えば牛車や自転車などエンジンのない極端に遅い車両が「移動する障害物」となり、後続の車両が、追い越せるまで長い車列を形成するPlatoon生成が見られます\cite{phogat2020study}。
%また特に2輪車の自己組織化はliteratureでよく研究がされています
In heterogeneous disordered traffic, various vehicle types exhibit self-organization behavior, which has been reported in several studies. For example, non-motorized vehicles such as bullock carts and bicycles, which move at significantly lower speeds, act as ``moving obstacles.'' As a result, following vehicles tend to accumulate behind them, forming a long queue until they find an opportunity to overtake. This phenomenon leads to the formation of traffic platoons, as observed in previous research~\cite{phogat2020study}.
Apart from platoon formation, another widely studied form of self-organization in mixed traffic is the collective behavior of motorcycles. Lee~\cite{lee2007agent} observed that motorcycles in mixed traffic conditions can demonstrate herd behavior, cycling through phases of filtering, gathering, and dispersing. This tendency enables motorcycles to self-organize dynamically within the traffic flow. Furthermore, Lee et al.~\cite{lee2016agentbased} successfully replicated motorcycle queuing patterns at intersections using an agent-based model. Additionally, Das et al.~\cite{das2017modelling} demonstrated that motorcycles create dynamic virtual lanes and engage in filtering behavior, further contributing to their self-organization.

%筆者らはあらゆる車種やそれらが混在する車両群の自己組織化する傾向を、定量的に検知することに取り組んできた\cite{nagahama2021detection,nagahama2022certain,nagahama2024performance}
%具体的には、グラフネットワークとポアソン分布による統計的手法によって、どの車種組み合わせの車両群がどのような追従関係を維持しやすいかを検出することに成功し\cite{nagahama2022certain}
%その発生メカニズムの定性的モデルまで提案している\cite{nagahama2024performance}。
%これらの、追従関係を頻繁に結び、長時間維持しやすい車両群をfrequent sub-networks in standardized traffic  (FSST), もしくはGroupと呼び、本研究でもGroupと呼ぶ。
An understanding and evaluation of the macroscopic characteristics and dynamics of 2D mixed traffic requires an analysis of the grouping behavior of different vehicles, including vehicle types other than motorcycles. 
Few studies have focused on groups of vehicles other than motorcycles. We previously suggested that different vehicle types (apart from motorcycles) tend to exhibit grouping behavior~\cite{nagahama2021detection}. 
Furthermore, we have successfully detected specific groups in Indian traffic, where traffic flow is relatively smooth~\cite{nagahama2022certain}. 
To identify groups, i.e., patterns of leader--follower relations, we introduced a new concept called frequent subnetworks in standardized traffic~\cite{nagahama2022certain}, which is referred to as ``group'' in the present study.

%無秩序で管理がなされていない交通では交通容量が減少することが示唆される\cite{bhardwaj2023understanding}一方で
%別の研究ではheterogeneous disorderd trafficにおける、車両群の自己組織化が交通量を改善する可能性が示唆されています。
%例えば特定の（複数）車種が交通中のどの場所にどのような追従関係をもって偏在し自己組織化するかによって
%交通特性が変化することが示唆されています。
%\cite{liu2016modeling,chen2016car,mason1997car}の研究は、heterogeneous でlane-based なtrafficに限ってcomposition or sequence of vehicle types in a traffic platoon---vehicles following one another in a leader--follower relationship---has been shown to influence traffic stability and the formation of stop-and-go wavesを示しました。
%これは\cite{sarvi2013heavy,aghabayk2011examining,sayer2000effect,nagahama2020response,nagahama2017dependence}で調べられた車種毎の異なる挙動に加えて、追従関係を結ぶ車種の組み合わせによって、マクロな交通特性が変化することを、lane-baseでありつつも示しています。
%さらに最近の研究\cite{li2022equilibrium,chen2025evidence}では、異なる車種群がそれぞれ協調的な行動を示しているか定量的に測定する試みも進みつつあり、heterogeneous trafficにおける車種を交えた自己組織化と交通流改善の関連に着目が集まりつつある。
It has been suggested that in unmanaged and highly disordered traffic conditions, overall traffic capacity tends to decrease~\cite{bhardwaj2023understanding}. However, other studies have indicated that the self-organization of vehicle formations in heterogeneous disordered traffic may contribute to improved traffic flow. Specifically, the spatial distribution and leader-follower relationships of particular vehicle types can influence overall traffic characteristics.
Previous studies have demonstrated that, at least in lane-based heterogeneous traffic, the composition and sequence of vehicle types within a traffic platoon---where vehicles follow one another in a leader-follower relationships---affect traffic stability and the emergence of stop-and-go waves~\cite{liu2016modeling,chen2016car,mason1997car}. These findings suggest that, in addition to the differences in driving behavior among vehicle types investigated in prior research \cite{sarvi2013heavy,aghabayk2011examining,sayer2000effect,nagahama2020response,nagahama2017dependence}, the specific combinations of vehicle types engaged in leader-follower interactions can also influence macroscopic traffic characteristics, in lane-based heterogeneous traffic.
Furthermore, recent studies~\cite{li2022equilibrium,chen2025evidence} have attempted to quantitatively assess the extent to which different vehicle types exhibit cooperative behavior. This growing body of research highlights the increasing attention toward the relationship between inter-vehicle self-organization and traffic flow improvements in heterogeneous traffic.

%筆者らの先行研究\cite{nagahama2021detection,nagahama2022certain}は、このような偏在する自己組織化した車種群を、groupとしてno lane-based heterogeneous trafficに定量的に見出しました。
%しかしながら、groupのheterogeneous disordered traffic交通特性への影響は明らかになっていません。
%本研究は、heterogeneous disorderd trafficのmacroな交通特性に、groupの多寡が影響するか、するのであればどのようにするかを明らかにすることを目的とする。
%そのために、heterogenietyの影響を最小限に抑えながら交通特性を評価できるPassenger car unit（PCU）を活用してfundamental diagramを解析する。
Our previous studies~\cite{nagahama2021detection,nagahama2022certain} quantitatively identified such spatially non-uniform self-organized vehicle formations, referred to as groups, in no-lane-based heterogeneous traffic. 
However, the impact of these groups on the traffic characteristics of heterogeneous disordered traffic remains unclear.
The objective of this study is to determine whether the prevalence of groups influences the macroscopic traffic characteristics of heterogeneous disordered traffic, and if so, to elucidate the nature of this influence. 
To achieve this, we utilize several types of passenger car unit (PCU) to evaluate the fundamental diagram, enabling the evaluation of traffic characteristics while minimizing the influence of heterogeneity.

%本研究によって heterogeneous disordered traffic における group の交通流への影響が明らかになれば、pragmatic driving~\cite{king2015traffic} やドライバー間の暗黙的な調整（たとえば信号のない交差点における相互適応~\cite{kumar2006self}）による自己組織化が、stop-and-go 現象が発生するような道路環境においても、交通流の改善に寄与する可能性があることが示唆される。この知見は交通工学において重要な意味を持ち、heterogeneous disordered traffic のミクロシミュレーションでは空間的な偏在や自己組織化の正確な再現が必要であることを補強する。また、空間的偏在に起因する交通特性の変化に加えて、レーン分離や停止位置分離~\cite{shiomi2011evaluation}、ITS技術を活用した運転支援などと組み合わせることで、最小限の交通規制下でもボトムアップ型の交通流改善が実現できる可能性がある。さらに広義には、このような分散的・自己組織化的な交通システムが、ローカルな相互作用を通じて効率性を向上させる仕組みを理解することは、社会制度の設計に関する新たな視座を提供する。トップダウンの規則制御に依存せず、個々の自律的な行動が全体として有益な結果をもたらすというパラダイムは、交通に限らず、回復力と適応性を備えた社会システムの構築にとっても重要な示唆となる
If this study successfully clarifies the influence of groups on the traffic flow of heterogeneous disordered traffic, it would suggest that self-organization driven by pragmatic driving~\cite{king2015traffic} and mutual adjustments among drivers---as seen at unsignalized intersections~\cite{kumar2006self}, where drivers engage in implicit coordination and mutual adaptation---could also contribute to traffic flow improvements on roads where stop-and-go waves can occur. This finding has several significant implications for the field of traffic engineering. First, it reinforces the necessity of accurately reproducing spatially non-uniform vehicle distributions and self-organization in microscopic simulations of heterogeneous disordered traffic~\cite{nagahama2022prototype}. Additionally, by leveraging changes in traffic characteristics due to spatial clustering, in combination with lane or stop position separation~\cite{shiomi2011evaluation} and driving support technologies enabled by Intelligent Transportation Systems (ITS), it may become feasible to achieve bottom-up traffic flow improvement even under minimal traffic regulations and predominantly pragmatic driving behaviors.
More broadly, understanding how such decentralized, self-organizing traffic systems can improve efficiency through local interactions may offer insights into alternative models of societal governance, e.g., public space utilization, mobility, energy use, etc. Rather than relying solely on top-down, rule-based control, these findings support a paradigm in which local, autonomous behavior leads to emergent global benefits—a concept relevant not only to transportation but also to the broader design of resilient, adaptive social systems.

%%%%%%%%%%%%20250319いまここまでできた！！次手法に入る
%全体書いてから

%本稿の構成は以下のとおりである。
%第\ref{sec:datamethod}節では、交通観測の方法、各車両の軌跡データの抽出方法、traffic situation の分類、group の抽出技術の概説、および本研究で使用する PCU の導入について述べる。
%第\ref{sec:result}節では、segment-wise median に基づくスムーズな分析と、詳細な scatter plot を用いた可視化という2つの手法により、group割合に対する一般的傾向および割合ごとの局所的変動・分布の違いを調査する。各小節には結果と議論を含めて構成している。
%最後に第\ref{sec:concl}節では、本研究の結論を述べる。
The remainder of this paper is structured as follows.
Section~\ref{sec:datamethod} provides an overview of the video observation method, the extraction of vehicle trajectories, the classification of traffic situations, the technique for detecting vehicle groups, and the introduction of PCU definitions used in this study.
Section~\ref{sec:result} investigates both general trends across group proportions and local variations within each group proportion using two complementary approaches: a smoothed analysis based on segment-wise medians and a detailed scatter plot visualization. 
We also examine whether the degree of vehicle-type mixing within a group is related to the flow performance in this section.
Each subsection in this part includes both results and discussion.
Section~\ref{subsec:consensus_discussion_final} provides an overall discussion of the results presented in Section~\ref{sec:result}.
Finally, Section~\ref{sec:concl} presents the concluding remarks.

\section{Data and Preprocessing}
\label{sec:datamethod}

\subsection{Observed Data and Vehicle Trajectory}
\label{ssec:data}

The present study utilizes the same dataset employed in our prior work~\cite{nagahama2022certain}, which was collected through traffic observation conducted in Mumbai, India, over a four-day period from January 18 to 21, 2017. For detailed analysis, we selected a segment of video footage recorded on January 19 between 11:34 AM and 2:34 PM. The recording was made with a Sony HDR-CX670 video camera mounted on the second-floor balcony of a shopping mall.

\begin{figure}[hbt]
\centering
\includegraphics[width=7cm]{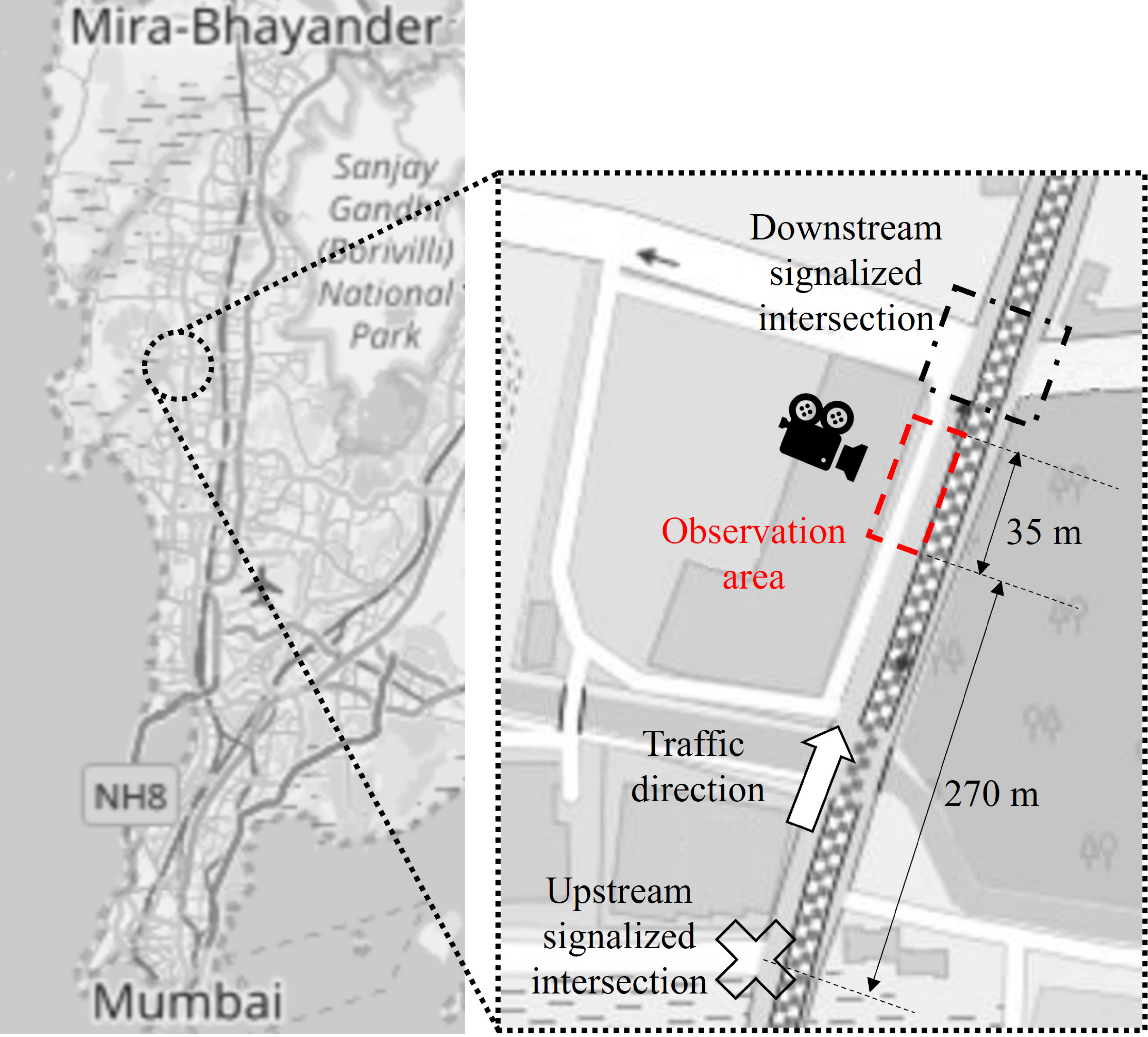}
\caption{Traffic-observation site used for data acquisition. The downstream signal, upstream intersection, and observation section are highlighted using a black chained rectangle, black cross mark, and red dashed rectangle, respectively. The map image is based on OpenStreetMap.}
\label{fig:obsMap}
\end{figure}

Figure~\ref{fig:obsMap} presents a configuration of the observed road. The road section analyzed spans 35 meters in length and is enclosed by a red dashed rectangle. The location of the downstream traffic signal is indicated by a black chained rectangle just beyond the observation area.

Vehicle positions in the image coordinate system were tracked semi-automatically using the Multiple Instance Learning (MIL) tracker from OpenCV~\cite{babenko2009visual}. Vehicle types were manually classified into four categories: motorcycles (``m''), auto-rickshaws (three-wheelers, ``r''), passenger cars (``c''), and heavy vehicles (``h'').

\subsection{Classification of Traffic Situations}
\label{ssec:trafSitu}
In the present study, we analyze a wide range of traffic situations, classified as follows:
\begin{itemize}
\item \textbf{All-flow (AF):} All traffic in the observation segment is steadily flowing.
\item \textbf{Downstream-jammed (DJ):} Traffic is congested in the downstream area but remains flowing upstream.
\item \textbf{All-jammed (AJ):} Traffic is congested throughout the entire observed segment.
\item \textbf{Other (OT):} Traffic does not fall under the above categories (e.g., flow is beginning to recover downstream).
\end{itemize}

\begin{figure}[hbt]
\centering
\includegraphics[width=10cm]{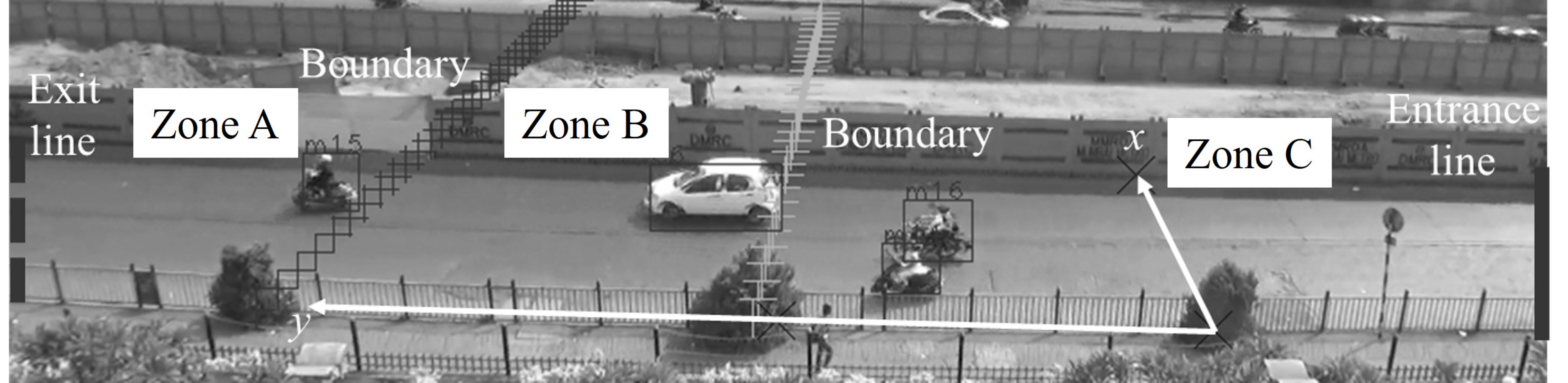}
\caption{Division of the observation segment into three zones (A, B, and C), arranged from downstream to upstream.}
\label{fig:zoneDiv}
\end{figure}

To classify the traffic situations, we utilized densities of several road segments. As illustrated in Fig.~\ref{fig:zoneDiv}, the observation segment was divided into three zones: A, B, and C. These zones were delineated based on the positions of trees adjacent to the road. Traffic conditions within each zone were evaluated by calculating vehicle density, denoted as $\rho_{\mathrm{A}}$, $\rho_{\mathrm{B}}$, and $\rho_{\mathrm{C}}$ for Zones A, B, and C, respectively. Densities were computed for each video frame using the number of vehicles and the length of each zone.

Traffic situations were classified according to the following logic:
\begin{itemize}
\item \textbf{AF:} All densities are at or below the threshold $\rho_{\mathrm{th}} = 0.6\,\mathrm{vehicles/m}$.
\item \textbf{DJ:} Downstream densities exceed the threshold, while upstream densities are at or below the threshold.
\item \textbf{AJ:} All densities exceed the threshold.
\item \textbf{OT:} Conditions that do not meet any of the above definitions.
\end{itemize}

For example, a situation is classified as DJ when $\rho_{\mathrm{A}} > \rho_{\mathrm{th}}$, $\rho_{\mathrm{B}} \leq \rho_{\mathrm{th}}$, and $\rho_{\mathrm{C}} \leq \rho_{\mathrm{th}}$. Conceptually, AF corresponds to completely free-flowing traffic. DJ reflects partial congestion, primarily in downstream areas. AJ represents full congestion across the observed section, while OT is the situation wherein vehicles downstream begin to move freely while others are still in congestion.

The threshold $\rho_{\mathrm{th}} = 0.6\,\mathrm{vehicles/m}$ is not an arbitrary choice but corresponds to the critical density of this road segment: the flow--density diagram constructed for the entire observation area shows that flow begins to decline beyond this value~\cite{nagahama2025grouping}. The same threshold has been adopted consistently in our previous studies based on this dataset.

Because the length of the observation section analyzed in this study is 35~m, the traffic situations defined here should be interpreted as ones inferred from the observation section.

\subsection{Detection of Groups}
\label{ssec:groupDetection}

\begin{figure}[h]
\centering
\includegraphics[width=10cm]{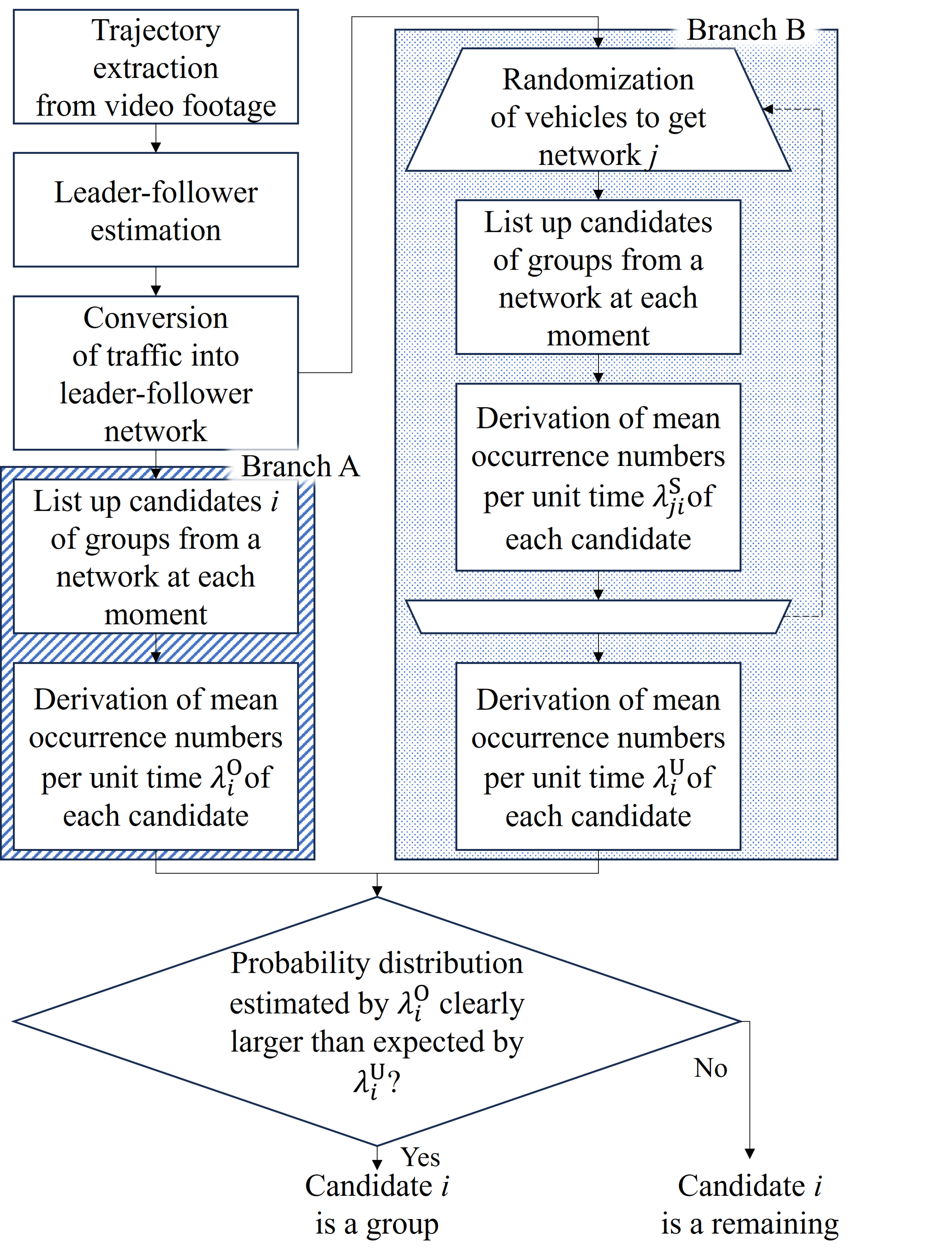}
\caption{Flowchart for classifying vehicles into groups and others.}
\label{fig:groupFlow}
\end{figure}

To distinguish groups from non-group vehicles, we extracted groups from traffic data following the methodology established in our previous work~\cite{nagahama2022certain}. While the full procedure is described in detail in that study, we provide a concise overview here. Figure~\ref{fig:groupFlow} presents a flowchart summarizing the group detection process.

The procedure begins by estimating each vehicle's trajectory from the video data and identifying leader-follower relationships---defined as instances where the acceleration or deceleration of a leading vehicle influences that of a following vehicle. In both the current and prior studies, this relationship is determined based on two criteria: lateral proximity within a predefined threshold and adjacency of Voronoi cells.

By representing vehicles as nodes and their leader-follower relationships as directed edges, the overall traffic structure can be modeled as a directed graph, which we refer to as the leader-follower network.

The process then diverges into two branches. In Branch A (highlighted with diagonal shading in Fig.~\ref{fig:groupFlow}), candidate groups, i.e., subnetworks frequently appearing in the leader-follower network, are extracted for each time frame. The frequency of each candidate subnetwork $i$ is quantified as $\lambda^{\mathrm{O}}_{i}$, the average number of occurrences per unit time across the entire observation period.

Branch B (highlighted with a dotted background) introduces a randomized baseline. Here, the vehicle nodes of the observed leader-follower network are randomly shuffled, preserving vehicle-type labels (``m,'' ``r,'' ``c,'' ``h'') to retain type-specific characteristics. This process is repeated multiple times to construct randomized networks, from which the average occurrence frequency of subnetwork $i$, denoted as $\lambda^{\mathrm{U}}_{i}$, is computed. In Fig.~\ref{fig:groupFlow}, $\lambda^{\mathrm{S}}_{ji}$ indicates the average occurrence of $i$ in the $j$-th randomized network.

Using the observed $\lambda^{\mathrm{O}}_{i}$ and the randomized $\lambda^{\mathrm{U}}_{i}$, we construct two Poisson distributions representing the expected number of occurrences per unit time for each subnetwork. If the distribution derived from $\lambda^{\mathrm{O}}_{i}$ is statistically higher than that from $\lambda^{\mathrm{U}}_{i}$, subnetwork $i$ is inferred to exhibit a significant tendency toward group formation and is thus classified as a \textit{group}.
Importantly, this statistical classification itself does not require any additional cutoff on $\lambda^{\mathrm{O}}_{i}$. In the subsequent analysis, we introduce explicit thresholds on $\lambda^{\mathrm{O}}_{i}$ only for interpretive purposes, namely, to distinguish highly frequent groups from moderately frequent ones and to examine how the observed tendencies depend on group frequency.

\begin{figure}[h]
\centering
\includegraphics[width=8cm]{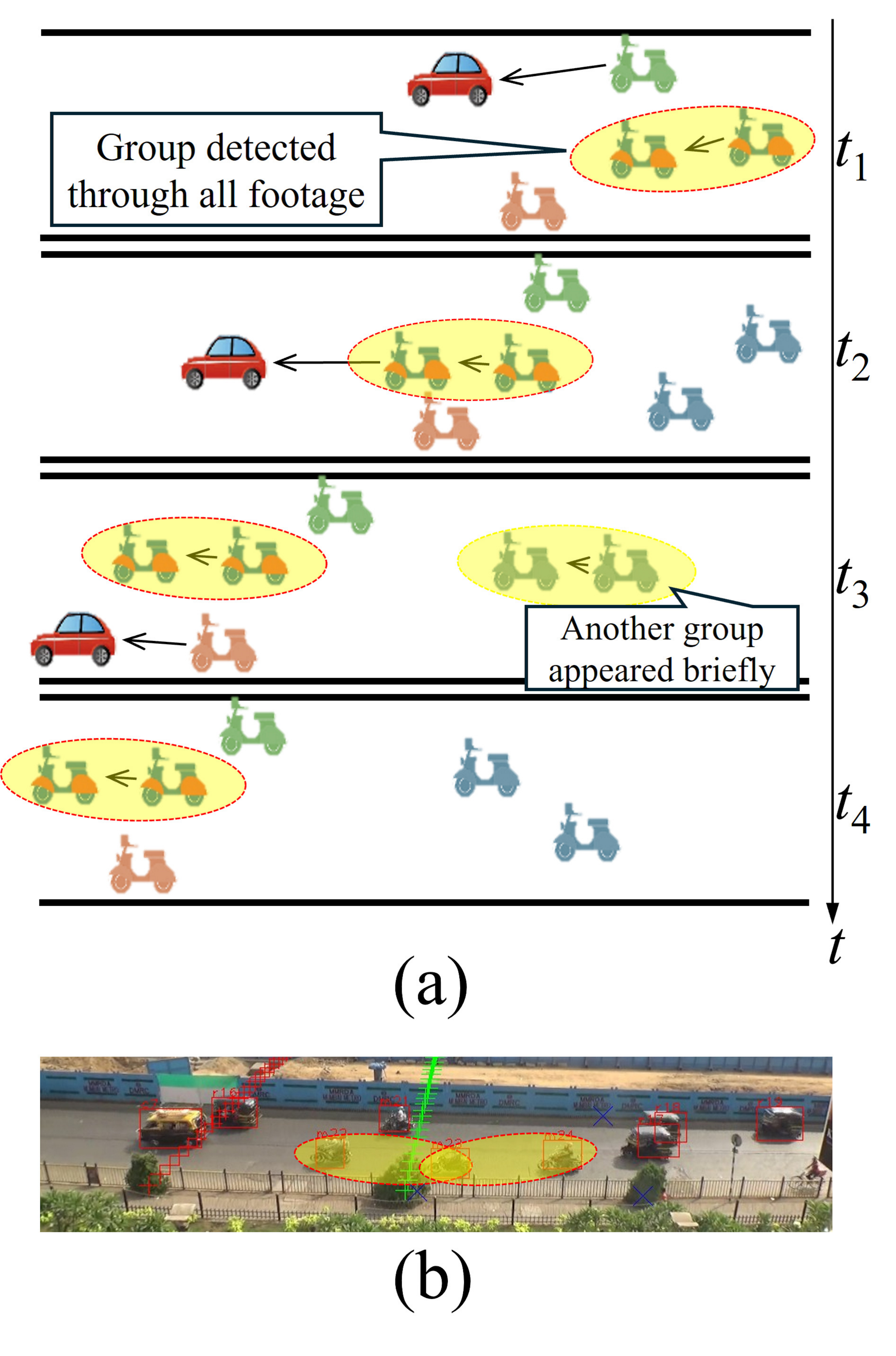}
\caption{(a) Conceptual image of a group in traffic: an example where two motorcycles maintaining a leader--follower relationship are detected as a group. (b) Actual traffic scene where a group of two motorcycles is detected.}
\label{fig:groupPonchi}
\end{figure}

For example, according to a previous study~\cite{nagahama2025grouping}, a pair of motorcycles maintaining a leader--follower relationship is considered a typical group in the AF. In such cases, as illustrated in Figure~\ref{fig:groupPonchi}~(a), the leader--follower relationship between two motorcycles tends to be sustained over a long period and is frequently generated. The arrows in Figure~\ref{fig:groupPonchi}~(a) represent leader--follower relationships. Even a momentary leader--follower interaction between two upstream motorcycles at time $t_3$ is identified as a group.
Figure~\ref{fig:groupPonchi}~(b) shows an actual snapshot in which two motorcycles with a leader--follower relationship are detected as a group in real traffic. In this example, although three motorcycles are arranged in a leader--follower sequence, the two motorcycles at the front and the two at the rear form separate groups based on their respective leader--follower relationships.

Note that the detection procedure summarized above embeds spatial and temporal constraints, in addition to the leader--follower frequency and the statistical test. Leader--follower relations are identified only between spatially proximate vehicles, namely pairs whose lateral offset is within a predefined threshold and whose Voronoi cells are adjacent, following the empirically validated criteria established in our earlier detection framework~\cite{nagahama2022certain}. In addition, the occurrence rate $\lambda^{\mathrm{O}}_{i}$ is derived from the frequency-time value of subnetwork $i$~\cite{nagahama2022certain}, which accumulates the product of the frequency and the duration of each candidate formation over the observation period, so that temporal persistence is reflected in the statistical classification. Speed synchronization is not modeled as an explicit criterion; however, its effect is still captured, because a subnetwork is classified as a group only when its leader--follower relations are maintained frequently and over long durations, which cannot be sustained without a substantial degree of mutual speed adjustment among the vehicles involved.

\clearpage

%All vehicles not part of any such identified group are considered to belong to the \textit{remains} in the leader-follower network.

\subsection{Applied PCUs}
\label{ssec:PCUmethod}

%Sharma ら\cite{sharma2021estimation}が概観しているように、PCU を推定する方法はさまざまに提案されている。
%しかし、異種混合・無秩序交通においては、どの1つの PCU 定義だけでも交通の異質性のすべてを表現できるわけではない。というのも、手法ごとに、車種間の「等価性」の異なる側面を重視しているためである。
%本研究では、{単一の手法に依存するのではなく、概念的基盤の異なる3つの PCU 推定法を相補的に用いることにした。
%具体的には、(1) Indian Highway Capacity Manual\cite{chandra2017indian} に記載された補間法（以下 IP）、(2) 重回帰法（MLR）\cite{sharma2021estimation}、(3) 各車種の投影面積と速度に基づく方法（以下 SB）\cite{Chandra1995DYNAMICPA} を用いる。
%IP 法は、インドの混合交通を対象とした highway-capacity 研究に根ざした実務的な基準を与え、観測された車種構成比に応じた等価性を反映する\cite{chandra2017indian}。MLR 法は、各車種が passenger car の速度低下にどのように寄与するかという観点から operational equivalence を表現するものであり、traffic situation に依存する相互作用を表すのに有用である\cite{sharma2021estimation}。SB 法は、物理的な道路占有空間と速度という観点から等価性を表現する\cite{Chandra1995DYNAMICPA}。これら3手法による結果を比較することで、本研究では、group proportion と flow--density structure の関係が異なる vehicle equivalence の概念の下でも頑健に観察されるかを評価するとともに、採用する PCU 概念によって解釈がどのように変化するかを明らかにすることを目指す。
%本節では、以上の各 PCU 推定法の概要を述べる。
Various methods for estimating PCU have been proposed, as reviewed by Sharma et al.~\cite{sharma2021estimation}. 
In heterogeneous disordered traffic, however, no single PCU definition can fully represent all aspects of traffic heterogeneity, because different methods emphasize different dimensions of equivalence among vehicle types.
In this study, we therefore selected three PCU estimation methods with complementary conceptual bases, rather than relying on a single method.
Specifically, we use (1) the interpolation-based method described in the Indian Highway Capacity Manual~\cite{chandra2017indian} (hereinafter referred to as IP), (2) the multiple linear regression method (MLR)~\cite{sharma2021estimation}, and (3) the method based on the projected area and speed of each vehicle type (hereinafter referred to as Speed-based, SB)~\cite{Chandra1995DYNAMICPA}. 
The IP method provides a practical benchmark rooted in highway-capacity studies of Indian mixed traffic and reflects composition-dependent equivalence based on observed vehicle proportions~\cite{chandra2017indian}. The MLR method captures operational equivalence in terms of how each vehicle type contributes to the reduction of passenger-car speed, and is therefore useful for representing traffic-situation-dependent interactions~\cite{sharma2021estimation}. The SB method reflects equivalence from the viewpoint of physical road-space occupancy combined with speed~\cite{Chandra1995DYNAMICPA}. By comparing results obtained from these three methods, we aim to evaluate whether the observed relationship between group proportion and flow--density structure is robust across different notions of vehicle equivalence, while also clarifying how the interpretation may vary depending on the PCU concept adopted.
In this section, we provide an overview of each PCU estimation method.

\subsubsection{IP method}
\label{subsubsec:IP}
%IP法は, ある車種が交通に含まれる占有率$x$\%から, PCU値$y$\,[PCU] を内挿によって求める推定法である. 
The IP method estimates the PCU value $y$\,(PCU) of a given vehicle type by interpolating from its proportion $x$\% in the traffic flow.

%内挿を求める際に, 用いられるパラメータを表\ref{table-Indian}に示す\cite{chandra2017indian}. $x_1$は, 各車種が交通で観測されるであろう占有率の下限値であり, ${x_2}$は上限値である. また, 
%${y_1}$は各車種の占有率が${x_1}$\%であった際のPCU推定値を指し, ${y_2}$は各車種の占有率が${x_2}$\%であった際のPCU推定値を指す.
%本研究で分類する車種｛m, r, c, h｝はそれぞれ表\ref{table-Indian}中の, 二輪車（TW）, オートリクシャ（Auto）, 乗用車（SC）, トラック（TAT）として扱った.
The parameters used for interpolation are shown in Table~\ref{table-Indian}~\cite{chandra2017indian}. $x_1$ represents the lower bound of the expected proportion of each vehicle type in traffic, and $x_2$ represents the upper bound. Similarly, $y_1$ denotes the estimated PCU value when the proportion of a given vehicle type is $x_1$\%, and $y_2$ denotes the estimated PCU value when the proportion is $x_2$\%.
In this study, the vehicle types \{m, r, c, h\} are treated as two-wheelers (TW), auto-rickshaws (Auto), small cars (SC), and trucks (TAT), respectively, as defined in Table~\ref{table-Indian}.

\begin{table}[h]
\centering
\begin{tabular}{c|c|c|c|c}
Vehicle Type& ${y_1}$ & ${y_2} $ & ${x_1}$ & ${x_2}$\\ \hline\hline
Standard Car(SC) & 1.0 & 1.0 & 6 & 30 \\
Big Car(BC) & 1.1 & 2.5 & 5 & 16 \\
Motorized Two Wheeler(TW) & 0.2 & 0.5 & 17 & 64 \\
Auto rickshaw(Auto) & 1.1 & 2.0 & 5 & 19 \\
Bus(B) & 2.8 & 4.8 & 5 & 10 \\
Light Commercial Vehicle(LCV) & 2.0 & 5.0 & 2 & 18 \\
Two /Three Axle Truck(TAT) & 3.0 & 5.5 & 5 & 20 \\
Multi Axle Truck / Vehicle(MAT) & 4.6 & 14.6 & 2 & 11 \\
Tractor Trailer(TT) & 5.0 & 8.0 & 2 & 5 \\ \hline 
\end{tabular}
\caption{Parameters utilized in the IP method~\cite{chandra2017indian}}
\label{table-Indian}
\end{table}

%さらに、1ビデオFrameごとの各車種の占有率$x$\%に基づいて内挿によってPCUを推定する式を式\ref{eq_PCU_est}に示す.
Furthermore, the interpolation formula used to estimate the PCU value based on the proportion $x$\% of each vehicle type in each video frame is shown in Equation~\ref{eq_PCU_est}.

\begin{equation}\label{eq_PCU_est}
y = y_1 + \biggl(\frac{y_2 -y_1}{x_2 -x_1}\biggr) \times {(x -x_1)} 
\end{equation}
where \\
$y$ = the estimated PCU value of a certain vehicle type, \\
$x$ = the proportion of a certain vehicle type in the traffic, \\
$x_1$ = lower boundary of proportion of a certain vehicle type, \\
$x_2$ = upper boundary of proportion of a certain vehicle type, \\
$y_1$ = lower boundary of PCU value of a certain vehicle type, \\
$y_2$ = upper boundary of PCU value of a certain vehicle type. \\

%なお, 車種占有率が下限値を下回った場合は${y_1}$をそのフレームにおける車種のPCU推定値とした. また, 上限値を上回った場合は${y_2}$をそのフレームにおける車種のPCU推定値とした. またIP法によるPCU推定で用いられる車種占有率は台ベースで計算される。
If the vehicle type proportion in a frame falls below the lower bound $x_1$, the estimated PCU value for that frame is set to $y_1$. Conversely, if the proportion exceeds the upper bound $x_2$, the estimated PCU value is set to $y_2$. Note that in the IP method, vehicle type proportions are calculated based on the number of vehicles (i.e., vehicle count-based).

\subsubsection{MLR method}
\label{subsubsec:PCUMLR}
%MLR method~\cite{sharma2021estimation}ではまず, 乗用車の平均速度と各車種の交通量から, 式\ref{eq_mlr_est}を用いて各車種に対する回帰係数$a_j$を重回帰分析により求める.
In the MLR method~\cite{sharma2021estimation}, the regression coefficient $a_j$ for each vehicle type is first obtained through multiple linear regression analysis using the average speed of passenger cars and the traffic flow of each vehicle type, as expressed in Equation~\ref{eq_mlr_est}.

\begin{equation}\label{eq_mlr_est}
V_{\mathrm{c}} = a_{\mathrm{0}} + \sum_{j = \mathrm{m, r, c, h}}a_j Q_j 
\end{equation}
where \\
$V_{\mathrm{c}}$: average speed of passenger cars (km/h), \\
$Q_j$: flow of vehicle type $j$ (vehicle/h), \\
$j$: index representing vehicle types. $j=\mathrm{m, r, c, h}$ for motorcycles, auto rickshaws, passenger cars, and heavy vehicles, respectively, \\
$a_{\mathrm{0}} $= intercept, \\
$a_j$ = regression coefficients for vehicle type $j$,\\
%$a_{\mathrm{c}}$ = regression coefficient for passenger cars.\\

%一般に交通では交通量が増加すると車両の速度は減少する. 回帰係数$a_j$は車種$j$が交通の中で乗用車をどの程度減速させるのに寄与しているかを表している. この回帰分析では$a_0$および$a_j$の合計5つの値を決定するが, 先述の回帰係数の意味を踏まえると係数$a_j$は負の値になることが想定されている. そして, $a_j$の絶対値の大きさが大きいほど, 車種$j$の車両が乗用車および他車両に対して与える影響が大きいということを示す. 
In general, as traffic flow increases, vehicle speed tends to decrease. The regression coefficient $a_j$ represents the extent to which vehicle type $j$ contributes to the reduction in the speed of passenger cars in traffic. This regression analysis determines a total of five parameters, including the intercept $a_0$ and the four coefficients $a_j$. Given the interpretation of these coefficients, each $a_j$ is expected to be negative. Moreover, the larger the absolute value of $a_j$, the greater the impact that vehicle type $j$ has on the speed of passenger cars and other vehicles in the traffic flow.

%最終的に各車種のPCU値は, 回帰分析で求めた回帰係数から以下の式\ref{mlr_pcu}で算出される.
Finally, the PCU value for each vehicle type is calculated from the regression coefficients obtained through the regression analysis, using Equation~\ref{mlr_pcu}.

\begin{equation}\label{mlr_pcu}
PCU_j = \frac{a_j}{a_{\mathrm{c}}}
\end{equation}

%この手法では, 着目しているFrameにおいて乗用車の平均速度$V_{\mathrm{c}}$を求める必要がある. 本論文では淺井\cite{Asai}の研究で用いられていた手法を採用している. 具体的な計算方法としては, まず通過した乗用車1台毎に対して, 1Frame毎の速度を計算し, その乗用車が観測された総Frame数で平均をとることで, 着目しているFrameの前後16.5秒以内の計測時間での1台毎の平均速度を求めることができる. この平均速度を計測時間に観測されたすべての乗用車に対して平均をとり, 計測時間内のすべての乗用車の平均速度を求めることができる. この値を$V_{\mathrm{c}}$とした. 

%また, 観測時間の前後16.5秒間の間に観測範囲に出入りしてきた車両の速度に関しては, 観測可能な時間の範囲のデータを用いて速度の算出を行った.

%次に, それぞれの車種の交通量$Q_j$については, 着目しているFrameの前後16.5秒で観測領域を通過した車両の台数を1ms毎に除算し, この値を60000倍することによって1時間あたりの交通量とした.

%上記のように求めた乗用車の平均速度$V_{\mathrm{c}}$とそれぞれの車種の交通量$Q_j$から, Scikit-learnのライブラリにある線形重回帰モデルを用いて, 重回帰分析を行った.なお, Scikit-learnのモデルでは, 最小二乗法を用いて回帰係数を決定する\cite{MLR}.

%また, このMLR methodでは交通量及び速度を用いるため, 交通の状態によって回帰係数が異なる可能性が考えられる. 本研究では交通状態を\ref{subsubsec:DetectionP}節で示した4状態でデータを分けて回帰分析を行い, それぞれの回帰係数およびPCU値を得た.
Moreover, since the MLR method uses both traffic flow and speed, the regression coefficients may vary depending on the traffic situation. In this study, we divided the data into the four traffic situations defined in Section~\ref{ssec:trafSitu}, and performed regression analysis separately for each situation to obtain the corresponding regression coefficients and PCU values.
Calculated PCU values are listed in Table~\ref{tab:MLR_PCUvalue}.
To assess the statistical support for the regression-based PCU estimates, we additionally examined the significance of the regression coefficients in Eq.~(2). All regression coefficients used to derive the MLR-based PCU values were statistically significant at the 5\% level, and most were significant at the 0.1\% level. The corresponding significance levels are indicated in Table~\ref{tab:MLR_PCUvalue}.

%（和訳）なおここでのMLR-PCUは，車両の物理的な大きさや占有空間を表す指標ではなく，
%（和訳）乗用車速度に対する各車種流量の「周辺効果（marginal effect）」の比として定義される運用的な換算係数である。
%（和訳）そのため，交通状況によっては PCU が 1 を超えるなど，車両の直感的な大小関係と一致しない値が得られうる。
%（和訳）また，係数比に基づく推定は不安定になり得るため，本研究ではIP法およびSB法も併用し，手法依存性を低減して結果の頑健性を確認する。
Note that the MLR-based PCU in this study is an operational equivalence factor defined as a ratio of marginal effects on passenger-car speed, rather than a measure of physical size or space occupancy. Therefore, depending on the traffic situation, the estimated PCU values may exceed one or deviate from intuitive size-based expectations. In addition, because coefficient-ratio-based estimates can be unstable, we also employ the IP and SB methods to reduce method dependence and to check the robustness of the findings.

\begin{table}[t]
\centering
\caption{MLR-based PCU values for respective vehicle types in respective traffic situations. Significance markers refer to the p-values of the underlying regression coefficients used to calculate the PCU values, not to the PCU values themselves. $^{*}p<0.05$, $^{**}p<0.01$, $^{***}p<0.001$.}
\label{tab:MLR_PCUvalue}
\begin{tabular}{lcccc}
\hline
Traffic situation & c & m & r & h \\
\hline
AF & 1.00$^{**}$ & 5.17$^{***}$ & 2.77$^{***}$ & 1.35$^{*}$ \\
DJ & 1.00$^{***}$ & 2.10$^{***}$ & 1.16$^{***}$ & 4.44$^{***}$ \\
AJ & 1.00$^{***}$ & 0.62$^{***}$ & 0.72$^{***}$ & 1.41$^{***}$ \\
OT & 1.00$^{***}$ & 0.16$^{*}$ & 1.32$^{***}$ & 0.63$^{*}$ \\
\hline
\end{tabular}
\end{table}

%\subsubsection{PCU estimation based on vehicle projected area and speed}
\subsubsection{SB method}
\label{subsubsec:PCUSB}
%Speed Based \cite{sharma2021estimation}は, 各車種の投影面積及び速度に着目して, 式\ref{eq_SB_est}に示すような計算でPCUを推定する方法である. 
The SB method~\cite{sharma2021estimation} estimates the PCU based on the projected area and speed of each vehicle type, using the calculation shown in Equation~\ref{eq_SB_est}.
\begin{equation}\label{eq_SB_est}
PCU_j = \frac{V_{\mathrm{c}}/V_{j}}{A_{\mathrm{c}}/A_{j}}
\end{equation}
%ただし, \\
%$V_{j} , V_{\mathrm{c}} $ = 車種$j$及び乗用車の平均速度 \\
%$A_{j} , A_{\mathrm{c}} $ = 道路における車種$j$及び乗用車の投影面積\\
%である. 
where\\
$V_{j}, V_{\mathrm{c}}$ = average speed of vehicle type $j$ and passenger cars, respectively\\
$A_{j}, A_{\mathrm{c}}$ = projected area of vehicle type $j$ and passenger cars on the road, respectively

%式\ref{eq_SB_est}の分子は乗用車の平均速度に対する推定する車種の平均速度の比として表しており, 分母は乗用車の投影面積に対する推定する車種の投影面積の平均値を表している.
%投影面積を求める際, ビデオ画像座標系の各車種の車長は, ホモグラフィー行列を用いて実世界座標に変換した. また各車種の幅は, 乗用車は1.7\,m,大型車は2.5\,m,二輪車は0.8\,m, そしてオートリキシャは1.3\,mと一律に設定した\cite{car,motorcycle,large,rikisho}. 観測ビデオは, 道路を斜め上方向から撮影しており, 車幅が正確に測定できないためである. 
%あるFrameにおける各車種の投影面積は, Frameのなかに含まれる各車種の投影面積の総和から, 1台当たりの投影面積を算出し, その値を各車種の投影面積とした.
The numerator in Equation~\ref{eq_SB_est} represents the ratio of the average speed of passenger cars to that of the target vehicle type, while the denominator represents the ratio of the projected area of passenger cars to that of the target vehicle type.
To calculate the projected area, the length of each vehicle in the image coordinate system was converted to real-world coordinates using a homography matrix. The vehicle widths were uniformly set to 1.7\,m for passenger cars, 2.5\,m for heavy vehicles, 0.8\,m for two-wheelers, and 1.3\,m for auto-rickshaws, since the video was recorded from an oblique angle above the road, making it difficult to accurately measure vehicle widths.
The projected area of a vehicle is calculated as the vehicle width multiplied by the length.
For each frame, the projected area of each vehicle type was calculated by averaging the total projected area of all vehicles of that type in the frame, and this average was used as the representative projected area of that vehicle type.

%各車種の平均速度は, \ref{subsubsec:PCUMLR}項に記した, 着目Frameの前後16.5秒の速度データを用いる手法を転用した. \ref{subsubsec:PCUMLR}項では乗用車の平均速度を求めたが, 車種を変化させて同じ操作を行い, 各車種の平均速度$V_j$を求めた. 

%%%%%%
%以降のresult sectionでは、以上のPCUに基づいて密度ならびに流量を求め基本図を描く。その際交通に閉めるgroup割合を（台ベースもしくはPCUベースで）計算し、それぞれの割合に応じて基本図上で区別がつくように描くことで、\ref{sec:result}節以降の解析を進める
In the subsequent sections, we calculate traffic density and flow based on the PCU values described above and construct fundamental diagrams. The proportion of groups in the traffic (either on a vehicle-count basis or a PCU-weighted basis) is calculated for each frame, and the data points are distinguished accordingly in the diagrams. This approach forms the basis for the analyses presented in Section~\ref{sec:result} and beyond.

%%先結果をかく20250403%
\section{Effect of Group Proportion on the Flow-Density Structure}
\label{sec:result}
To investigate how the proportion of groups in traffic affects macroscopic traffic behavior, we analyze the relationship between flow and density under varying group proportions. Two definitions of group proportion are considered: the vehicle-count-based proportion, which reflects the number of vehicles, and the PCU-based proportion, which accounts for the PCU of each vehicle type. 
For example, let us assume that the PCU of a motorcycle is 0.5, while that of a car is 1.0. In the traffic situation depicted at the top of Figure~\ref{fig:groupPonchi}(a), the PCU-based group proportion is $1.0/3.0$ because the total PCU of all vehicles in the traffic is 3.0, and the group itself accounts for 1.0 PCU. Meanwhile, the vehicle-count-based group proportion is $2/5$, as there are five vehicles in total, two of which belong to the group.
By comparing flow-density relationships across these proportions, we aim to reveal how the presence of groups influences traffic characteristics in terms of flow contribution.

This section presents two complementary visualizations. First, a smoothed analysis using segment-wise medians highlights general trends across group proportions. Then, a more detailed scatter plot visualization exposes local variations and distributional differences within each group proportion. Together, these analyses provide both a macroscopic overview and a microscopic perspective on how group composition affects traffic flow dynamics.
In addition, we examine whether the degree of vehicle-type mixing within a group is related to the flow performance.
To convey the temporal variation of the group proportion more intuitively, its time series over the entire observation period is presented in Appendix~\ref{app:timeseries}.

\subsection{Overview of Flow-Density Structure via Segment-Wise Medians}
\label{subsec:median}
\subsubsection{Result}
\label{subsubsec:medianResult}

\begin{figure}[h]
\begin{center}
\subfigure[Groups $\lambda^{\mathrm{O}}_{i}\geq 0.5$~(/frame) are considered.]{
\includegraphics[width=70mm]
{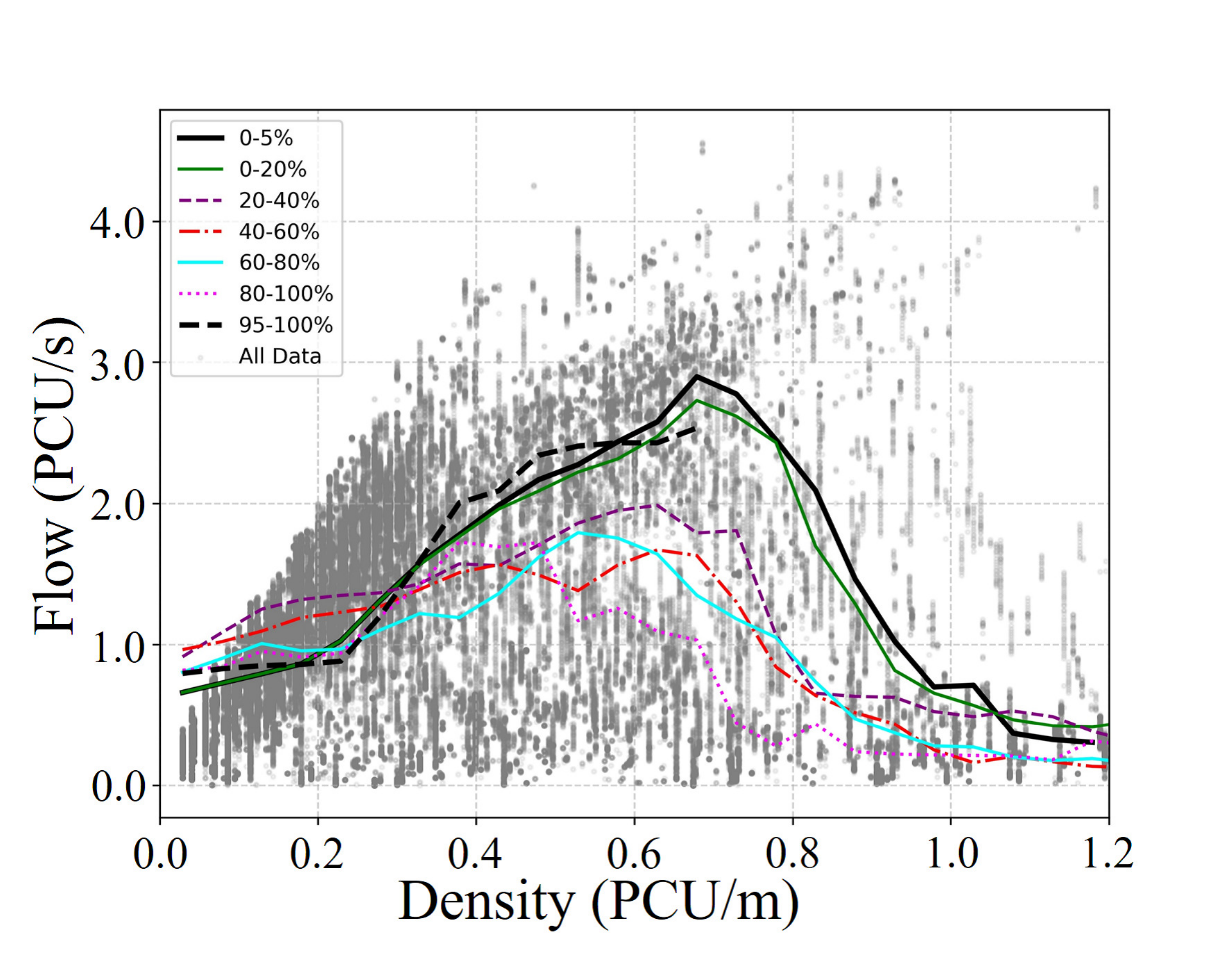}
\label{subfig:median_indo_5e_dai}}
\subfigure[Groups $\lambda^{\mathrm{O}}_{i}\geq 0.1$~(/frame) are considered.]{
\includegraphics[width=70mm]
{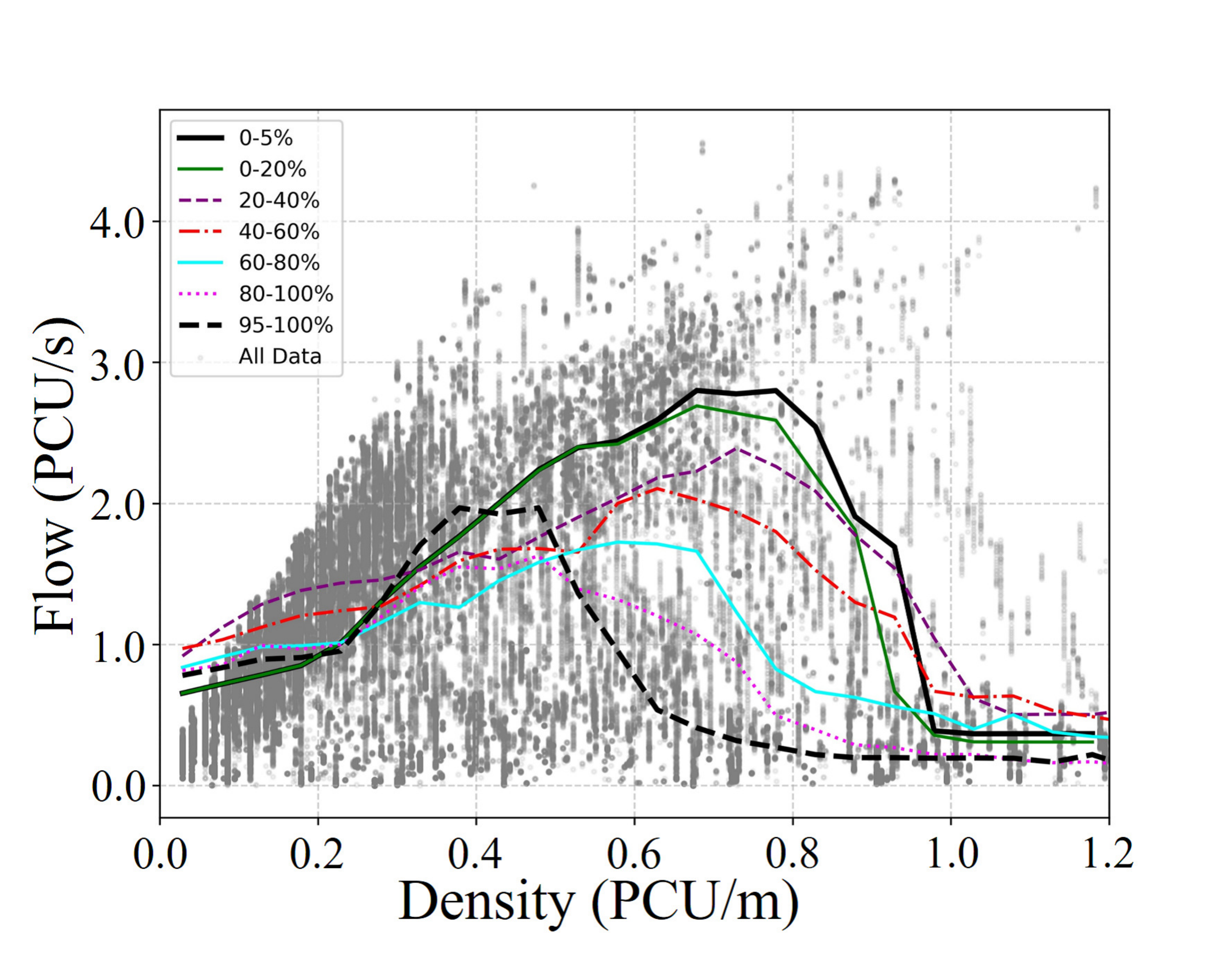}
\label{subfig:median_indo_1e_dai}}
\end{center}
\caption{Medians of flow in PCU (IP method) for respective group proportion by number of vehicles. The percentage indicates the proportion of group vehicles in the traffic flow. $\lambda_i^{\mathrm{O}}$ is the threshold of the occurrence frequency used to detect groups.}
\label{median_indo_dai}
\end{figure}

%図\ref{median_indo_dai}は交通中に含まれるgroupの割合毎に、flow-densityの関係を各種の線で示している。
%背景の灰色のプロットは全てのデータ点を示している。
%flowおよびdensityはIndian HCMに基づくPCUで計算している。
%flowの値は、あるdensityに対して$\pm 0.15$~(PCU/m)以内のflowの値のmedian値を計算した。
%この操作を0.05~(PCU/m)ずつdensityを移動させながらプロットしている。
%なおmedian値をとるwindow内にデータが10より少ない場合はmedian値を計算せずflow-densityの線を描いていない。
%また、図\ref{subfig:median_indo_5e_dai}は特に出現頻度$\lambda^{\mathrm{O}}_{i}$が0.5~(/frame)以上であったgroupに関して、
%図\ref{subfig:median_indo_1e_dai}は出現頻度$\lambda^{\mathrm{O}}_{i}$がより少ない0.1~(/frame)以上であったgroupまで範囲を広げている。
Figure~\ref{median_indo_dai} shows the flow-density relationship for each group proportion in traffic, illustrated with various lines.  
The background gray dots represent all the data points.  
Flow and density are calculated based on the PCU values defined by the IP method.  
For each density value, the flow value shown is the median of flow values within $\pm 0.15$~(PCU/m).  
This computation is repeated by shifting the density window in increments of 0.05~(PCU/m).  
When the number of data points within a window used to compute the median is fewer than 10, the median is not calculated and the corresponding flow–density curve is not plotted.
The various lines represent these median values, and the percentage shown in the legend denotes the proportion of group vehicles in the traffic flow.

Figure~\ref{subfig:median_indo_5e_dai} specifically focuses on groups whose occurrence frequency $\lambda^{\mathrm{O}}_{i}$ was at least 0.5~(/frame),  
while Figure~\ref{subfig:median_indo_1e_dai} expands the range to include groups with lower occurrence frequency, down to 0.1~(/frame).  
Although the statistical detection of groups itself is based on the comparison between $\lambda^{\mathrm{O}}_{i}$ and $\lambda^{\mathrm{U}}_{i}$ and does not require an additional cutoff on $\lambda^{\mathrm{O}}_{i}$, we introduce these two thresholds here for analytical interpretation. The threshold $\lambda^{\mathrm{O}}_{i}\geq 0.5$~(/frame) is intended to capture highly frequent groups, whereas $\lambda^{\mathrm{O}}_{i}\geq 0.1$~(/frame) includes moderately frequent groups as well. By comparing results across these two thresholds, we also assess how sensitive the observed tendencies are to group frequency.
Figure~\ref{subfig:median_indo_5e_dai} therefore shows the flow–density characteristics of vehicle groups that appear relatively frequently and are statistically concluded to be likely to form leader–follower relationships, whereas Figure~\ref{subfig:median_indo_1e_dai} also includes less frequent but still statistically supported groups.

%図\ref{subfig:median_indo_5e_dai}からは以下の\label{median_char1}から\label{median_char3}の特徴が見て取れる。
%図\ref{subfig:median_indo_1e_dai}からは\label{median_char4}の特徴も見ることができる。
%\begin{enumerate}
%\item Group割合が特に小さいときすべての流量中最大の流量を実現する\label{median_char1}
%\item Group割合が特に大きいときすべての流量中、比較的高い流量を実現する\label{median_char2}
%\item 中程度の密度では、Group割合が高くなるに従い流量が下がる\label{median_char3}
%\item 密度の特に高い領域では、ある程度group割合がある、20-40%もしくは40-60%の流量が比較的大きい\label{median_char4}
%\end{enumerate}
From Figure~\ref{subfig:median_indo_5e_dai}, we can observe the following characteristics, labeled from (\ref{median_char1}) to (\ref{median_char3}).  
Additionally, Figure~\ref{subfig:median_indo_1e_dai} reveals the characteristic described in (\ref{median_char4}).
\begin{enumerate}
\renewcommand{\labelenumi}{(\arabic{enumi})}
\item When the group proportion is particularly small, the highest median flow among all conditions is observed. \label{median_char1}
\item When the group proportion is particularly large, relatively high flow values are observed across all conditions. \label{median_char2}
\item In the medium density range, the flow decreases as the group proportion increases. \label{median_char3}
\item In the high-density range, relatively large flow values are observed when the group proportion is moderate (20–40\% or 40–60\%). \label{median_char4}
\end{enumerate}

\clearpage

\begin{figure}[h]
\begin{center}
\subfigure[Groups $\lambda^{\mathrm{O}}_{i}\geq 0.5$~(/frame) are considered.]{
\includegraphics[width=70mm]
{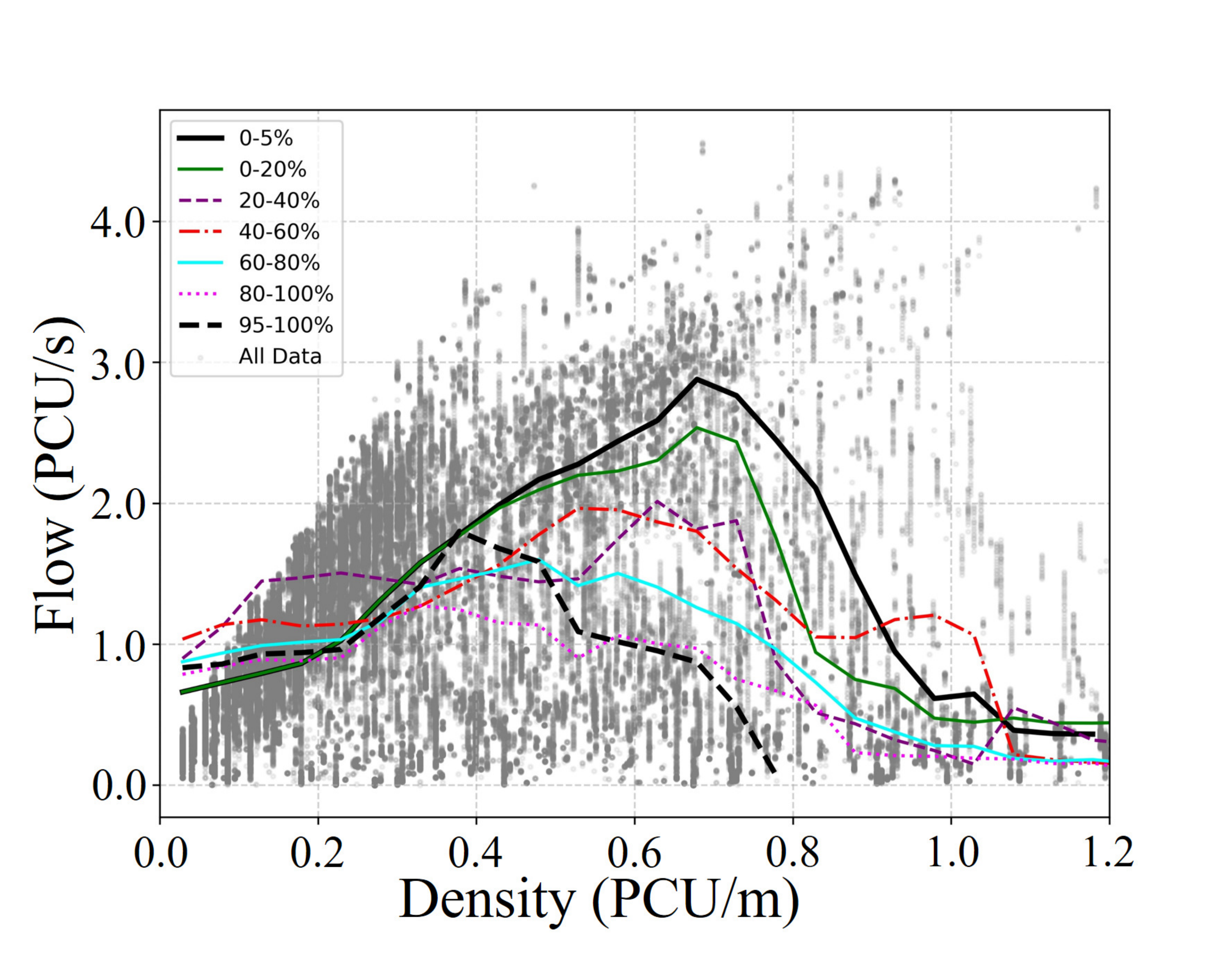}
\label{subfig:median_indo_5e_PCU}}
\subfigure[Groups $\lambda^{\mathrm{O}}_{i}\geq 0.1$~(/frame) are considered.]{
\includegraphics[width=70mm]
{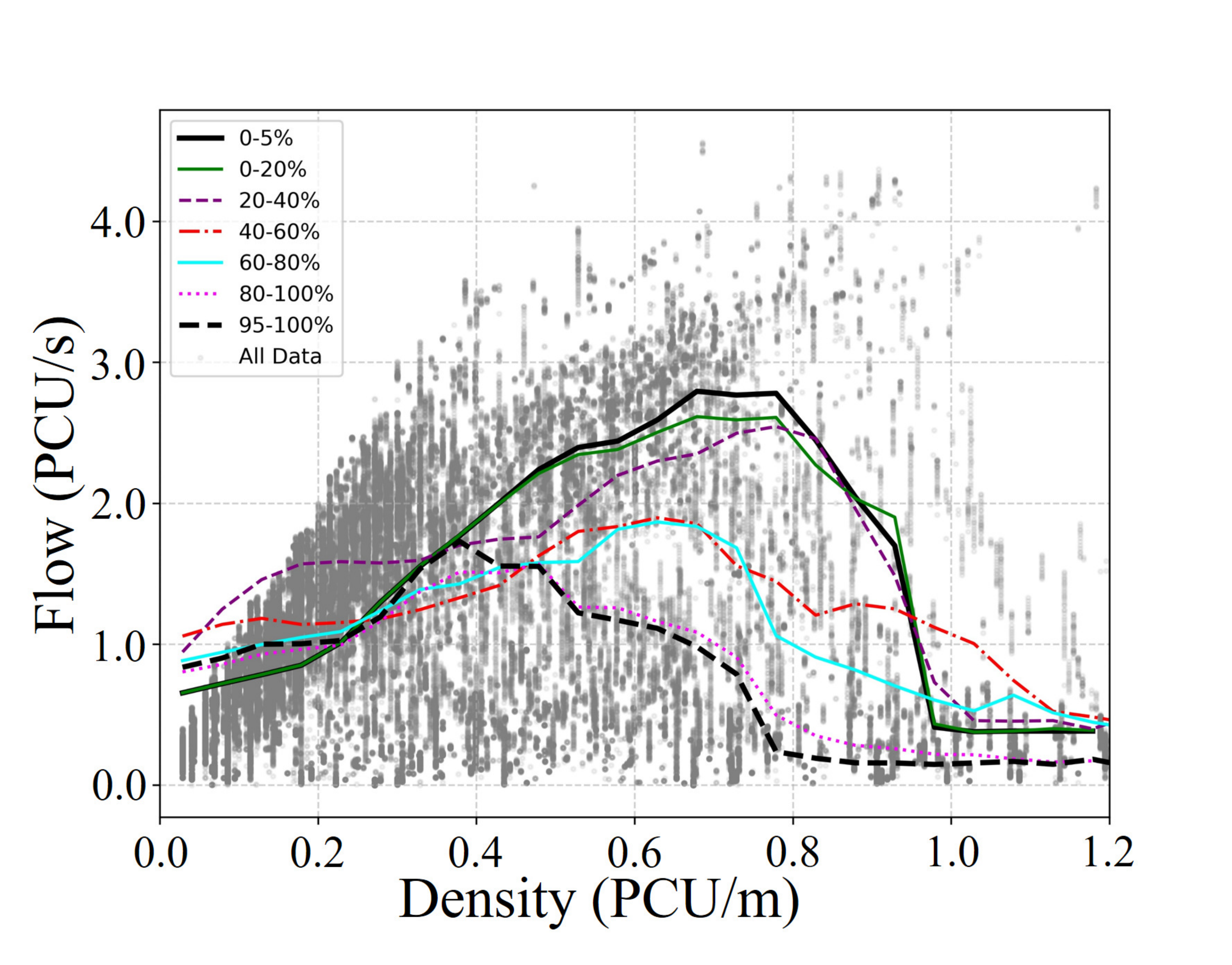}
\label{subfig:median_indo_1e_PCU}}
\end{center}
\caption{Medians of flow in PCU (IP method) for respective group proportion by PCU values.}
\label{median_indo_PCU}
\end{figure}

%図\ref{median_indo_PCU}は先に示した図\ref{median_indo_dai}のgroup割合をPCUで計算したものである。
Figure~\ref{median_indo_PCU} presents a counterpart to Figure~\ref{median_indo_dai} shown earlier, with group proportions computed based on PCU values.

\clearpage

\begin{figure}[h]
\begin{center}
\subfigure[Groups $\lambda^{\mathrm{O}}_{i}\geq 0.5$~(/frame) are considered.]{
\includegraphics[width=70mm]
{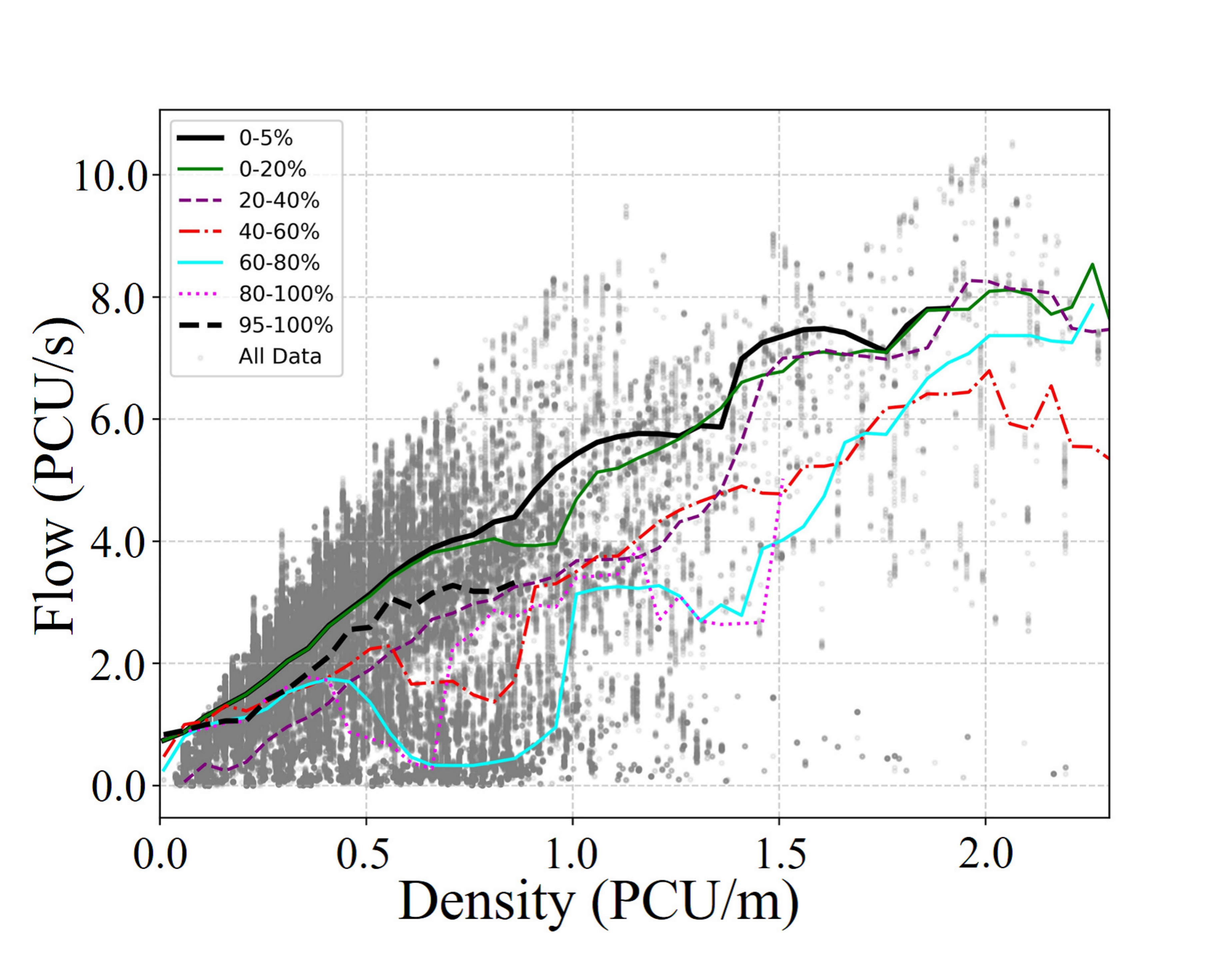}
\label{subfig:median_MLR_5e_dai}}
\subfigure[Groups $\lambda^{\mathrm{O}}_{i}\geq 0.1$~(/frame) are considered.]{
\includegraphics[width=70mm]
{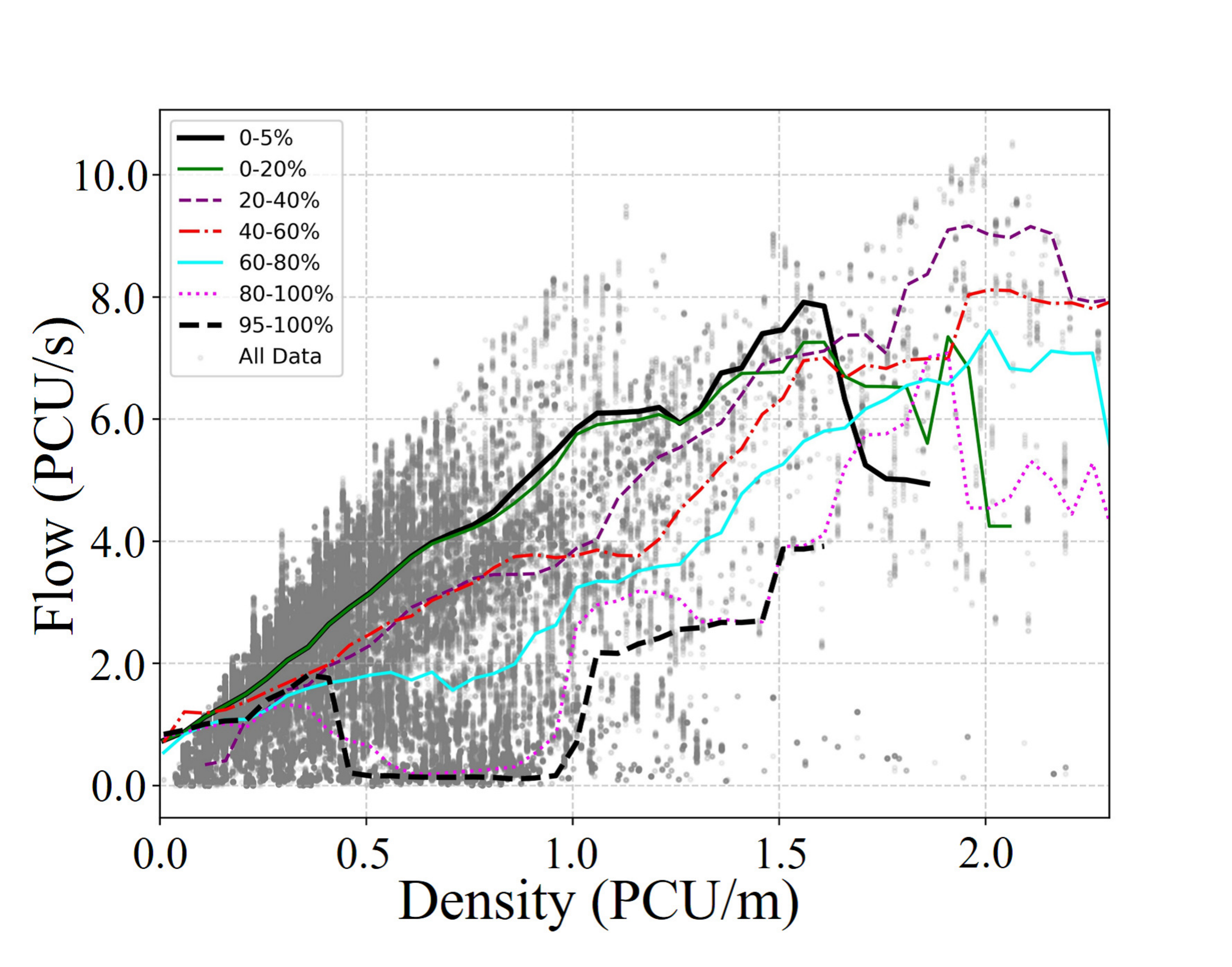}
\label{subfig:median_MLR_1e_dai}}
\end{center}
\caption{Medians of flow in PCU (MLR method) for respective group proportion by number of vehicles.}
\label{median_MLR_dai}
\end{figure}

\begin{figure}[h]
\begin{center}
\subfigure[Groups $\lambda^{\mathrm{O}}_{i}\geq 0.5$~(/frame) are considered.]{
\includegraphics[width=70mm]
{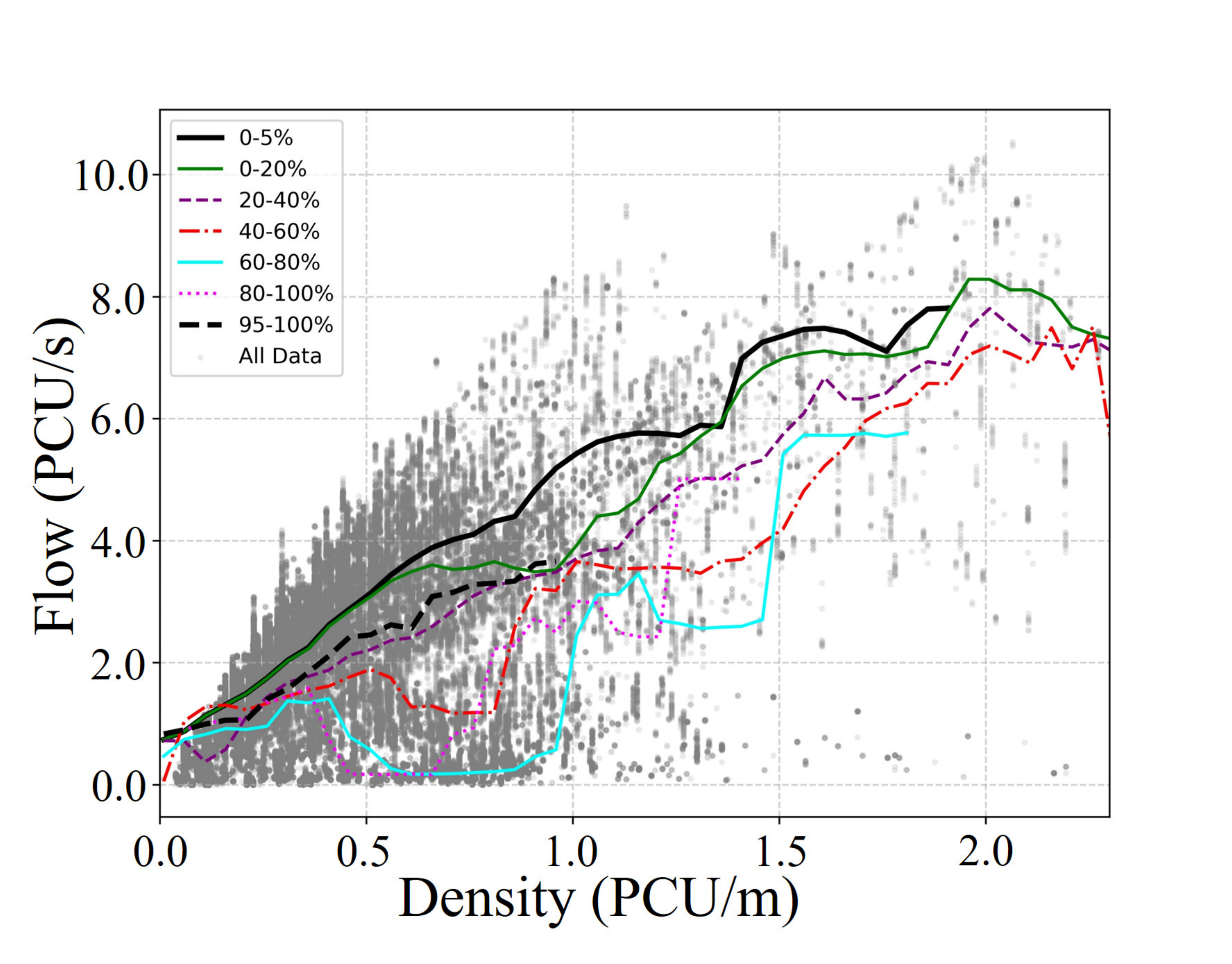}
\label{subfig:median_MLR_5e_PCU}}
\subfigure[Groups $\lambda^{\mathrm{O}}_{i}\geq 0.1$~(/frame) are considered.]{
\includegraphics[width=70mm]
{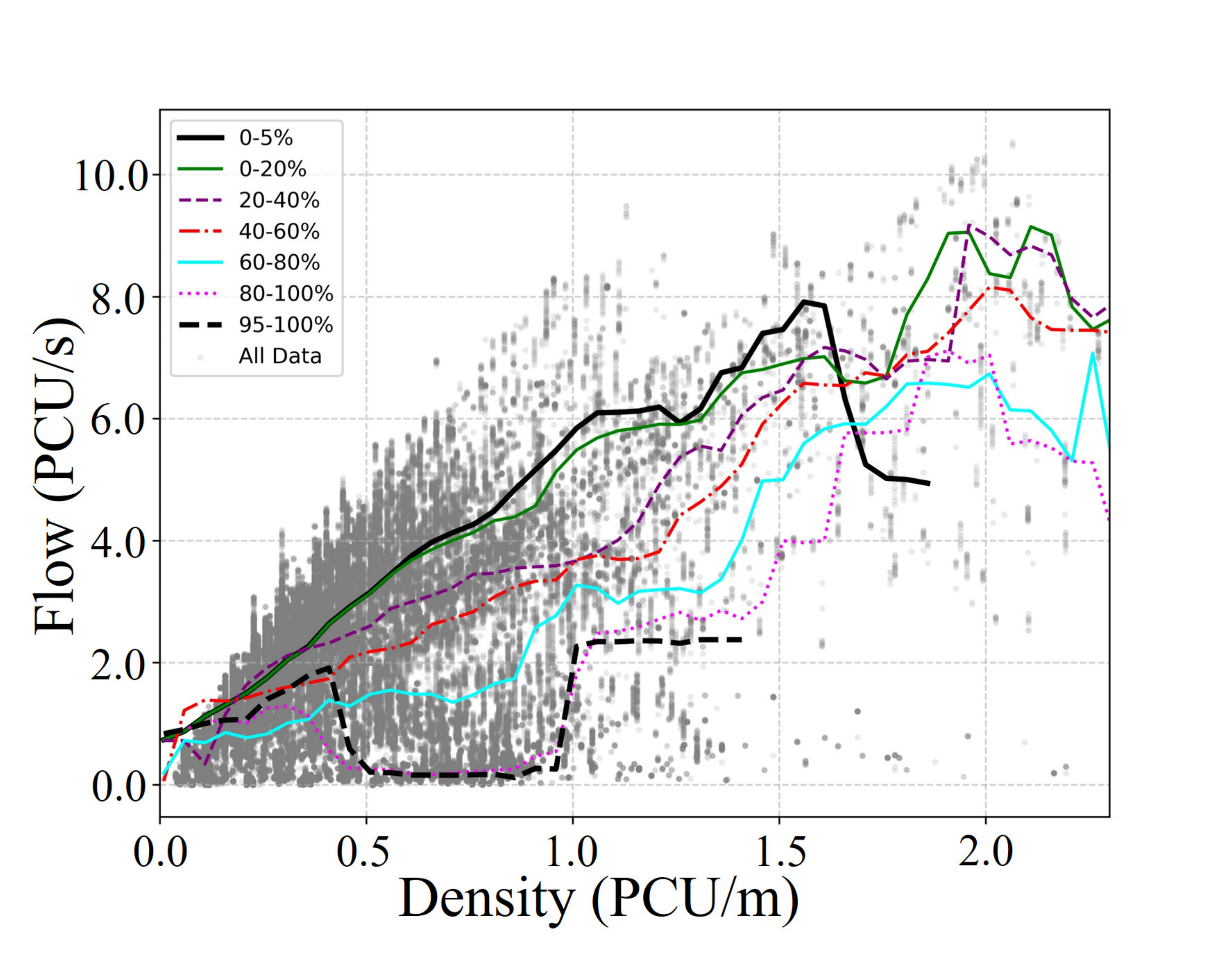}
\label{subfig:median_MLR_1e_PCU}}
\end{center}
\caption{Medians of flow in PCU (MLR method) for respective group proportion by PCU values.}
\label{median_MLR_PCU}
\end{figure}

\begin{figure}[h]
\begin{center}
\subfigure[Groups $\lambda^{\mathrm{O}}_{i}\geq 0.5$~(/frame) are considered.]{
\includegraphics[width=70mm]
{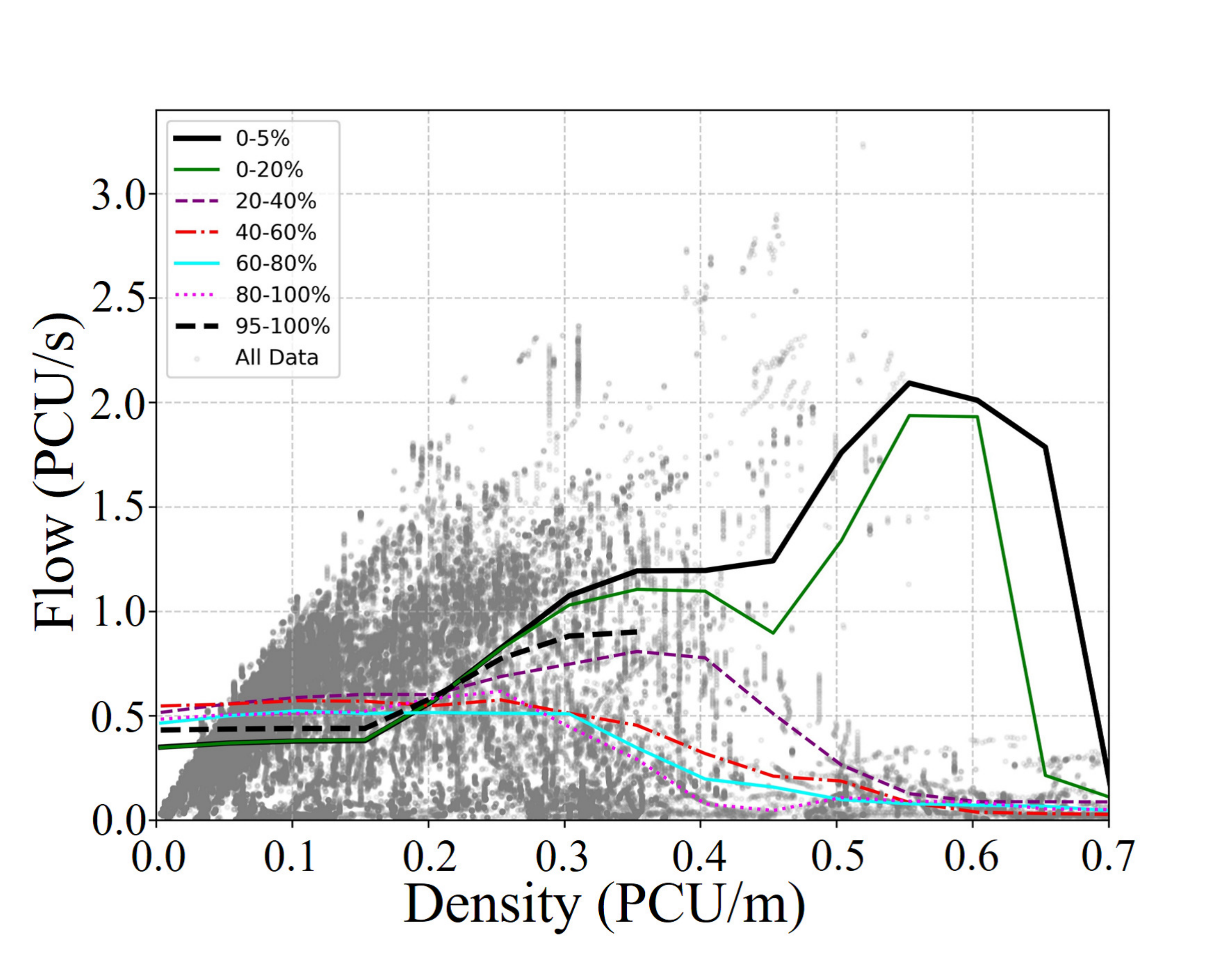}
\label{subfig:median_SB_5e_dai}}
\subfigure[Groups $\lambda^{\mathrm{O}}_{i}\geq 0.1$~(/frame) are considered.]{
\includegraphics[width=70mm]
{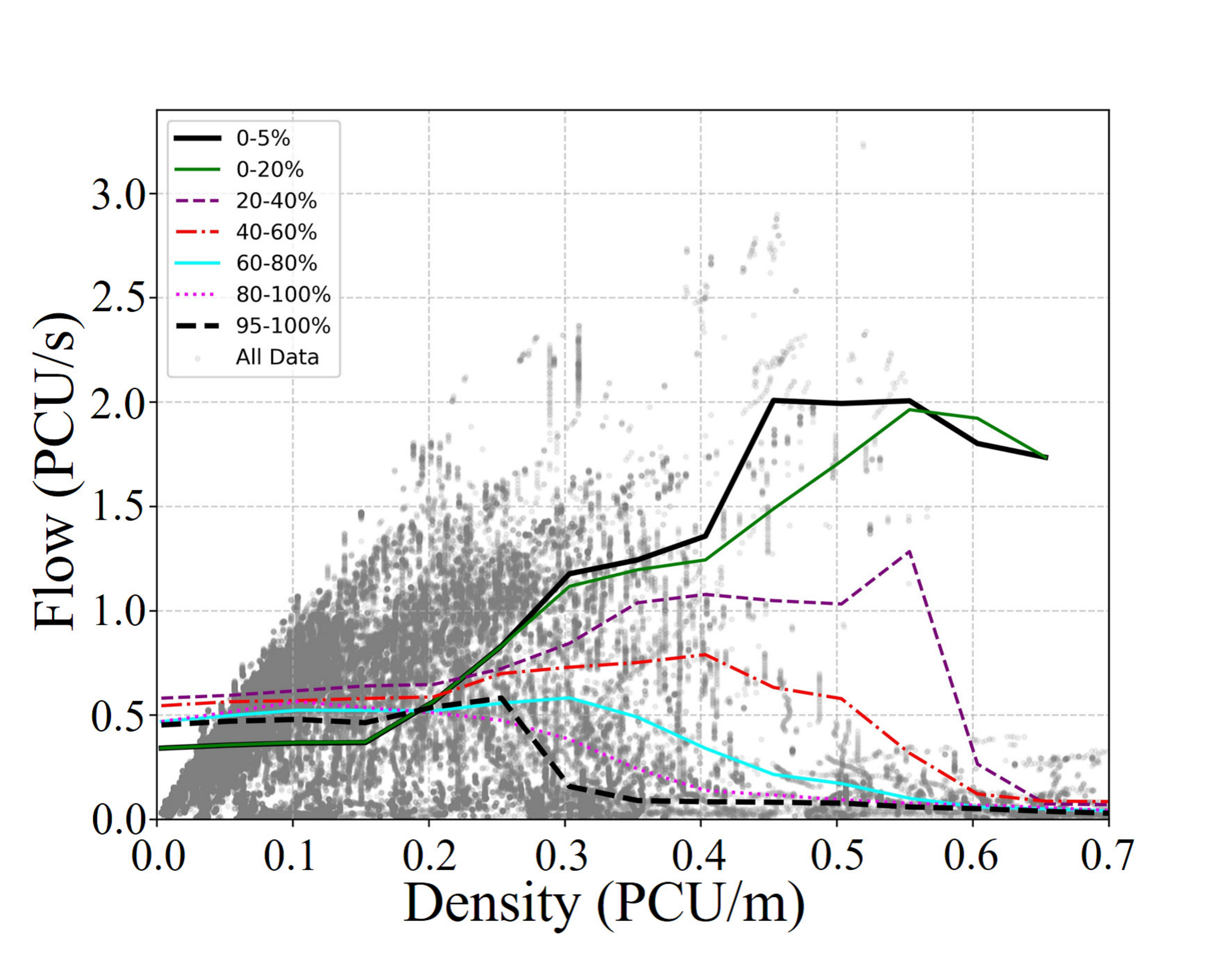}
\label{subfig:median_SB_1e_dai}}
\end{center}
\caption{Medians of flow in PCU (SB method) for respective group proportion by number of vehicles.}
\label{median_SB_dai}
\end{figure}

\begin{figure}[h]
\begin{center}
\subfigure[Groups $\lambda^{\mathrm{O}}_{i}\geq 0.5$~(/frame) are considered.]{
\includegraphics[width=70mm]
{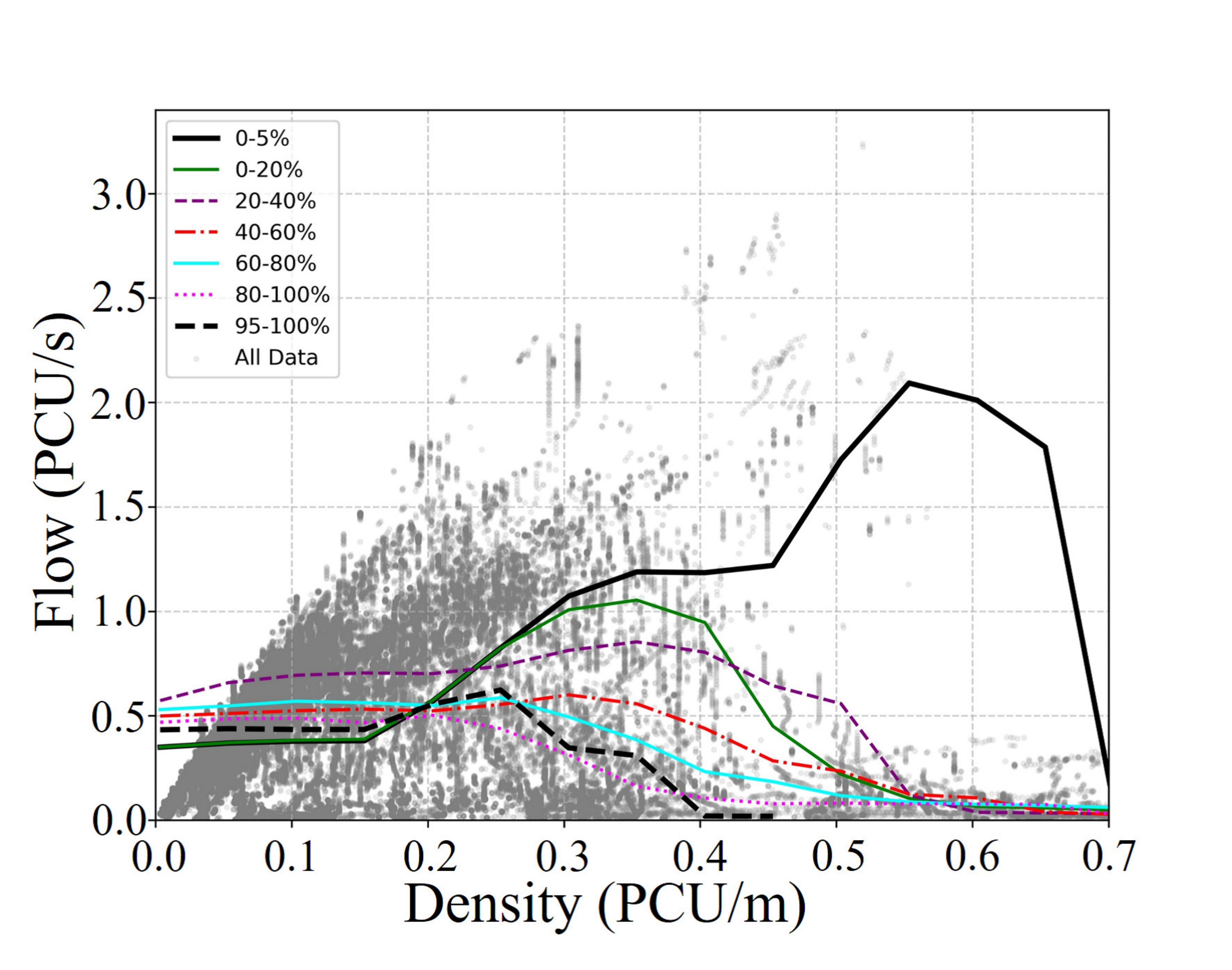}
\label{subfig:median_SB_5e_PCU}}
\subfigure[Groups $\lambda^{\mathrm{O}}_{i}\geq 0.1$~(/frame) are considered.]{
\includegraphics[width=70mm]
{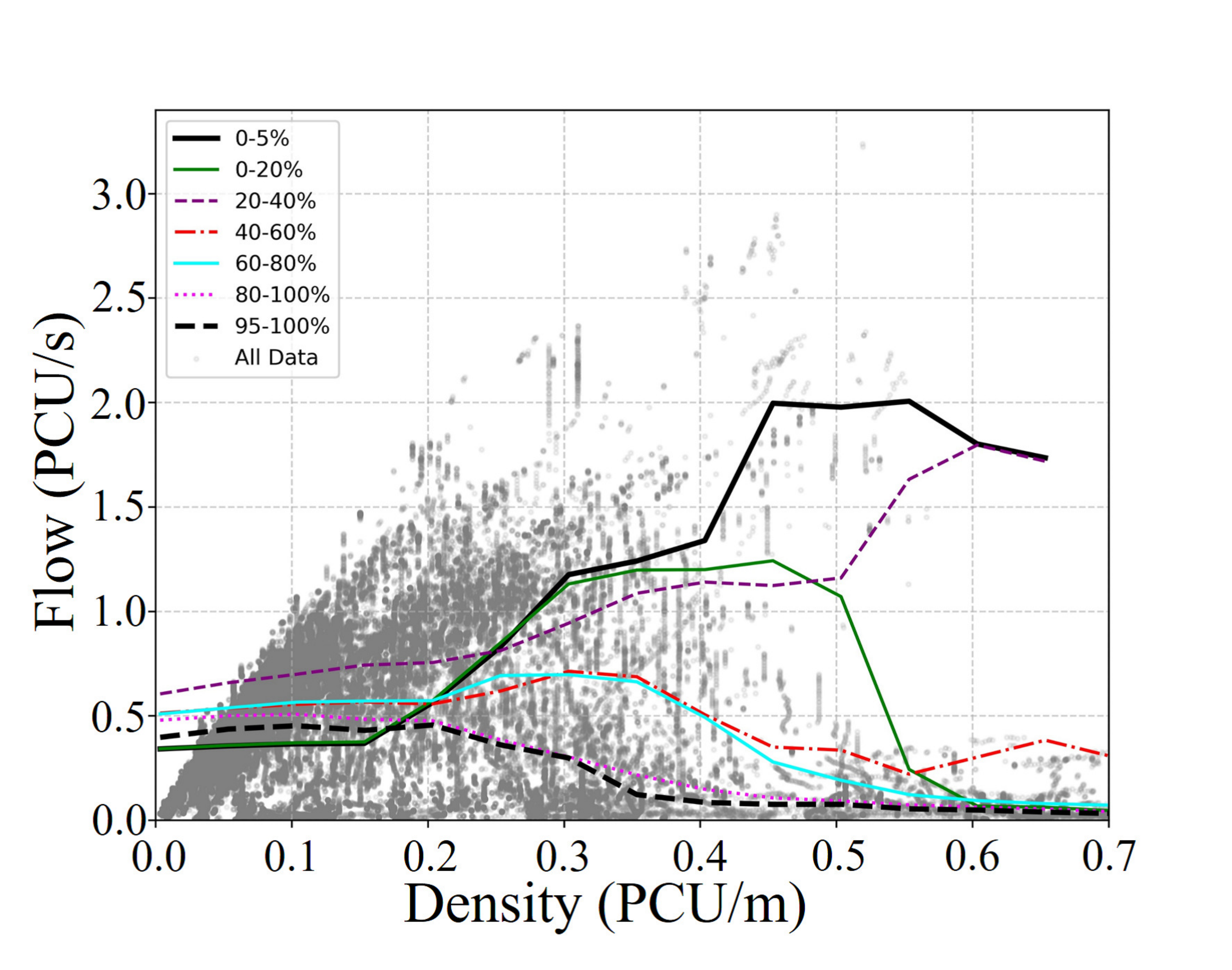}
\label{subfig:median_SB_1e_PCU}}
\end{center}
\caption{Medians of flow in PCU (SB method) for respective group proportion by PCU values.}
\label{median_SB_PCU}
\end{figure}

%さらに図\ref{median_MLR_dai}および図\ref{median_MLR_PCU}にMLR methodによってPCUを計算した場合のflow-densityの関係を
%図\ref{median_SB_dai}および図\ref{median_SB_PCU}にSB methodによってPCUを計算した場合のflow-densityの関係を示す。
Furthermore, Figures~\ref{median_MLR_dai} and~\ref{median_MLR_PCU} show the flow-density relationships when PCU is calculated using the MLR method, and Figures~\ref{median_SB_dai} and~\ref{median_SB_PCU} show those obtained using the SB method.


\clearpage

%図\ref{median_indo_dai}から図~\ref{median_SB_PCU}で確認できる特徴をTable~\ref{tab:group_props_vehicle_5e7}からTable\ref{tab:group_props_pcu_1e7}にまとめた
The characteristics observed in Figures~\ref{median_indo_dai} through~\ref{median_SB_PCU} are summarized in Tables~\ref{tab:group_props_vehicle_5e7} through~\ref{tab:group_props_pcu_1e7}.
In the tables, the symbols indicate the level of support for each characteristic as follows:  
$\bigcirc$~denotes that the characteristic is clearly supported by the data;  
\ding{55}~indicates that the characteristic is not supported;  
$\bigtriangleup$~represents partial or ambiguous support, often due to limited data or variability.  
The superscripts provide additional notes on specific limitations or conditions under which the support is observed.  
The symbol \textsuperscript{-} indicates that the result is confirmed but with some limitations,  
while \textsuperscript{+} indicates that the characteristic is generally unsupported, though minor exceptions may exist.

\begin{table}[h]
\centering
\caption{Flow-density characteristics under vehicle-count-based group proportions ($\lambda^{\mathrm{O}}_{i} \geq 0.5$~(/frame)).}
\label{tab:group_props_vehicle_5e7}
\begin{tabular}{p{9cm} | c c c c}
\hline
\textbf{Feature} & IP & MLR & SB & Consensus \\
\hline
(1) When the group proportion is particularly small, the highest flow is observed among all conditions. 
& $\bigcirc$ & $\bigtriangleup$\textsuperscript{a} & $\bigcirc$ & $\bigcirc$\textsuperscript{-} \\

(2) When the group proportion is particularly large, relatively high flow values are observed. 
& $\bigcirc$ & \ding{55} & \ding{55} & \ding{55}\textsuperscript{+} \\

(3) In the medium-density range, flow decreases as the group proportion increases. 
& $\bigcirc$\textsuperscript{b} & $\bigtriangleup$\textsuperscript{c} & $\bigcirc$\textsuperscript{b} & $\bigcirc$\textsuperscript{-} \\

(4) In the high-density range, relatively large flows are observed for moderate group proportions (20--40\% or 40--60\%). 
& \ding{55} & $\bigtriangleup$\textsuperscript{d} & \ding{55} & \ding{55}\textsuperscript{+} \\
\hline
\end{tabular}
\vspace{0.8em}
\begin{flushleft}
\textbf{Notes:} \\
\textsuperscript{a}~No median data in very high density regions ($> 1.8$~PCU/m). \\
\textsuperscript{b}~Excluding extremely large group proportions (95-100\%). \\
\textsuperscript{c}~The order is not strictly monotonic. \\
\textsuperscript{d}~Based on limited data. \\
\textsuperscript{+}~Majority vote, leaning toward support. \quad
\textsuperscript{-}~Majority vote, leaning toward non-support.
\end{flushleft}
\end{table}

\begin{table}[h]
\centering
\caption{Flow-density characteristics under vehicle-count-based group proportions ($\lambda^{\mathrm{O}}_{i} \geq 0.1$~(/frame)).}
\label{tab:group_props_vehicle_1e7}
\begin{tabular}{p{9cm} | c c c c}
\hline
\textbf{Feature} & IP & MLR & SB & Consensus \\
\hline
(1) When the group proportion is particularly small, the highest flow is observed among all conditions. 
& \ding{55} & \ding{55} & $\bigcirc$ & \ding{55}\textsuperscript{+} \\
(2) When the group proportion is particularly large, relatively high flow values are observed. 
& \ding{55} & \ding{55} & \ding{55} & \ding{55} \\
(3) In the medium-density range, flow decreases as the group proportion increases. 
& $\bigcirc$ & $\bigcirc$ & $\bigcirc$ & $\bigcirc$ \\
(4) In the high-density range, relatively large flows are observed for moderate group proportions (20--40\% or 40--60\%). 
& $\bigtriangleup$\textsuperscript{d} & $\bigtriangleup$\textsuperscript{d} & $\bigtriangleup$\textsuperscript{e} & $\bigtriangleup$ \\
\hline
\end{tabular}

\vspace{0.8em}
\begin{flushleft}
\textbf{Notes:} Same as in Table~\ref{tab:group_props_vehicle_5e7}.\\
\textsuperscript{e}~Observed only in specific density intervals.
\end{flushleft}
\end{table}

\begin{table}[h]
\centering
\caption{Flow-density characteristics under PCU-based group proportions ($\lambda^{\mathrm{O}}_{i} \geq 0.5$~(/frame)).}
\label{tab:group_props_pcu_5e7}
\begin{tabular}{p{9cm} | c c c c}
\hline
\textbf{Feature} & IP & MLR & SB & Consensus \\
\hline
(1) When the group proportion is particularly small, the highest flow is observed among all conditions. 
& $\bigcirc$ & $\bigtriangleup$\textsuperscript{a} & $\bigcirc$ & $\bigcirc$\textsuperscript{-} \\
(2) When the group proportion is particularly large, relatively high flow values are observed. 
& \ding{55} & \ding{55} & \ding{55} & \ding{55} \\
(3) In the medium-density range, flow decreases as the group proportion increases. 
& $\bigcirc$ & $\bigtriangleup$\textsuperscript{c} & $\bigcirc$ & $\bigcirc$\textsuperscript{-} \\
(4) In the high-density range, relatively large flows are observed for moderate group proportions (20--40\% or 40--60\%). 
& $\bigtriangleup$\textsuperscript{e} & $\bigtriangleup$\textsuperscript{d} & $\bigtriangleup$\textsuperscript{e} & $\bigtriangleup$ \\
\hline
\end{tabular}

\vspace{0.8em}
\begin{flushleft}
\textbf{Notes:} \\
\textsuperscript{a}~No median data in very high density regions ($> 1.8$~PCU/m). \\
\textsuperscript{c}~The order is not strictly monotonic. \\
\textsuperscript{d}~Based on limited data. \\
\textsuperscript{e}~Observed only in specific density intervals. \\
\textsuperscript{+}~Majority vote, leaning toward support. \quad
\textsuperscript{-}~Majority vote, leaning toward non-support.
\end{flushleft}
\end{table}

\begin{table}[h]
\centering
\caption{Flow-density characteristics under PCU-based group proportions ($\lambda^{\mathrm{O}}_{i} \geq 0.1$~(/frame)).}
\label{tab:group_props_pcu_1e7}
\begin{tabular}{p{9cm} | c c c c}
\hline
\textbf{Feature} & IP & MLR & SB & Consensus \\
\hline
(1) When the group proportion is particularly small, the highest flow is observed among all conditions. 
& \ding{55} & \ding{55} & $\bigcirc$ & \ding{55}\textsuperscript{+} \\
(2) When the group proportion is particularly large, relatively high flow values are observed. 
& \ding{55} & \ding{55} & \ding{55} & \ding{55} \\
(3) In the medium-density range, flow decreases as the group proportion increases. 
& $\bigcirc$ & $\bigcirc$ & $\bigtriangleup$\textsuperscript{c} & $\bigcirc$\textsuperscript{-} \\
(4) In the high-density range, relatively large flows are observed for moderate group proportions (20--40\% or 40--60\%). 
& $\bigtriangleup$\textsuperscript{d} & $\bigtriangleup$\textsuperscript{d} & \ding{55} & $\bigtriangleup$\textsuperscript{-} \\
\hline
\end{tabular}

\vspace{0.8em}
\begin{flushleft}
\textbf{Notes:} Same as in Table~\ref{tab:group_props_pcu_5e7}.
\end{flushleft}
\end{table}

\clearpage

\subsubsection{Discussion}
\label{subsubsec:medianDiscus}

%まず特徴1は特に交通中で頻出するGroup($0.5 \leq \lambda^{\mathrm{O}}_{i})で成立すると言える。最大流量を取る交通には、頻出groupはなく、$0.1 \leq \lambda^{\mathrm{O}}_{i} < 0.5$~(/frame)の頻出ではないgroupが一定数含まれ、特別少ないわけではない、事を示している。
Feature~(\ref{median_char1}) is predominantly observed under conditions where frequently appearing groups are absent from traffic, i.e., groups with $\lambda^{\mathrm{O}}_{i} \geq 0.5$ are not present. Instead, the maximum flow tends to be achieved in the presence of a certain proportion of less frequent groups with $0.1 \leq \lambda^{\mathrm{O}}_{i} < 0.5$~(/frame).

%ここで~\cite{nagahama2025grouping}より、$0.1 \leq \lambda^{\mathrm{O}}_{i}$に属するgroupは$0.5 \leq \lambda^{\mathrm{O}}_{i}$に属するgroupより概ね台数の多い大型なgroupが多いことが分かっている。ここから特徴1は、ある程度密度が上がり頻出ではない大型groupが現れている、という点を表しており自然な結果である。
%一方で特徴2はすべての条件を通して、成立するとはいいがたい。特徴1と2を合わせると、ただgroup割合が交通で増えると流量が増えるわけではないことになる。ここから、$0.1 \leq \lambda^{\mathrm{O}}_{i} < 0.5$~(/frame)といった頻出ではないgroupが交通にある程度含まれることで、最大流量が実現している可能性が示唆される。
As reported in~\cite{nagahama2025grouping}, groups falling within the range $\lambda^{\mathrm{O}}_{i} \geq 0.1$ are generally composed of larger vehicles and exhibit greater member counts compared to those in the $\lambda^{\mathrm{O}}_{i} \geq 0.5$ range. Accordingly, Feature~(\ref{median_char1}) can be interpreted as a reflection of traffic in which increased density facilitates the emergence of such less frequent, larger groups—resulting in enhanced flow—thereby constituting a plausible outcome.
On the other hand, Feature~(\ref{median_char2}) does not consistently hold across all conditions. Taken together with Feature~(\ref{median_char1}), the findings suggest that an increase in group proportion alone does not guarantee a corresponding increase in traffic flow. Rather, the presence of moderately frequent groups, specifically those with $0.1 \leq \lambda^{\mathrm{O}}_{i} < 0.5$~(/frame), appears to play a crucial role in enabling maximum flow conditions.

%特徴3はgroup割合計算方法、Groupの観測頻度に関わらず成立する特徴である。中程度の密度における減速波によって生成されたクラスターが、groupとして観測されていると考えれば、自然な結果であると言えよう。
Feature~(\ref{median_char3}) is observed robustly, irrespective of the method used to compute group proportion or the frequency of group occurrence. This robustness can be attributed to the formation of vehicle clusters generated by deceleration waves at moderate densities, which are likely being captured as groups in the analysis in~\cite{nagahama2025grouping}.

%最後に，Characteristic~(\ref{median_char4}) は部分支持にとどまる。中程度のグループ比率が高密度領域で比較的大きな median flow を示す傾向は，group proportion を PCU ベースで算出した場合にやや明瞭に現れるが，すべての設定で一様に観察されるわけではない。なお，Sec.~3.2 でさらに論じるように，scatter-based analysis は，moderate group proportions がより高密度な交通条件のもとで peak/high-flow cases とより一貫して関連していることを示唆している。
Lastly, Feature~(\ref{median_char4}) receives only partial support. The tendency for moderate group proportions to exhibit relatively large median flow in high-density regions appears somewhat more clearly when group proportions are calculated based on PCU, but it is not uniformly observed across all settings. As further discussed in Sec.~\ref{subsec:scatter}, the scatter-based analysis suggests that moderate group proportions are more consistently associated with peak/high-flow cases under denser traffic conditions.

\subsection{Detailed Visualization of Group-Wise Variability in Flow-Density Data}
\label{subsec:scatter}

\subsubsection{Result}
\label{subsubsec:scatResult}
%%%%%%%%%%%%%%20250408ここから
%図\ref{scat_indo_dai}は交通中に含まれるgroupの割合毎に、flow-densityの関係性をオレンジの散布図で示したものである。
%group割合はnumber of vehiclesで求めた。
%背景の黒色のプロットはすべてのデータ点を示しており、flowおよびdensityはIP methodに基づくPCUで計算している。
%特に図\ref{subfig:scat_indo_5e_dai}は出現頻度$\lambda^{\mathrm{O}}_{i}$が0.5~(/frame)以上であったgroupのflow-densityの関係性を示しており、図\ref{subfig:scat_indo_1e_dai}は$\lambda^{\mathrm{O}}_{i}$が0.1~(/frame)以上であったgroupのflow-densityの関係性を示している。

\begin{figure}[h]
\begin{center}
\subfigure[Groups $\lambda^{\mathrm{O}}_{i}\geq 0.5$~(/frame) are considered.]{
\includegraphics[width=90mm]
{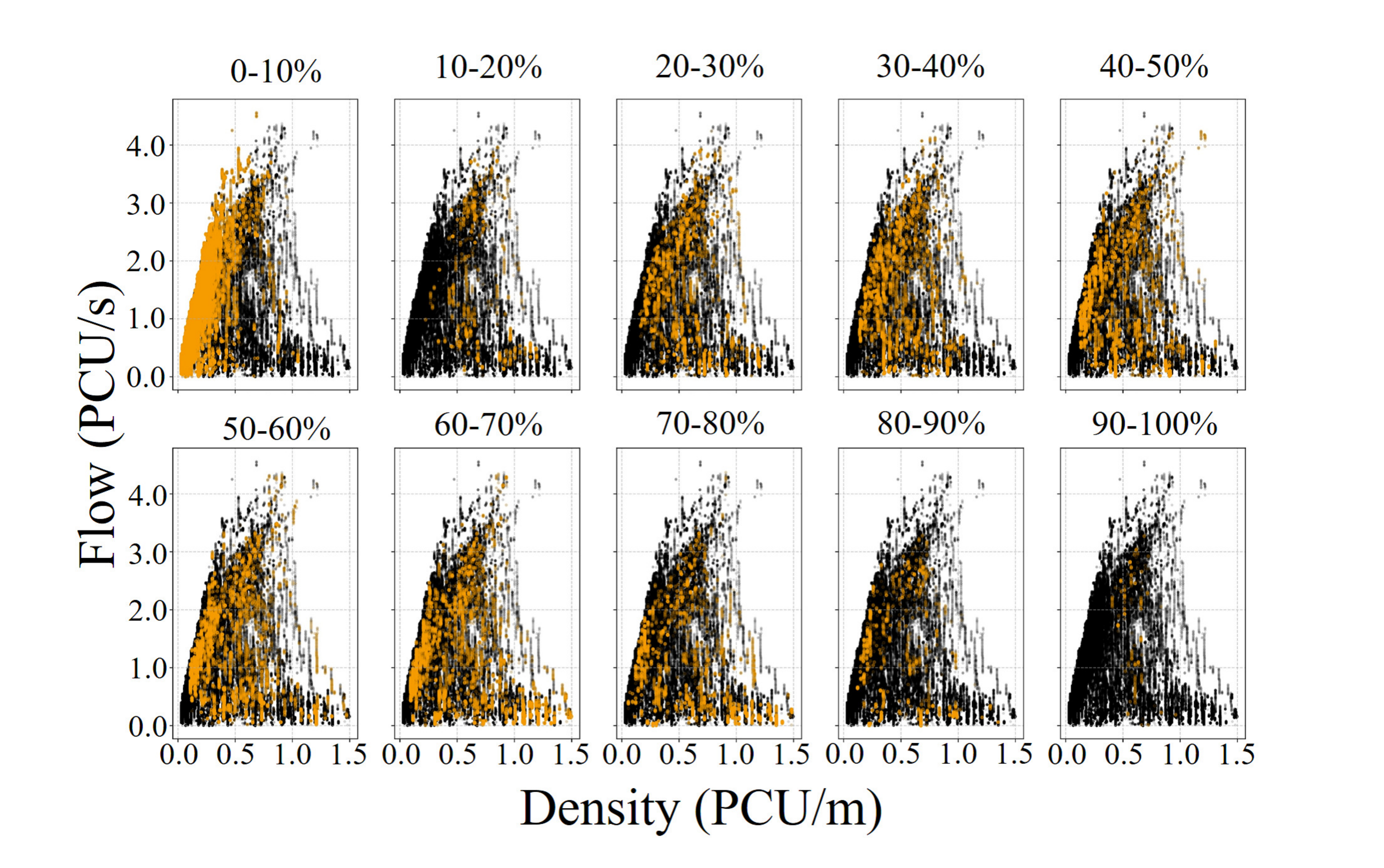}
\label{subfig:scat_indo_5e_dai}}
\subfigure[Groups $\lambda^{\mathrm{O}}_{i}\geq 0.1$~(/frame) are considered.]{
\includegraphics[width=90mm]
{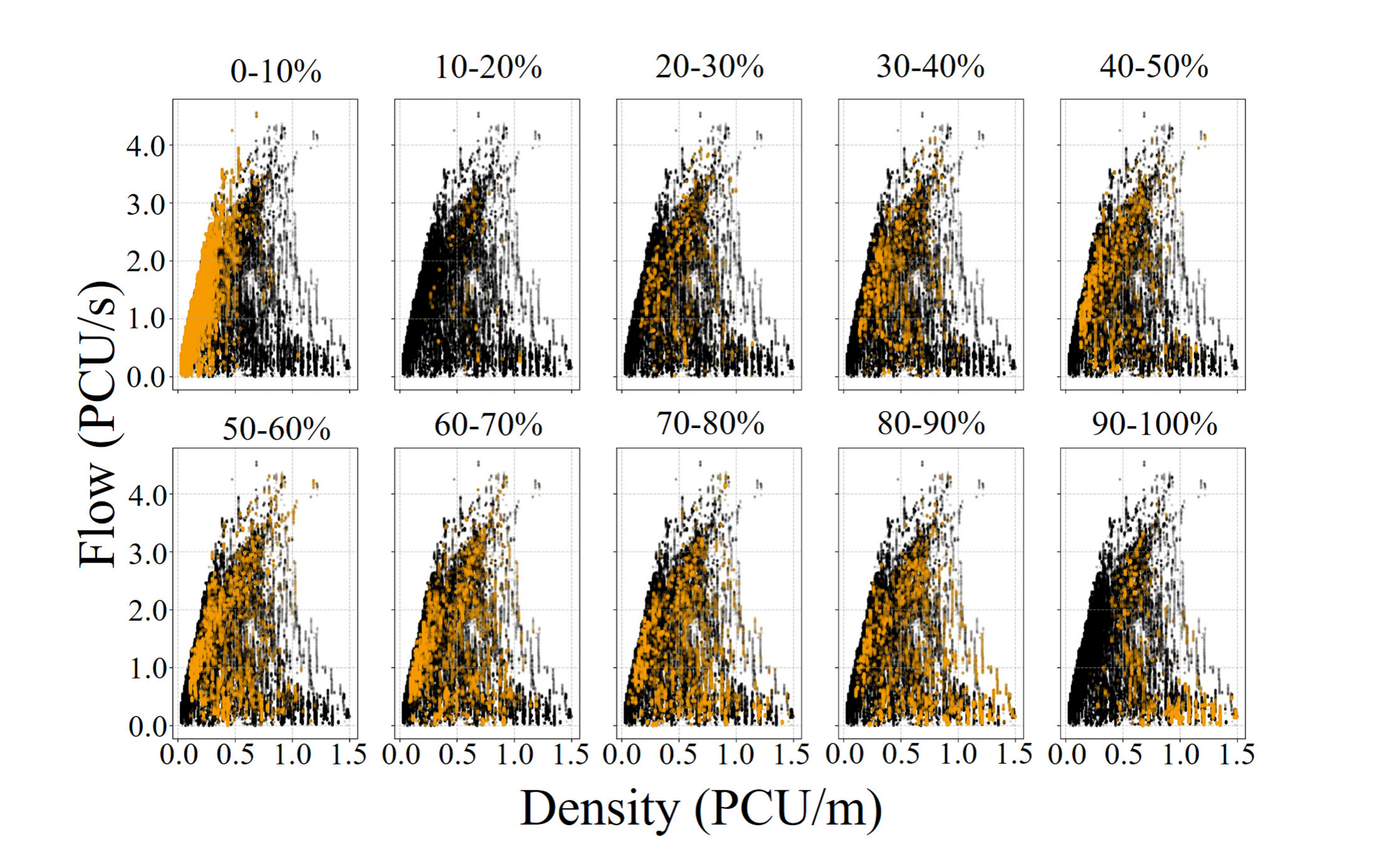}
\label{subfig:scat_indo_1e_dai}}
\end{center}
\caption{Scatter plot of flow in PCU (IP method) for respective group proportion by number of vehicles.}
\label{scat_indo_dai}
\end{figure}

%図\ref{scat_indo_dai}からは以下の\label{scat_char1}から\label{scat_char7}の特徴が見て取れる。
%\begin{enumerate}
%\item Group割合が0-10%のときfree flowの高い流量にデータが集中する\label{scat_char1}
%\item Group割合が0-10%のときは高いdensityが実現しにくい\label{scat_char2}
%\item Group割合が10-20%において低いdensityのデータ点が減少する\label{scat_char3}
%\item Group割合が中程度30-60%のとき、最も高い流量が実現する\label{scat_char4}
%\item Group割合が50%から増加するに従い、中密度かつ中流量のデータが減少する\label{scat_char5}
%\item Group割合が90-100%において高い密度が実現しない\label{scat_char6}
%\item Group割合が90-100%において低い密度が実現しない\label{scat_char7}
%\end{enumerate}
Figure~\ref{scat_indo_dai} illustrates the flow-density relationships for different ranges of group proportions, shown as orange scatter plots. The group proportions are calculated based on the number of vehicles.
The black background dots represent all available data points. Both flow and density are computed using PCUs derived from the IP method.
Specifically, Figure~\ref{subfig:scat_indo_5e_dai} displays the flow-density relationships for groups with an occurrence frequency of $\lambda^{\mathrm{O}}_{i} \geq 0.5$~(/frame), whereas Figure~\ref{subfig:scat_indo_1e_dai} covers groups with $\lambda^{\mathrm{O}}_{i} \geq 0.1$~(/frame).

The following characteristics, labeled from \ref{scat_char1} to \ref{scat_char7}, are identified from Figure~\ref{scat_indo_dai}:
\begin{enumerate}[label=(\Alph*)]
    \item When the group proportion is between 0--10\%, the data are concentrated in the high-flow, free-flow regime. \label{scat_char1}
    \item When the group proportion is between 0--10\%, high-density states are rarely observed. \label{scat_char2}
    \item In the 10--20\% group proportion range, data points at low densities are reduced. \label{scat_char3}
    \item The highest flow values are achieved when the group proportion lies in the moderate range of 30--60\%. \label{scat_char4}
    \item As the group proportion increases beyond 50\%, the number of data points in the medium-density and medium-flow region decreases. \label{scat_char5}
    \item In the 90--100\% group proportion range, high-density states do not occur. \label{scat_char6}
    \item In the 90--100\% group proportion range, low-density states do not occur. \label{scat_char7}
\end{enumerate}

\begin{figure}[h]
\begin{center}
\subfigure[Groups $\lambda^{\mathrm{O}}_{i}\geq 0.5$~(/frame) are considered.]{
\includegraphics[width=90mm]
{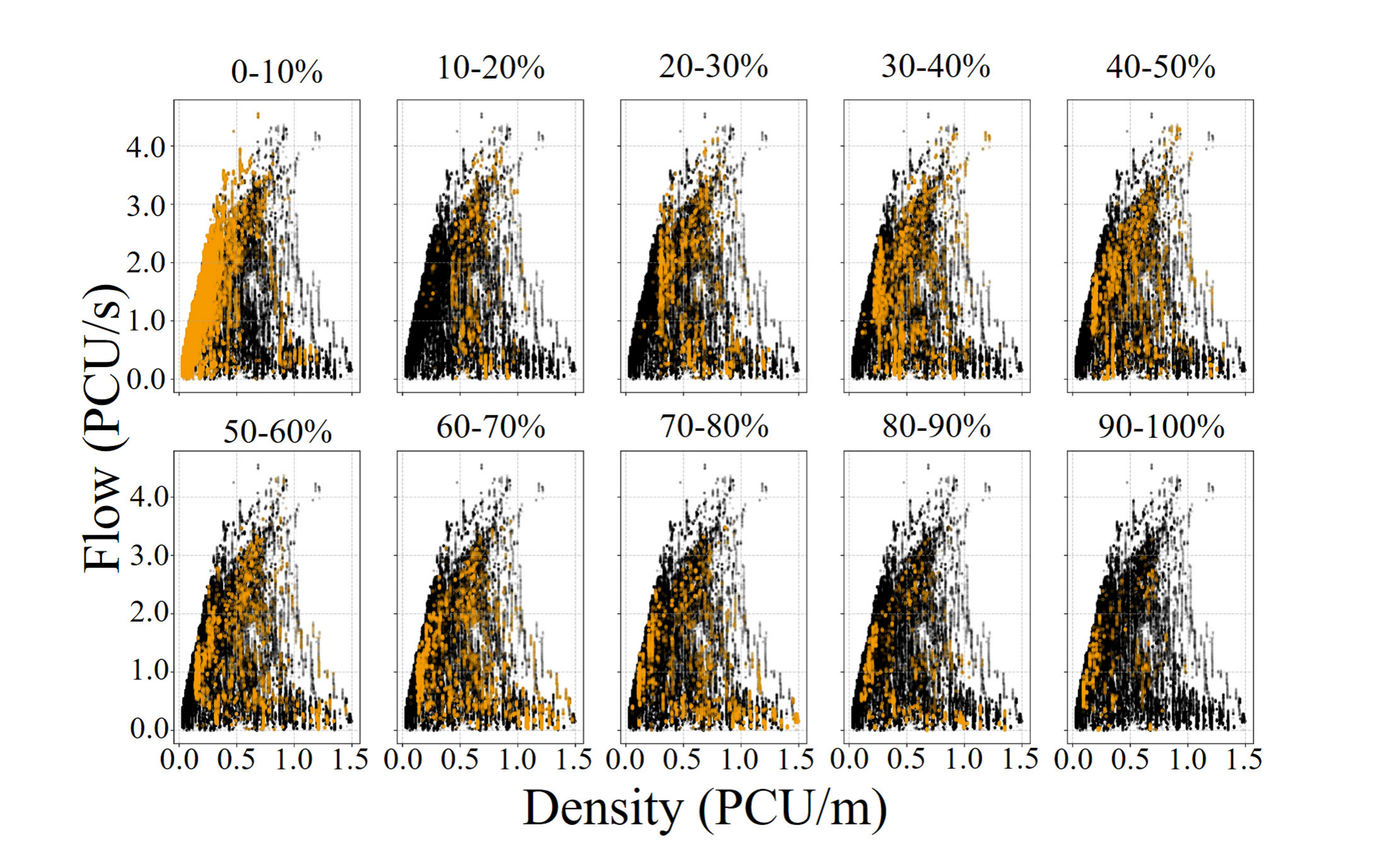}
\label{subfig:scat_indo_5e_PCU}}
\subfigure[Groups $\lambda^{\mathrm{O}}_{i}\geq 0.1$~(/frame) are considered.]{
\includegraphics[width=90mm]
{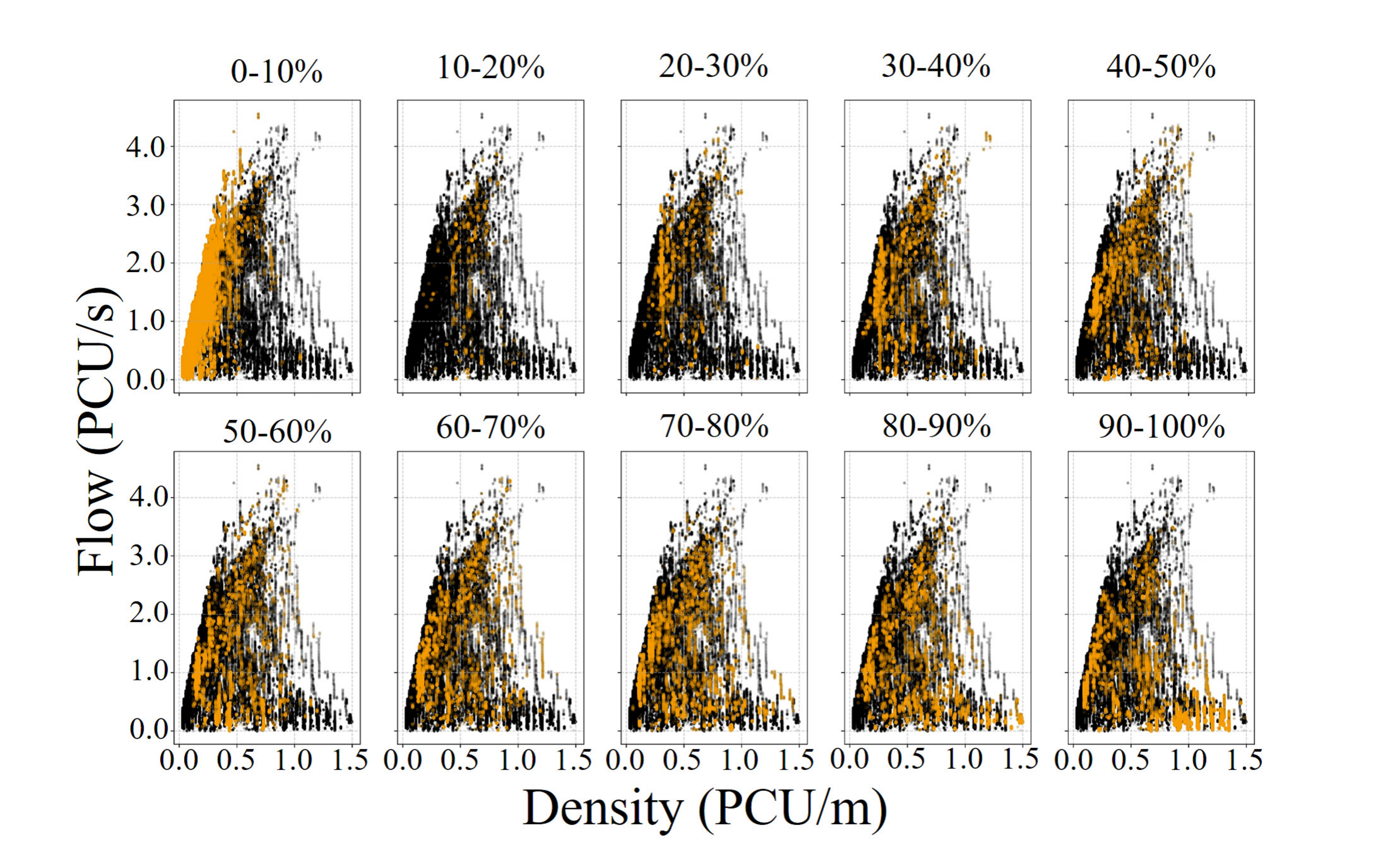}
\label{subfig:scat_indo_1e_PCU}}
\end{center}
\caption{Scatter plot of flow in PCU (IP method) for respective group proportion by PCU.}
\label{scat_indo_PCU}
\end{figure}

%図\ref{scat_indo_PCU}は先に示した図\ref{scat_indo_dai}のgroup割合をPCUで計算したものである。
Figure~\ref{scat_indo_PCU} presents the same group proportion analysis as shown in Figure~\ref{scat_indo_dai}, but with group proportions calculated based on PCU values obtained using the IP method.

\begin{figure}[h]
\begin{center}
\subfigure[Groups $\lambda^{\mathrm{O}}_{i}\geq 0.5$~(/frame) are considered.]{
\includegraphics[width=90mm]
{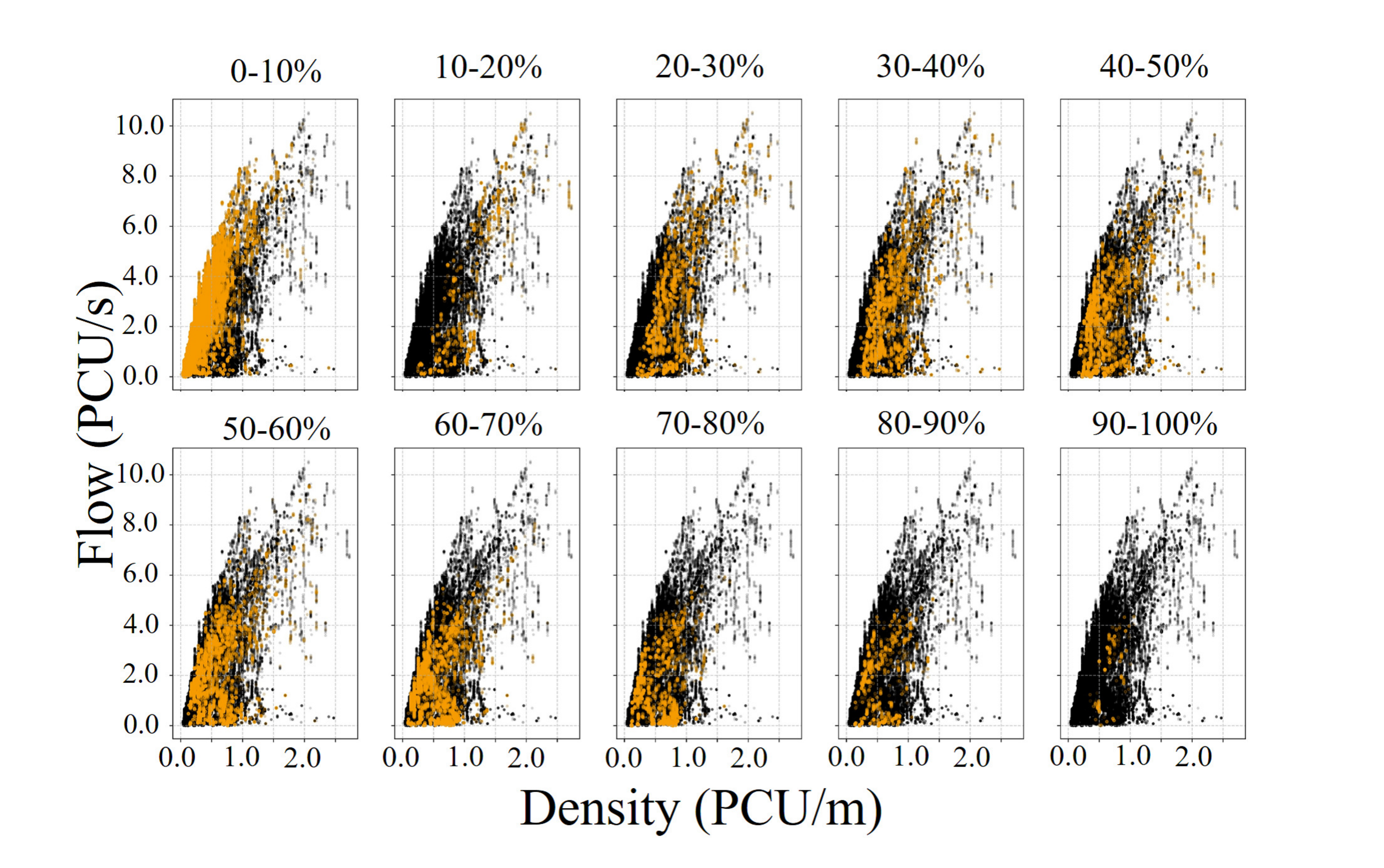}
\label{subfig:scat_MLR_5e_dai}}
\subfigure[Groups $\lambda^{\mathrm{O}}_{i}\geq 0.1$~(/frame) are considered.]{
\includegraphics[width=90mm]
{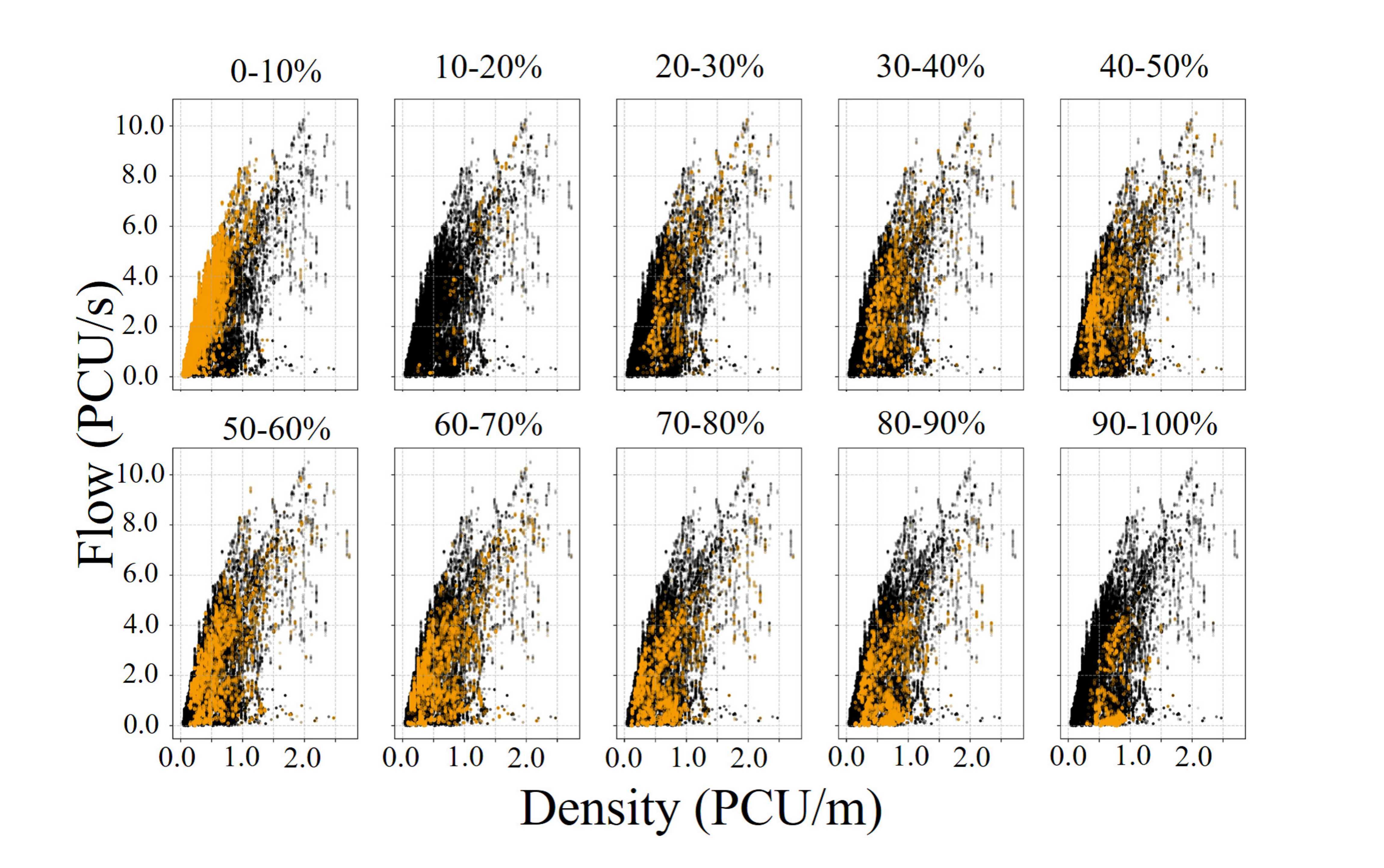}
\label{subfig:scat_MLR_1e_dai}}
\end{center}
\caption{Scatter plot of flow in PCU (MLR method) for respective group proportion by number of vehicles.}
\label{scat_MLR_dai}
\end{figure}

\begin{figure}[h]
\begin{center}
\subfigure[Groups $\lambda^{\mathrm{O}}_{i}\geq 0.5$~(/frame) are considered.]{
\includegraphics[width=90mm]
{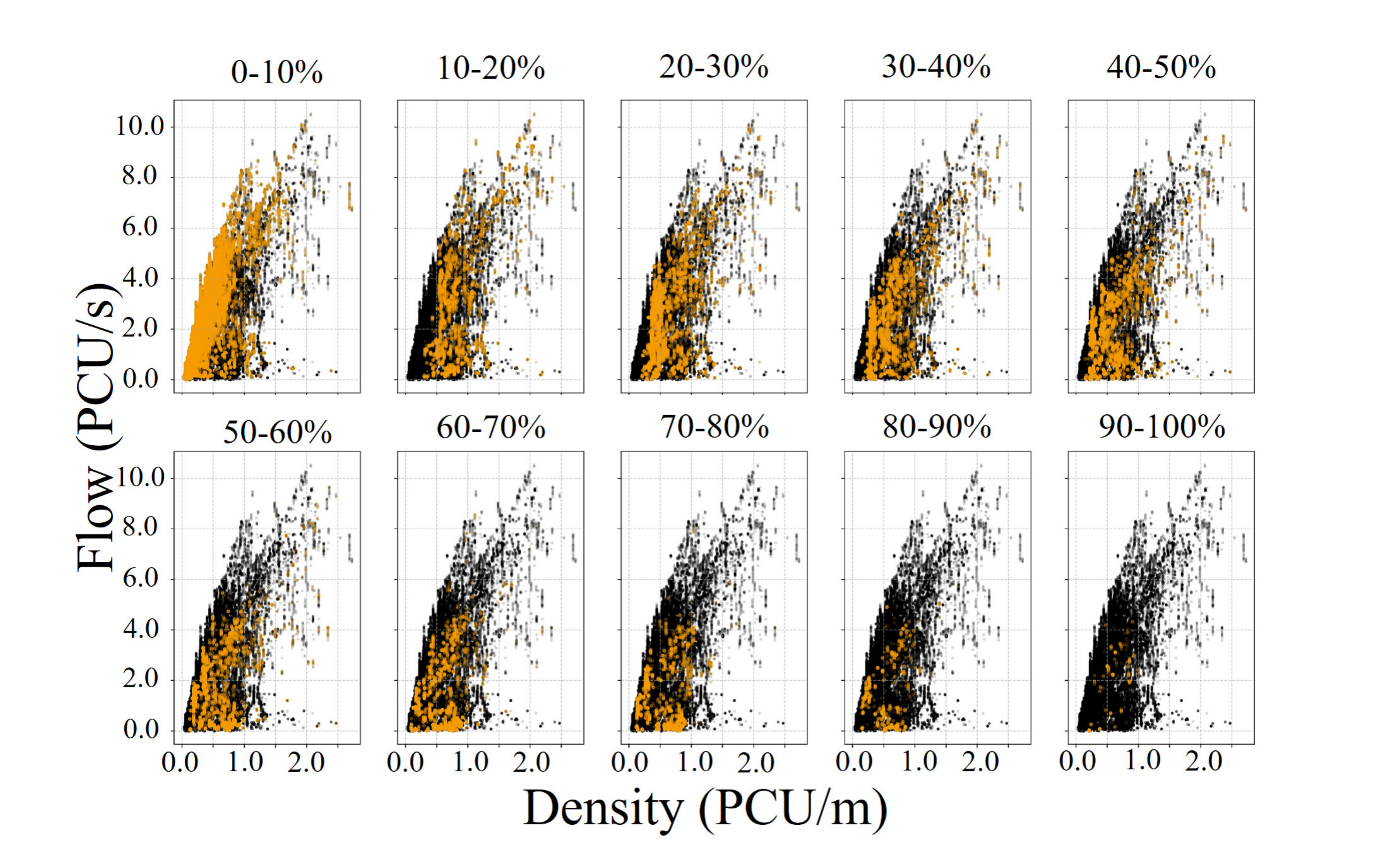}
\label{subfig:scat_MLR_5e_PCU}}
\subfigure[Groups $\lambda^{\mathrm{O}}_{i}\geq 0.1$~(/frame) are considered.]{
\includegraphics[width=90mm]
{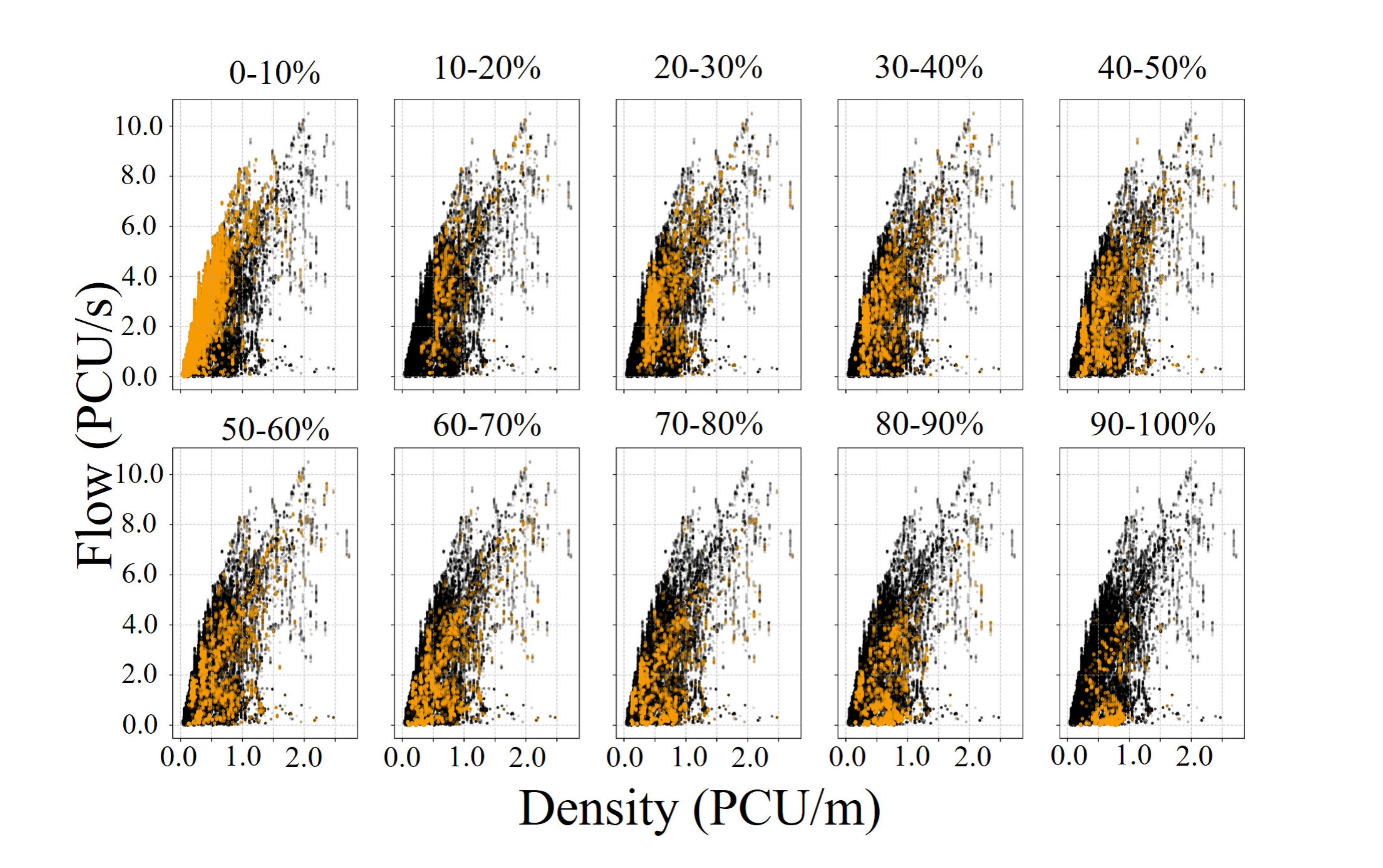}
\label{subfig:scat_MLR_1e_PCU}}
\end{center}
\caption{Scatter plot of flow in PCU (MLR method) for respective group proportion by PCU.}
\label{scat_MLR_PCU}
\end{figure}

\begin{figure}[h]
\begin{center}
\subfigure[Groups $\lambda^{\mathrm{O}}_{i}\geq 0.5$~(/frame) are considered.]{
\includegraphics[width=90mm]
{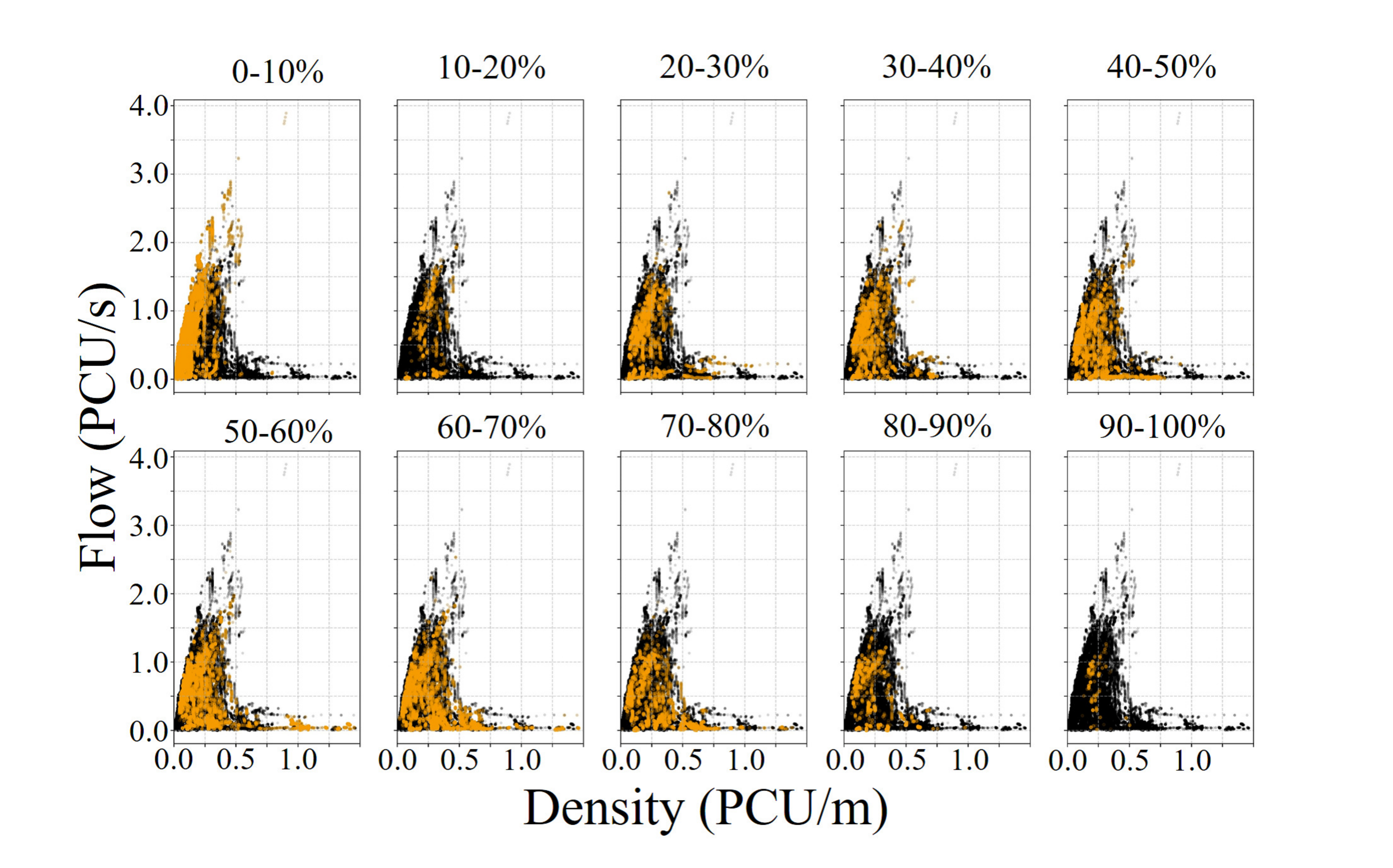}
\label{subfig:scat_SB_5e_dai}}
\subfigure[Groups $\lambda^{\mathrm{O}}_{i}\geq 0.1$~(/frame) are considered.]{
\includegraphics[width=90mm]
{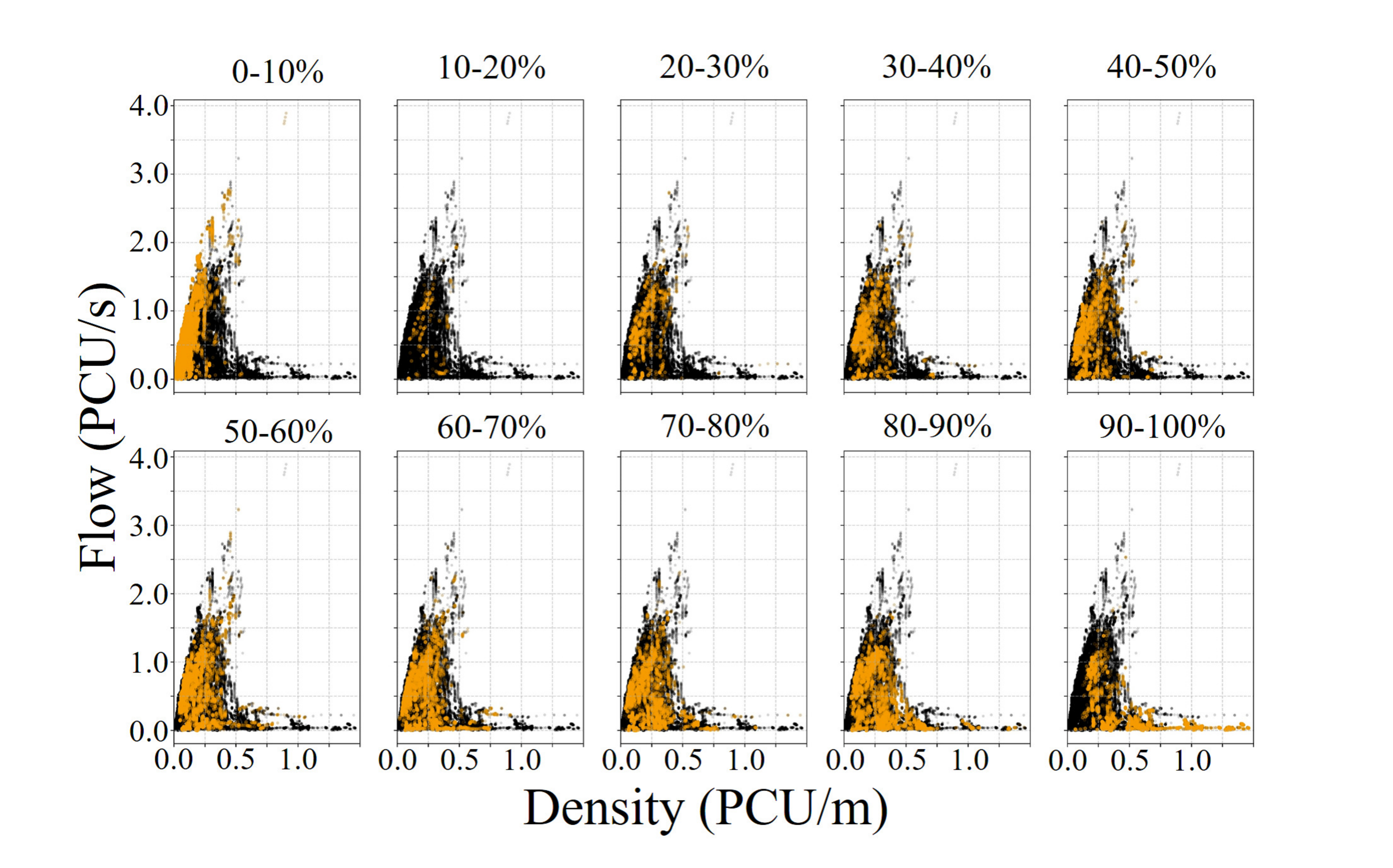}
\label{subfig:scat_SB_1e_dai}}
\end{center}
\caption{Scatter plot of flow in PCU (SB method) for respective group proportion by number of vehicles.}
\label{scat_SB_dai}
\end{figure}

\begin{figure}[h]
\begin{center}
\subfigure[Groups $\lambda^{\mathrm{O}}_{i}\geq 0.5$~(/frame) are considered.]{
\includegraphics[width=90mm]
{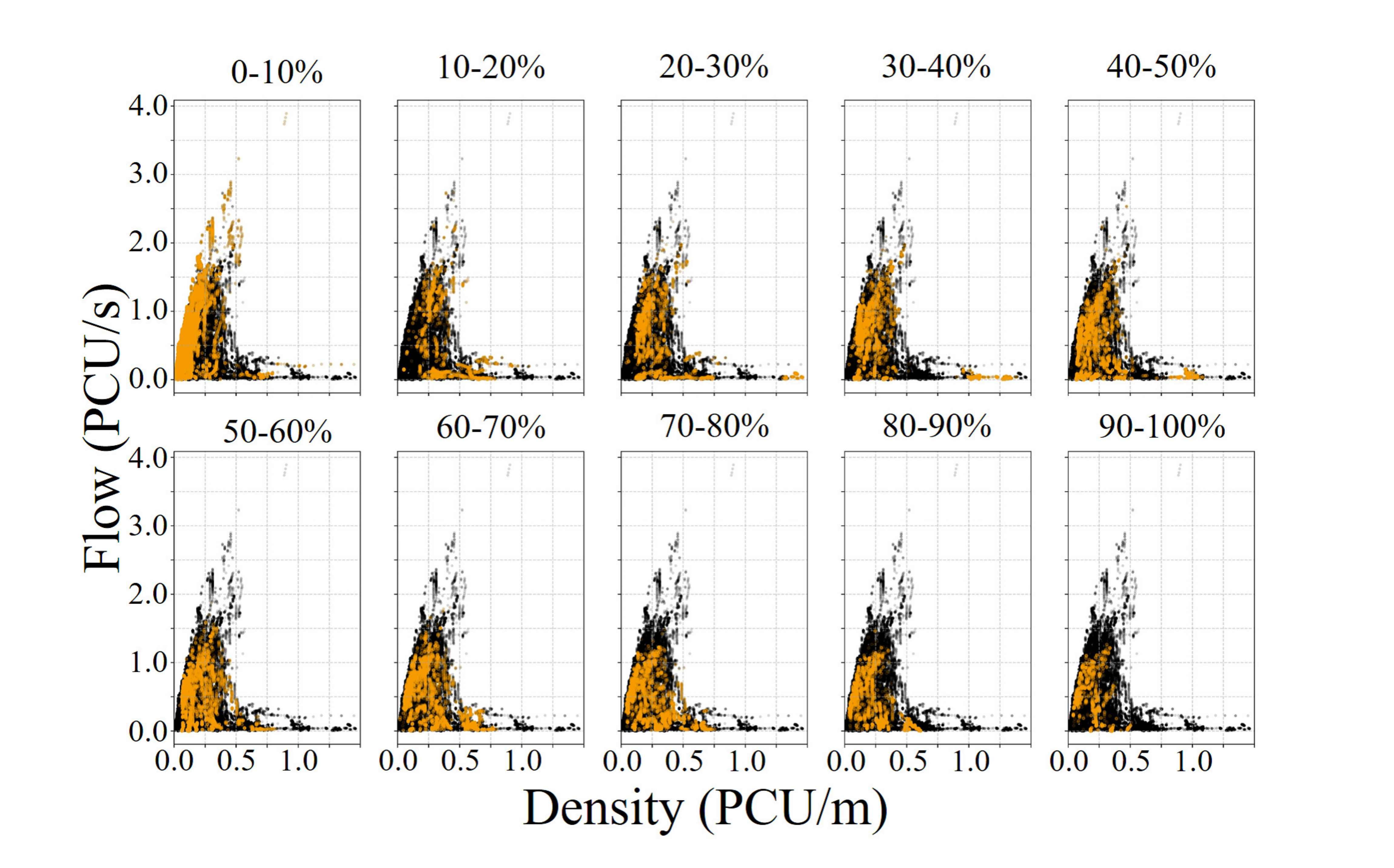}
\label{subfig:scat_SB_5e_PCU}}
\subfigure[Groups $\lambda^{\mathrm{O}}_{i}\geq 0.1$~(/frame) are considered.]{
\includegraphics[width=90mm]
{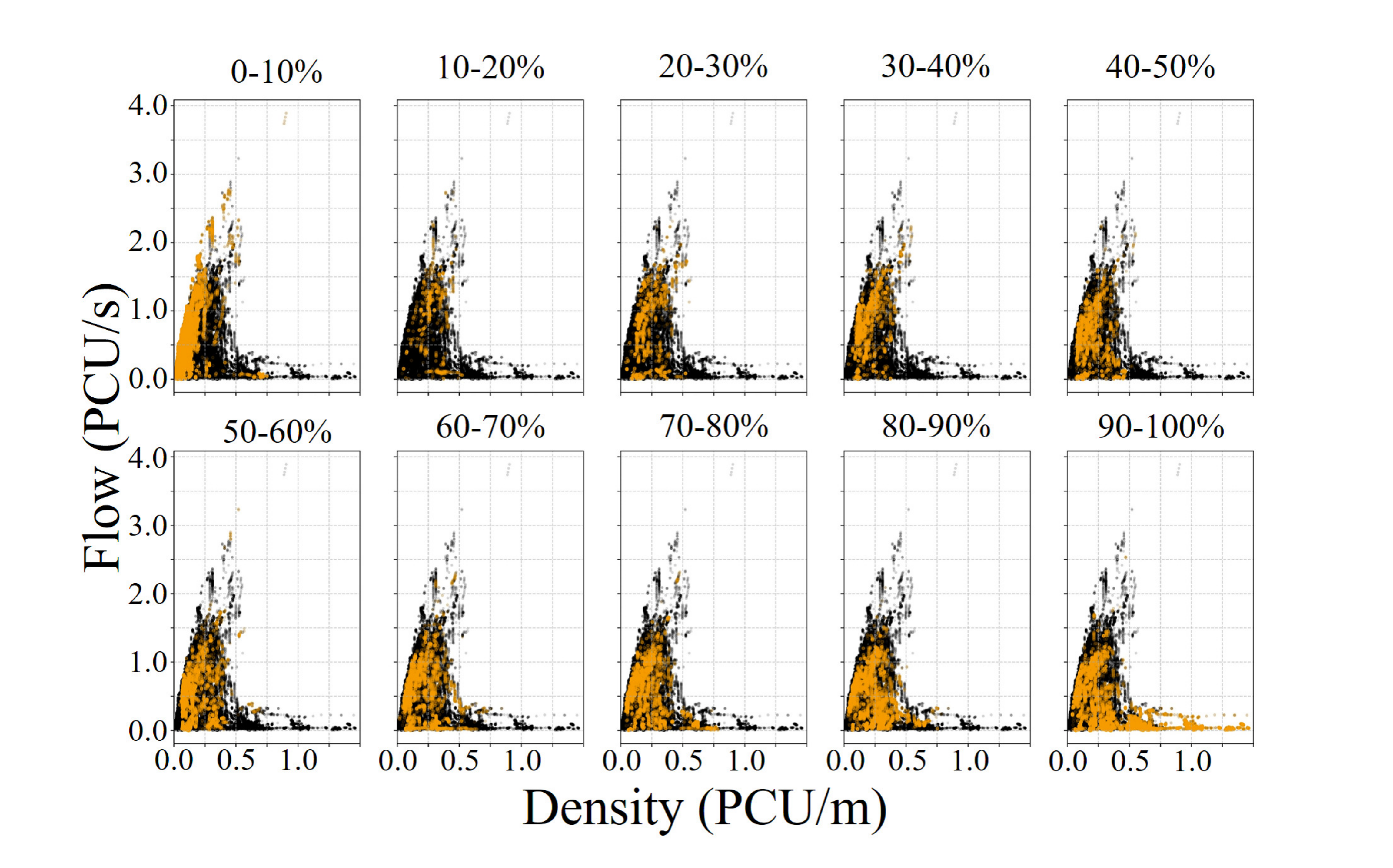}
\label{subfig:scat_SB_1e_PCU}}
\end{center}
\caption{Scatter plot of flow in PCU (SB method) for respective group proportion by PCU.}
\label{scat_SB_PCU}
\end{figure}

%さらに図\ref{scat_MLR_dai}および図\ref{scat_MLR_PCU}にMLR methodによってPCUを計算した場合のflow-densityの関係を
%図\ref{scat_SB_dai}および図\ref{scat_SB_PCU}にSB methodによってPCUを計算した場合のflow-densityの関係を、散布図で示す。
Furthermore, Figures~\ref{scat_MLR_dai} and~\ref{scat_MLR_PCU} show the flow-density relationships based on PCU values computed using the MLR method, while Figures~\ref{scat_SB_dai} and~\ref{scat_SB_PCU} present the corresponding relationships derived using the SB method, all illustrated as scatter plots.

\clearpage

%図\ref{scat_indo_dai}から図~\ref{scat_SB_PCU}で確認できる特徴をTable~\ref{tab:group_props_scatter_vehicle_5e7}からTable\ref{tab:group_props_scatter_pcu_1e7}にまとめた
The characteristics observed in Figures~\ref{scat_indo_dai} through~\ref{scat_SB_PCU} are summarized in Tables~\ref{tab:group_props_scatter_vehicle_5e7} through~\ref{tab:group_props_scatter_pcu_1e7}.
In the tables, the symbols indicate the level of support for each characteristic as follows:  
$\bigcirc$~denotes that the characteristic is clearly supported by the data;  
\ding{55}~indicates that the characteristic is not supported;  
$\bigtriangleup$~represents partial or ambiguous support, often due to limited data or variability.  
In the ``Consensus'' column, we indicate whether each characteristic is universally supported ($\bigcirc$), partially supported ($\bigtriangleup$), or considered unsupported (\ding{55}) by a majority vote based on the results of the three methods (IP, MLR, and SB). We adopt the symbol that appears most frequently among the three outcomes; if all three symbols are different, we assign $\bigtriangleup$. In addition, when the three outcomes are \{$\bigcirc$, $\bigcirc$, \ding{55}\} or \{$\bigcirc$, $\bigtriangleup$, $\bigcirc$\}, we denote the Consensus entry as $\bigcirc$\textsuperscript{-} to indicate that the characteristic is not fully supported. Conversely, when the three outcomes are \{$\bigcirc$, \ding{55}, \ding{55}\} or \{\ding{55}, $\bigtriangleup$, \ding{55}\}, we denote the Consensus entry as \ding{55}\textsuperscript{+} to indicate that the characteristic is not entirely unsupported.

\begin{table}[h]
\centering
\caption{Flow-density characteristics under vehicle-count-based group proportions (scatter plot analysis) ($\lambda^{\mathrm{O}}_{i} \geq 0.5$~(/frame)).}
\label{tab:group_props_scatter_vehicle_5e7}
\begin{tabular}{p{9cm} | c c c c}
\hline
\textbf{Feature} & IP & MLR & SB & Consensus \\
\hline
(A) When the group proportion is 0--10\%, data concentrate in the high-flow, free-flow regime. 
& $\bigcirc$ & $\bigcirc$ & $\bigcirc$ & $\bigcirc$ \\
(B) When the group proportion is 0--10\%, high-density states are rarely observed. 
& $\bigcirc$ & $\bigcirc$ & $\bigcirc$ & $\bigcirc$ \\
(C) In the 10--20\% group proportion range, low-density data points are scarce. 
& $\bigcirc$ & $\bigcirc$ & $\bigcirc$ & $\bigcirc$ \\
(D) The highest flow values are observed when the group proportion is in the moderate range of 30--60\%. 
& $\bigcirc$ & $\bigcirc$\textsuperscript{f} & \ding{55} & $\bigcirc$\textsuperscript{-} \\
(E) As the group proportion increases beyond 50\%, medium-density and medium-flow data points decrease. 
& $\bigtriangleup$ & $\bigcirc$ & $\bigtriangleup$ & $\bigtriangleup$\textsuperscript{+} \\
(F) In the 90--100\% group proportion range, high-density states do not occur. 
& $\bigcirc$ & $\bigcirc$ & $\bigcirc$ & $\bigcirc$ \\
(G) In the 90--100\% group proportion range, low-density states do not occur. 
& $\bigcirc$ & $\bigcirc$ & $\bigcirc$ & $\bigcirc$ \\
\hline
\end{tabular}

\vspace{0.8em}
\begin{flushleft}
\textbf{Notes:} \\
\textsuperscript{f}~The highest flow values are also observed within the 10--20\% group proportion range. \\
\textsuperscript{+}~Majority vote, leaning toward support. \quad
\textsuperscript{-}~Majority vote, leaning toward non-support.
\end{flushleft}
\end{table}

\begin{table}[h]
\centering
\caption{Flow-density characteristics under vehicle-count-based group proportions (scatter plot analysis) ($\lambda^{\mathrm{O}}_{i} \geq 0.1$~(/frame)).}
\label{tab:group_props_scatter_vehicle_1e7}
\begin{tabular}{p{9cm} | c c c c}
\hline
\textbf{Feature} & IP & MLR & SB & Consensus \\
\hline
(A) When the group proportion is 0--10\%, data concentrate in the high-flow, free-flow regime. 
& $\bigcirc$ & $\bigcirc$ & $\bigcirc$ & $\bigcirc$ \\
(B) When the group proportion is 0--10\%, high-density states are rarely observed. 
& $\bigcirc$ & $\bigcirc$ & $\bigcirc$ & $\bigcirc$ \\
(C) In the 10--20\% group proportion range, low-density data points are scarce. 
& $\bigcirc$ & $\bigcirc$ & $\bigcirc$ & $\bigcirc$ \\
(D) The highest flow values are observed when the group proportion is in the moderate range of 30--60\%. 
& $\bigcirc$ & $\bigcirc$\textsuperscript{f} & \ding{55} & $\bigcirc$\textsuperscript{-} \\
(E) As the group proportion increases beyond 50\%, medium-density and medium-flow data points decrease. 
& $\bigtriangleup$ & $\bigtriangleup$ & \ding{55} & $\bigtriangleup$\textsuperscript{-} \\
(F) In the 90--100\% group proportion range, high-density states do not occur. 
& \ding{55} & $\bigcirc$ & \ding{55} & \ding{55}\textsuperscript{+} \\
(G) In the 90--100\% group proportion range, low-density states do not occur. 
& $\bigcirc$ & $\bigcirc$ & $\bigcirc$ & $\bigcirc$ \\
\hline
\end{tabular}

\vspace{0.8em}
\begin{flushleft}
\textbf{Notes:} \\
\textsuperscript{f}~In some cases, the highest flow values are also observed within the 10--20\% group proportion range. \\
\textsuperscript{+}~Majority vote, leaning toward support. \quad
\textsuperscript{-}~Majority vote, leaning toward non-support.
\end{flushleft}
\end{table}

\begin{table}[h]
\centering
\caption{Flow-density characteristics under PCU-based group proportions (scatter plot analysis) ($\lambda^{\mathrm{O}}_{i} \geq 0.5$~(/frame)).}
\label{tab:group_props_scatter_pcu_5e7}
\begin{tabular}{p{9cm} | c c c c}
\hline
\textbf{Feature} & IP & MLR & SB & Consensus \\
\hline
(A) When the group proportion is 0--10\%, data concentrate in the high-flow, free-flow regime. 
& $\bigcirc$ & $\bigcirc$ & $\bigcirc$ & $\bigcirc$ \\
(B) When the group proportion is 0--10\%, high-density states are rarely observed. 
& $\bigcirc$ & $\bigtriangleup$ & $\bigtriangleup$ & $\bigtriangleup$\textsuperscript{+} \\
(C) In the 10--20\% group proportion range, low-density data points are scarce. 
& $\bigcirc$ & $\bigcirc$ & $\bigcirc$ & $\bigcirc$ \\
(D) The highest flow values are observed when the group proportion is in the moderate range of 30--60\%. 
& $\bigcirc$ & $\bigcirc$\textsuperscript{g} & \ding{55} & $\bigcirc$\textsuperscript{-} \\
(E) As the group proportion increases beyond 50\%, medium-density and medium-flow data points decrease. 
& $\bigtriangleup$ & $\bigcirc$ & $\bigtriangleup$ & $\bigtriangleup$\textsuperscript{+} \\
(F) In the 90--100\% group proportion range, high-density states do not occur. 
& $\bigcirc$ & $\bigcirc$ & $\bigcirc$ & $\bigcirc$ \\
(G) In the 90--100\% group proportion range, low-density states do not occur. 
& \ding{55} & $\bigcirc$ & \ding{55} & \ding{55}\textsuperscript{+} \\
\hline
\end{tabular}

\vspace{0.8em}
\begin{flushleft}
\textbf{Notes:} \\
\textsuperscript{g}~The highest flow values are also observed within the 0--20\% group proportion range. \\
\textsuperscript{+}~Majority vote, leaning toward support. \quad
\textsuperscript{-}~Majority vote, leaning toward non-support.
\end{flushleft}
\end{table}

\begin{table}[h]
\centering
\caption{Flow-density characteristics under PCU-based group proportions (scatter plot analysis) ($\lambda^{\mathrm{O}}_{i} \geq 0.1$~(/frame)).}
\label{tab:group_props_scatter_pcu_1e7}
\begin{tabular}{p{9cm} | c c c c}
\hline
\textbf{Feature} & IP & MLR & SB & Consensus \\
\hline
(A) When the group proportion is 0--10\%, data concentrate in the high-flow, free-flow regime. 
& $\bigcirc$ & $\bigcirc$ & $\bigcirc$ & $\bigcirc$ \\
(B) When the group proportion is 0--10\%, high-density states are rarely observed. 
& $\bigcirc$ & $\bigcirc$ & $\bigcirc$ & $\bigcirc$ \\
(C) In the 10--20\% group proportion range, low-density data points are scarce. 
& $\bigcirc$ & $\bigcirc$ & $\bigcirc$ & $\bigcirc$ \\
(D) The highest flow values are observed when the group proportion is in the moderate range of 30--60\%. 
& $\bigcirc$ & $\bigcirc$\textsuperscript{g} & \ding{55} & $\bigcirc$\textsuperscript{-} \\
(E) As the group proportion increases beyond 50\%, medium-density and medium-flow data points decrease. 
& $\bigtriangleup$ & $\bigtriangleup$ & \ding{55} & $\bigtriangleup$\textsuperscript{-} \\
(F) In the 90--100\% group proportion range, high-density states do not occur. 
& \ding{55} & $\bigcirc$ & \ding{55} & \ding{55}\textsuperscript{+} \\
(G) In the 90--100\% group proportion range, low-density states do not occur. 
& \ding{55} & $\bigcirc$ & \ding{55} & \ding{55}\textsuperscript{+} \\
\hline
\end{tabular}

\vspace{0.8em}
\begin{flushleft}
\textbf{Notes:} \\
\textsuperscript{g}~The highest flow values are also observed within the 0--20\% group proportion range. \\
\textsuperscript{+}~Majority vote, leaning toward support. \quad
\textsuperscript{-}~Majority vote, leaning toward non-support.
\end{flushleft}
\end{table}

\clearpage

\subsubsection{Discussion}
\label{subsubsec:scatDisc}
%特徴\ref{scat_char1}から\ref{scat_char4}は、group割合の計算方法とgroupの$\lambda^{\mathrm{O}}_{i}$に関わらず成立する特徴であった。
%特徴\ref{scat_char1}は、Freeflowにおいては、速度同調が少ない状態の方が流れが良いことを、
%特徴\ref{scat_char2}は、密度が高まると速度同調が増えGroup化が進むことを示しており、自然な結果である。
Features~\ref{scat_char1} to \ref{scat_char4} were consistently observed regardless of how group proportions were calculated or which threshold of $\lambda^{\mathrm{O}}_{i}$ was used for group identification. Feature~\ref{scat_char1} suggests that in the free-flow regime, traffic flow increases when there is less velocity synchronization, indicating looser interactions among vehicles. Feature~\ref{scat_char2} shows that as density increases, velocity synchronization strengthens and group formation becomes more prominent, which is a natural result.

%これに対して、Features~\ref{scat_char3} と \ref{scat_char4} は、非自明な統計的パターンを表している。
%Feature~\ref{scat_char3} は、10--20% の group-proportion range の外側では low-density data points が相対的に少なく観測されることを示しており、Feature~\ref{scat_char4} は、中程度の group proportions（30--60%）と highest flow rates との関連を示している。これらのパターンは、直接的な因果効果としてではなく、記述的な関連として解釈されるべきである。それでも、これらのパターンは traffic situation を分類したうえで、複数の PCU 設定および group 設定を比較しても観察されているため、group proportion が flow--density structure と非線形かつ traffic-situation-dependent に関係していることを示唆している。さらに、Feature~\ref{scat_char4} は median-based の結果を補完するものであり、moderate group proportions は、より高密度な交通条件のもとで peak/high-flow cases とより一貫して関連していることを示唆している。(1-4,3-1)
In contrast, Features~\ref{scat_char3} and \ref{scat_char4} represent nontrivial statistical patterns. 
Feature~\ref{scat_char3} indicates that low-density data points are less frequently observed outside the 10--20\% group-proportion range, whereas Feature~\ref{scat_char4} indicates an association between moderate group proportions (30--60\%) and the highest flow rates. These patterns should be interpreted as descriptive associations rather than as direct causal effects. Nevertheless, because these patterns are observed after classifying traffic by situation and comparing multiple PCU and group settings, they still suggest that group proportion is systematically related to the flow--density structure in a nonlinear and traffic-situation-dependent manner. Furthermore, Feature~\ref{scat_char4} complements the median-based results and suggests that moderate group proportions are more consistently associated with peak/high-flow cases under denser traffic conditions.

%特徴\ref{scat_char5}および\ref{scat_char6}は頻出するgroup ($\lambda^{\mathrm{O}}_{i} \geq 0.5$~(/frame))のみで成立する特徴である。
%特徴\ref{scat_char6}は、高い密度で見られる大型のGroupはあまり観測頻度が高くないことを示しており自然な結果である。
%一方で特徴\ref{scat_char5}は、group割合が50\%を超えて多い交通は、低密度もしくは高密度な交通、加えて、中密度ながら低流量もしくは高流量の交通、いずれかの極端な交通状況となることを示している。
%本研究の解析からはその原因は不明であるが、例えば、減速波でクラスター化し、Groupと検知されたものは速度が低下すると考えられるので中密度低流量な交通を発生させ、他の要因で自律的にgroup化したものが多い場合は、中密度でも高い流量を維持できるのではないだろうか。
Features~\ref{scat_char5} and \ref{scat_char6} were observed only in cases where frequently appearing groups ($\lambda^{\mathrm{O}}_{i} \geq 0.5$~(/frame)) were used. Feature~\ref{scat_char6} indicates that large groups typically seen in high-density traffic do not appear frequently, which is a natural outcome. 
In contrast, Feature~\ref{scat_char5} reveals that when the group proportion exceeds 50\%, the traffic tends to become extreme—either low- or high-density, or medium-density but with either very low or very high flow. While the exact cause of this remains unclear from the present analysis, one possible interpretation is that clusters formed by deceleration waves and detected as groups tend to result in low-speed and hence low-flow, medium-density traffic. Conversely, if the majority of group formation is due to autonomous coordination under other conditions than the deceleration waves, the traffic may be able to maintain high flow even at medium densities.

%最後に，特徴~\ref{scat_char7} は，group 割合を台数ベースで算出した場合にのみ観察された特徴である。台数ベースの group 割合が 90–100% となる状況は，実質的に，ほぼすべての車両が何らかの group に属していることを要求する。とりわけ，バイクのような PCU の小さな車両であっても，同時に(i) 安定した leader–follower 関係，(ii) 速度同調，(iii) 持続的な近接状態を満たしている必要がある。
%しかし，このような条件は一般に低密度交通下では維持が困難である。低密度状態では，車両間距離が大きく，小型車両はより自由に走行できるため，隊列的な同調行動が持続しにくいからである。したがって，台数ベースの group 割合が 90–100% に達した場合に低密度状態が欠落する傾向が見られるのは自然である。
%これに対して，PCU ベースの group 割合では，一部の小型車両が group に属していなくても（例えば，大型車が支配的な group が形成されている場合には），group 割合が 90–100% に達することがあり得る。そのため，PCU ベースの場合には，低密度状態が必ずしも欠落すると期待されるわけではない。
%もっとも，車間距離を比較的大きく保ちながらも集団的に走行するなど，低密度条件下でも group が維持され得る状況であれば，台数ベースの group 割合が 90–100\% であっても，原理的には低密度状態が出現し得る。
Finally, Feature~\ref{scat_char7} was observed only when the group proportion was computed on a vehicle-count basis.
A vehicle-based group proportion of 90--100\% effectively requires that almost all vehicles belong to some group;
in particular, it requires that even small-PCU vehicles such as motorcycles are continually engaged in detected leader--follower relations, which in practice entails spatial proximity and a substantial degree of behavioral coupling with surrounding vehicles.
Such conditions are generally difficult to maintain in low-density traffic,
where inter-vehicle distances are large, small vehicles can travel more freely,
and platoon-like synchronization is less likely to persist.
Therefore, it is natural that low-density states tend to be absent when the vehicle-count based group proportion reaches 90--100\%.
By contrast, a PCU-based group proportion can reach 90--100\% even when some small vehicles remain outside groups
(e.g., when groups are dominated by large-PCU vehicles), so the absence of low-density states is not necessarily expected.
That said, if groups can be maintained even at low density---for example, by preserving larger headways while still moving cohesively---then
low-density states could in principle occur even under vehicle-based 90--100\% grouping.

\subsection{Relationship between mean speed of groups and their entropy}
\label{subsec:entropy}

Section~\ref{subsec:median} and Section~\ref{subsec:scatter} suggested that maintaining an appropriate group proportion may be associated with better traffic performance.
% Section~\ref{subsec:median} および Section~\ref{subsec:scatter} では，適切な group 割合を維持することが交通パフォーマンスの向上と関連しうることが示唆された。
In the final part of this section, we examine whether the degree of vehicle-type mixing within a group---quantified by entropy---is related to the mean speed achieved by that group, which serves as a proxy for flow performance.
% 本研究の最後のパートでは，group 内の車種混合の程度（エントロピーで定量化）と，当該 group が達成した平均速度（流量パフォーマンスの代理指標）との関連を検討する。
This analysis is intended to clarify whether entropy alone provides explanatory power for speed differences and, if so, what types of group structures might be desirable for enhancing traffic performance.
% この分析は，エントロピー単体が速度差の説明力を持つかどうか，また（もし持つなら）交通パフォーマンス向上のために望ましい group 構造がどのようなものかを明らかにすることを目的とする。

% --- Entropy definition block (UNCHANGED as requested) ---
Note that, to quantify the degree of vehicle-type mixing within each group, entropy was used.
On the one hand, the entropy is zero when a group consists of only one vehicle type. 
On the other hand, the entropy reaches one when a group contains an equal number of vehicles from each of the four vehicle types.
The entropy is calculated using Equation~\ref{entropy}:
\begin{equation} \label{entropy}
{\mathrm{Entropy}}_{i} = -\sum_{j = 1}^{4} p_{ij}\log_4{ p_{ij} }
\end{equation}
where\\
$j$: Vehicle type $j$\\
$p_{ij}$: Occupation rate of type $j$ in group $i$, based on the number of vehicle type $j$.

By using the logarithmic base of four, which corresponds to the number of vehicle types, the maximum value of entropy is normalized to 1. In calculating the group entropy, the value of $p_{ij}$ is defined based on the number of vehicles of type $j$. This means, for example, that a group consisting of one motorcycle and one heavy vehicle is considered to have the same level of mixing as a group consisting of one motorcycle and one auto-rickshaw.
Although it is also possible to consider mixing based on PCU, this study focuses on how vehicle types themselves are mixed. Therefore, the degree of mixing is measured using the number of vehicles.
% --- End of unchanged block ---

\subsubsection{Result}
\label{subsubsec:entResult}

For each traffic situation, we estimated an OLS regression of speed on entropy using all observations, and computed cluster-robust inference by clustering standard errors by group ID (i.e., an identifier for each unique combination of vehicle-type composition and leader--follower network structure).
% 各 traffic state について，全観測点を用いて「速度をエントロピーで説明する」OLS回帰を推定し，group 構造レベル（group ID）でクラスタ化したロバスト推論を行った。
Table~\ref{tab:entropy_cluster_robust} reports the estimated slope for entropy, the 95\% confidence interval based on cluster-robust standard errors, and the corresponding p-value, together with the number of observations ($n_{\mathrm{points}}$) and clusters ($n_{\mathrm{cluster}}$).
% 表\ref{tab:entropy_cluster_robust} に，エントロピーの傾き推定値，クラスタ・ロバスト標準誤差に基づく95\%信頼区間，p値，および観測点数（$n_{\mathrm{points}}$）とクラスタ数（$n_{\mathrm{cluster}}$）を示す。

\begin{table}[h]
\centering
\caption{All-observation regressions of speed on entropy with cluster-robust inference (clustered by group ID).}
% 全観測点を用いた「速度 vs エントロピー」回帰の結果（group\_idでクラスタ化したロバスト推論）。
\label{tab:entropy_cluster_robust}
\begin{tabular}{l|rr|r|c|r||rr|r|c|r}
\hline
\multirow{2}{*}{State}
& \multicolumn{5}{c||}{$\lambda^{\mathrm{O}}_{i} \geq 0.5$~(/frame)}
& \multicolumn{5}{c}{$\lambda^{\mathrm{O}}_{i} \geq 0.1$~(/frame)} \\
\cline{2-11}
& $n_{\mathrm{points}}$ & $n_{\mathrm{cluster}}$ & Slope & 95\% CI & $p$
& $n_{\mathrm{points}}$ & $n_{\mathrm{cluster}}$ & Slope & 95\% CI & $p$ \\
\hline
AJ & 438{,}691 & 31 & $-0.205$ & $[-0.51,\ 0.10]$ & 0.184
   & 438{,}691 & 31 & $0.396$  & $[-0.16,\ 0.95]$ & 0.153 \\
OT & 129{,}130 & 17 & $-0.585$ & $[-1.39,\ 0.22]$ & 0.145
   & 129{,}130 & 17 & $-0.229$ & $[-0.85,\ 0.39]$ & 0.446 \\
AF & 209{,}109 & 4  & $-0.147$ & $[-4.67,\ 4.38]$ & 0.924
   & 209{,}109 & 4  & $-1.373$ & $[-3.63,\ 0.89]$ & 0.149 \\
DJ & 301{,}155 & 10 & $0.360$  & $[-0.29,\ 1.01]$ & 0.240
   & 301{,}155 & 10 & $-0.355$ & $[-0.66,\ -0.05]$ & 0.027 \\
\hline
\end{tabular}
\end{table}

Under $\lambda^{\mathrm{O}}_{i} \geq 0.5$, the 95\% confidence intervals of the entropy slopes include zero for all traffic states, and none of the slopes are statistically significant.
% $\lambda^{\mathrm{O}}_{i} \geq 0.5$ では，すべての traffic state において傾きの95\%信頼区間が0を含み，統計的に有意な傾きは得られなかった。
Under $\lambda^{\mathrm{O}}_{i} \geq 0.1$, DJ shows a negative slope with a 95\% confidence interval that does not include zero ($p=0.027$), whereas the confidence intervals for AJ, OT, and AF include zero.
% $\lambda^{\mathrm{O}}_{i} \geq 0.1$ では，DJ は95\%信頼区間が0を含まない負の傾きを示した（$p=0.027$）。一方，AJ・OT・AF はいずれも信頼区間が0を含む。

\subsubsection{Discussion}
\label{subsubsec:entDisc}

Table~\ref{tab:entropy_cluster_robust} shows that the estimated relationship between group entropy and speed is not robust across traffic situations or across the two frequency thresholds.
% 表\ref{tab:entropy_cluster_robust} より，group のエントロピーと速度の推定された関係は，traffic situation 間および2つの出現頻度閾値の間でロバストではないことが分かる。

Under $\lambda^{\mathrm{O}}_{i} \geq 0.5$~(/frame), the 95\% confidence intervals include zero for all traffic situations, indicating no statistically reliable association between entropy and speed at this threshold.
% $\lambda^{\mathrm{O}}_{i} \geq 0.5$~(/frame) では，すべての traffic situation において95\%信頼区間が0を含み，この閾値の下ではエントロピーと速度の統計的に信頼できる関連は確認されない。

Under $\lambda^{\mathrm{O}}_{i} \geq 0.1$~(/frame), DJ exhibits a negative slope whose 95\% confidence interval does not include zero ($p=0.027$), whereas AJ, OT, and AF remain non-significant with confidence intervals that include zero.
% $\lambda^{\mathrm{O}}_{i} \geq 0.1$~(/frame) では，DJのみが95\%信頼区間が0を含まない負の傾きを示した（$p=0.027$）。一方でAJ・OT・AFは信頼区間が0を含み，有意ではない。
However, this apparent DJ association does not replicate under $\lambda^{\mathrm{O}}_{i} \geq 0.5$~(/frame), where the estimated slope becomes positive and non-significant.
% ただし，このDJに見られる関連は $\lambda^{\mathrm{O}}_{i} \geq 0.5$~(/frame) では再現されず，この条件では推定された傾きは正となり，かつ有意ではない。

Moreover, the slope sign is not stable across thresholds in some traffic situations (e.g., AJ and DJ), further suggesting that entropy alone does not provide a consistent explanation for speed variations.
% さらに，一部の traffic situation（例：AJおよびDJ）では閾値間で傾きの符号が安定しておらず，エントロピー単体では速度変動を一貫して説明できないことが示唆される。

Taken together, the state-wise analyses based on all observations with cluster-robust inference indicate that any association between entropy and speed, if present, is not consistent across traffic situations or across frequency thresholds.
% 以上より，全観測点を用いクラスタ・ロバスト推論に基づく traffic situation 別分析の範囲では，エントロピーと速度の関連が存在するとしても，traffic situation 間や出現頻度閾値間で一貫しているとは言い難い。

Although entropy provides a compact summary of vehicle-type mixing within a group, richer structural descriptors will likely be necessary to explain traffic performance more robustly. Promising candidates include network-based characteristics within groups, such as in-/out-degree distributions, network length, and network density. These descriptors may further be refined by incorporating weights on leader--follower relations based on vehicle-type-specific dynamic and physical characteristics, as well as time-varying interaction strengths. In addition, the densities and speeds inside and outside groups, and the differences between them, may offer useful information for linking internal group structure to traffic performance.
%entropy は group 内の車種混合を簡潔に要約する指標ではあるが、交通性能をより頑健に説明するには、より豊かな structural descriptors が必要になると考えられる。有望な候補としては、group 内の in-/out-degree distributions、network length、network density などのネットワーク的特徴が挙げられる。さらに、これらの記述子は、車種ごとの動的・物理的特性に基づく leader–follower 関係の重み付けや、時間変動する interaction strength を組み込むことで、より精緻化できる可能性がある。加えて、group 内外の密度や速度、およびそれらの差も、内部構造と交通性能とを結びつけるうえで有用な情報になりうる。

\section{Discussion based on the overall results}
\label{subsec:consensus_discussion_final}

%表~\ref{tab:majority_median_vehicle}--\ref{tab:majority_scatter_pcu} は，セグメント別 median 解析および scatter 解析について，多数決（Consensus）の結果のみを整理したものである。median 解析は flow--density 構造の代表的傾向を，scatter 解析は分布の制約や極値的事例を捉える点で補完的であり，多数決結果を用いることで，両者に共通する頑健な特徴と，集約方法に依存して現れる特徴とを区別できる。
Tables~\ref{tab:majority_median_vehicle}--\ref{tab:majority_scatter_pcu}
summarize only the consensus (majority-vote) results obtained from
the segment-wise median analysis and scatter analysis.
While the median analysis captures the representative trends of the
flow--density structure,
the scatter analysis complements it by preserving distributional constraints
and extreme cases.
By focusing on the consensus results,
it becomes possible to distinguish features that are robust across both analyses
from those that emerge depending on the aggregation method.

%==============================
% Majority-vote-only tables (Median)
%==============================

\begin{table}[h]
\centering
\caption{Majority-vote summary of flow--density characteristics (segment-wise medians): vehicle-count-based group proportion.}
\label{tab:majority_median_vehicle}
\begin{tabular}{p{9.6cm} | c c}
\hline
\textbf{Feature} & $\lambda^{\mathrm{O}}_{i}\geq 0.5$ & $\lambda^{\mathrm{O}}_{i}\geq 0.1$ \\
\hline
(1) Particularly small group proportion yields the highest median flow. 
& $\bigcirc$\textsuperscript{-} & \ding{55}\textsuperscript{+} \\
(2) Particularly large group proportion yields relatively high median flow. 
& \ding{55}\textsuperscript{+} & \ding{55} \\
(3) In medium density, median flow decreases as group proportion increases. 
& $\bigcirc$\textsuperscript{-} & $\bigcirc$ \\
(4) In high density, moderate group proportions (20--40\% or 40--60\%) yield relatively large median flow. 
& \ding{55}\textsuperscript{+} & $\bigtriangleup$ \\
\hline
\end{tabular}

\vspace{0.6em}
\begin{flushleft}
\textbf{Legend:}
$\bigcirc$ supported; $\bigtriangleup$ partially/ambiguously supported; \ding{55} unsupported. 
$\textsuperscript{-}$: majority vote leans toward support but not fully consistent; 
$\textsuperscript{+}$: majority vote leans toward non-support but not entirely absent.
\end{flushleft}
\end{table}

\begin{table}[h]
\centering
\caption{Majority-vote summary of flow--density characteristics (segment-wise medians): PCU-based group proportion.}
\label{tab:majority_median_pcu}
\begin{tabular}{p{9.6cm} | c c}
\hline
\textbf{Feature} & $\lambda^{\mathrm{O}}_{i}\geq 0.5$ & $\lambda^{\mathrm{O}}_{i}\geq 0.1$ \\
\hline
(1) Particularly small group proportion yields the highest median flow. 
& $\bigcirc$\textsuperscript{-} & \ding{55}\textsuperscript{+} \\
(2) Particularly large group proportion yields relatively high median flow. 
& \ding{55} & \ding{55} \\
(3) In medium density, median flow decreases as group proportion increases. 
& $\bigcirc$\textsuperscript{-} & $\bigcirc$\textsuperscript{-} \\
(4) In high density, moderate group proportions (20--40\% or 40--60\%) yield relatively large median flow. 
& $\bigtriangleup$ & $\bigtriangleup$\textsuperscript{-} \\
\hline
\end{tabular}
\vspace{0.6em}
\begin{flushleft}
\textbf{Legend:} Same as Table~\ref{tab:majority_median_vehicle}.
\end{flushleft}
\end{table}

%==============================
% Majority-vote-only tables (Scatter)
%==============================

\begin{table}[h]
\centering
\caption{Majority-vote summary of flow--density characteristics (scatter plots): vehicle-count-based group proportion.}
\label{tab:majority_scatter_vehicle}
\begin{tabular}{p{9.6cm} | c c}
\hline
\textbf{Feature} & $\lambda^{\mathrm{O}}_{i}\geq 0.5$ & $\lambda^{\mathrm{O}}_{i}\geq 0.1$ \\
\hline
(A) 0--10\% group proportion concentrates in high-flow free-flow regime. 
& $\bigcirc$ & $\bigcirc$ \\
(B) 0--10\% group proportion rarely exhibits high density. 
& $\bigcirc$ & $\bigcirc$ \\
(C) 10--20\% group proportion has scarce low-density points. 
& $\bigcirc$ & $\bigcirc$ \\
(D) Highest flow is observed at moderate group proportion (30--60\%). 
& $\bigcirc$\textsuperscript{-} & $\bigcirc$\textsuperscript{-} \\
(E) Beyond 50\%, medium-density/medium-flow points decrease. 
& $\bigtriangleup$\textsuperscript{+} & $\bigtriangleup$\textsuperscript{-} \\
(F) At 90--100\%, high density does not occur. 
& $\bigcirc$ & \ding{55}\textsuperscript{+} \\
(G) At 90--100\%, low density does not occur. 
& $\bigcirc$ & $\bigcirc$ \\
\hline
\end{tabular}
\vspace{0.6em}
\begin{flushleft}
\textbf{Legend:} Same as Table~\ref{tab:majority_median_vehicle}.
\end{flushleft}
\end{table}

\begin{table}[h]
\centering
\caption{Majority-vote summary of flow--density characteristics (scatter plots): PCU-based group proportion.}
\label{tab:majority_scatter_pcu}
\begin{tabular}{p{9.6cm} | c c}
\hline
\textbf{Feature} & $\lambda^{\mathrm{O}}_{i}\geq 0.5$ & $\lambda^{\mathrm{O}}_{i}\geq 0.1$ \\
\hline
(A) 0--10\% group proportion concentrates in high-flow free-flow regime. 
& $\bigcirc$ & $\bigcirc$ \\
(B) 0--10\% group proportion rarely exhibits high density. 
& $\bigtriangleup$\textsuperscript{+} & $\bigcirc$ \\
(C) 10--20\% group proportion has scarce low-density points. 
& $\bigcirc$ & $\bigcirc$ \\
(D) Highest flow is observed at moderate group proportion (30--60\%). 
& $\bigcirc$\textsuperscript{-} & $\bigcirc$\textsuperscript{-} \\
(E) Beyond 50\%, medium-density/medium-flow points decrease. 
& $\bigtriangleup$\textsuperscript{+} & $\bigtriangleup$\textsuperscript{-} \\
(F) At 90--100\%, high density does not occur. 
& $\bigcirc$ & \ding{55}\textsuperscript{+} \\
(G) At 90--100\%, low density does not occur. 
& \ding{55}\textsuperscript{+} & \ding{55}\textsuperscript{+} \\
\hline
\end{tabular}
\vspace{0.6em}
\begin{flushleft}
\textbf{Legend:} Same as Table~\ref{tab:majority_median_vehicle}.
\end{flushleft}
\end{table}

%\clearpage

%\subsubsection{定義・閾値に依らず比較的頑健な特徴}
%\label{subsubsec:finalDisc_robust}
%median に基づく多数決結果からは，中密度域においてグループ割合が増加するほど median flow が低下する（特徴(\ref{median_char3})）という関係が，グループ割合の定義（台数割合／PCU 割合）やグループ抽出閾値 $\lambda^{\mathrm{O}}_{i}$ によらず概ね支持されることが分かる（表~\ref{tab:majority_median_vehicle}，表~\ref{tab:majority_median_pcu}）。
%これは，中密度域では，検出されるグループの増加が典型的には速度低下を伴い，median の意味での流量低下に結びつくことを示唆する。
%減速波によって形成される車群（クラスター）がグループとして検出される場合，速度低下を通じて flow が低下するという解釈は自然である。
%\subsubsection{Relatively robust features independent of proportion definitions and thresholds}
\paragraph{Relatively robust features independent of proportion definitions and thresholds}
\label{subsubsec:finalDisc_robust}
The majority-vote results based on the median analysis indicate that,
in the intermediate-density regime,
the relationship in which the median flow decreases as the group proportion increases
(feature~(\ref{median_char3}))
is generally supported regardless of
the definition of the group proportion (vehicle-based or PCU-based)
or the group extraction threshold $\lambda^{\mathrm{O}}_{i}$
(Tables~\ref{tab:majority_median_vehicle} and~\ref{tab:majority_median_pcu}).
This suggests that, in the intermediate-density regime,
an increase in the number of detected groups is typically accompanied by a reduction in speed,
which in turn leads to a decrease in flow in the median sense.
When vehicle clusters formed by deceleration are detected as groups,
it is natural to interpret the resulting decrease in flow
as a consequence of the associated speed reduction.

%一方，scatter に基づく多数決結果は，グループ割合に応じて実現しやすい密度域が制約されることを示す。
%特に，0--10\% および 10--20\% の割合域における分布上の特徴（特徴(\ref{scat_char1})--(\ref{scat_char3})）は，Section~\ref{subsec:scatter} で詳述した通り，多くの設定で一貫して観察されている（表~\ref{tab:majority_scatter_vehicle}--\ref{tab:majority_scatter_pcu}）。
%これらの結果は，グループ割合が flow の大小だけでなく，観測される density の範囲（低密度・高密度が出現しやすいかどうか）にも影響し得ることを示す。
%（i）密度状態が速度同調やクラスター形成を通じてグループ割合を規定する可能性
%ももちろん考えられる一方で、
%（ii）グループ割合が密度状態の可到達範囲を変化させる可能性，
%（iii）車種構成などの第三の要因が両者に同時に影響している可能性
%も考えられる。
In contrast, the majority-vote results based on the scatter analysis indicate that
the density regimes that are likely to be realized are constrained depending on
the group proportion.
In particular, the distributional features observed in the
0--10\% and 10--20\% proportion ranges
(Features~\ref{scat_char1}--\ref{scat_char3})
have been consistently observed across many settings,
as described in detail in Section~\ref{subsec:scatter}
(Tables~\ref{tab:majority_scatter_vehicle}--\ref{tab:majority_scatter_pcu}).
These results suggest that the group proportion may influence not only
the magnitude of flow,
but also the range of density that is observed,
that is, whether low-density or high-density states are likely to appear.
While it is certainly possible that
\begin{itemize}
\item the density state determines the group proportion
through speed synchronization or cluster formation,
it is also conceivable that
\item the group proportion alters the reachable range of density states,
or that
\item a third factor, such as vehicle composition,
\end{itemize}
simultaneously affects both.

%\subsubsection{median と scatter が分岐する点：典型構造と極値（高流量事例）の違い}
%\label{subsubsec:finalDisc_medianScatDif}
%最も重要な相違は，「最大流量がどの割合で現れるか」に関して現れる。scatter では，中程度のグループ割合（30--60\%）で最大流量が観測されやすい（特徴(\ref{scat_char4})）という傾向が一貫して $\bigcirc^{ - }$ として支持される。
%これは，\emph{ピーク性能}を示す（比較的まれな）高流量点が，中程度の割合で出現しやすいことを示唆する。
%\subsubsection{Where median and scatter diverge: differences between typical structures and extreme (high-flow) cases}
\paragraph{Where median and scatter diverge: differences between typical structures and extreme (high-flow) cases}
\label{subsubsec:finalDisc_medianScatDif}

The most important difference appears in terms of
which group proportion is associated with the maximum flow.
In the scatter analysis,
the tendency for the maximum flow to be observed at
moderate group proportions (30--60\%)
(Feature~\ref{scat_char4})
is consistently supported as $\bigcirc^{ - }$.
This suggests that high-flow points representing
peak performance, which are relatively rare, tend to emerge at intermediate group proportions.

%しかし median に基づく結果では，高密度域で中程度割合が相対的に大きな median flow を示す（特徴(\ref{median_char4})）ことはせいぜい部分支持にとどまり，さらに「グループ割合が極めて大きいと高い流量が得られる」（特徴(\ref{median_char2})）という主張は多くの条件で支持されない。
%これらを総合すると，scatter が捉える「高流量の極値的事例」と，median が捉える「典型的な flow--density 構造」は必ずしも一致しない。
However, in the median-based results,
the tendency for moderate group proportions to exhibit
relatively large median flow in the high-density regime
(Feature~(\ref{median_char4}))
receives at most partial support.
Moreover, the claim that
very large group proportions lead to high flow
(Feature~(\ref{median_char2}))
is not supported under many conditions.
Taken together, these findings indicate that
the extreme high-flow cases captured by the scatter analysis
and the typical flow--density structure captured by the median analysis
do not necessarily coincide.

%このことは，同一の密度・割合条件の下でも，交通が「頻出する低効率状態」と「希少だが高効率な状態」の二つの異なる挙動様式を取り得ることを示唆している。具体的には，中密度域では，減速に起因するグループ増加が支配的となり，典型的には流量低下が生じる一方で，中程度のグループ割合においては，より協調的な配置や運動が成立した場合に限って，高流量状態が点として観測されるのではないか。
This suggests that,
even under the same density and group proportion conditions,
traffic can exhibit two distinct modes of behavior:
a frequently observed low-efficiency state
and a rare but high-efficiency state.
More specifically, in the intermediate-density regime,
an increase in groups caused by deceleration tends to be dominant,
typically resulting in a reduction in flow.
In contrast, at moderate group proportions,
high-flow states may be observed as isolated points
only when more cooperative configurations and motions are realized.

%これらの知見は、本研究で採用した group-based perspective の付加価値を明確にするものでもある。ここで用いている group proportion は、単に密度や車種構成を言い換えたものではなく、自己組織化された leader–follower structures の prevalence、すなわち、どの程度の車両が統計的に有意な関係構造に組み込まれているかを表している。この視点から見ることで、本研究の分析は、従来の集約変数だけでは直接には表現されない traffic organization の側面を明示化している。とりわけ、本研究は、group prevalence と flow–density structure のあいだに、非線形かつ traffic-situation-dependent な関係が存在することを示し、さらに、代表的な median trend と、まれに現れる peak/high-flow cases とのずれを明らかにした。したがって、本研究の貢献は、従来の集約指標を置き換えることにあるのではなく、relation-based な organization が heterogeneous disordered traffic を理解するための、追加的かつ分析的に意味のある次元を与えることを示した点にある。
These findings also help clarify the added value of the group-based perspective adopted in this study. The group proportion used here is not simply a restatement of density or vehicle composition; rather, it represents the prevalence of self-organized leader--follower structures, that is, the extent to which vehicles are incorporated into statistically significant relational formations. From this perspective, the present analysis makes explicit aspects of traffic organization that are not directly expressed by conventional aggregate variables alone. In particular, it reveals a nonlinear and traffic-situation-dependent relationship between group prevalence and the flow--density structure, and it highlights a divergence between representative median trends and rare peak/high-flow cases. Thus, the contribution of the present study is not to replace conventional aggregate indicators, but to show that relation-based organization provides an additional and analytically meaningful dimension for understanding heterogeneous disordered traffic.

\paragraph{Sensitivity to the occurrence-frequency threshold $\lambda^{\mathrm{O}}_{i}$}
\label{subsubsec:finalDisc_sensitivity}
%この consensus の結果は、2つの出現頻度閾値をまたいで安定している知見と、閾値依存的な知見とがあることを示している。特に、median 分析において中密度域で group proportion の増加とともに flow が低下する傾向は、両閾値で概ね支持される。これに対して、高密度域で中程度の group proportion が比較的大きな flow を示すという median ベースの傾向は部分支持にとどまるが、group proportion を PCU ベースで定義した場合にやや明瞭に現れる傾向がある。一方、scatter 分析では、peak/high-flow cases が中程度の group proportion に現れやすいという傾向が、2つの閾値をまたいでより一貫して観察される。これらの解釈は Tables 12–15 の consensus summary に基づいている。
The consensus results indicate that some findings are stable across the two occurrence-frequency thresholds, whereas others are threshold-dependent. In particular, the median-based tendency for flow to decrease with increasing group proportion in the intermediate-density regime remains broadly supported across both thresholds. By contrast, the median-based tendency for moderate group proportions to exhibit relatively large flow in the high-density regime receives only partial support, although it tends to appear somewhat more clearly when group proportion is defined on a PCU basis. In the scatter analysis, however, the tendency for peak/high-flow cases to appear at moderate group proportions is more consistently observed across the two thresholds. These interpretations are based on the consensus summaries in Tables~\ref{tab:majority_median_vehicle}--\ref{tab:majority_scatter_pcu}.

%一方、scatter 分析において極端な group proportion 範囲（90–100%）で観察される分布端の特徴、たとえば高密度あるいは低密度状態が実現しないといった特徴は、出現頻度閾値 の選択に依存する傾向がある。
From the consensus results,
the distributional endpoint features observed in the scatter analysis
in extreme group proportion ranges (90--100\%)---such as the absence of
realized high-density or low-density states---tend to depend on the
choice of the occurrence-frequency threshold $\lambda^{\mathrm{O}}_{i}$.

%これは次のように理解できる。より厳しい閾値のもとでは、高頻度の group のみが抽出されるため、90–100% カテゴリは、ほとんどすべての車両が少数の非常に頻繁な group に属する交通状態によって支配される。そのような条件は、多くのまれな configuration も現れる最高密度域では起こりにくく、また leader–follower 関係がより流動的である最低密度域でも起こりにくい。これに対し、閾値を ≥0.1 に緩和すると、頻度は低いが統計的には支持される group も含まれるため、極端な group-proportion class に寄与する交通状態の範囲が広がる。したがって、このような極端割合領域における endpoint feature は、上で要約したより一般的な傾向に比べて、分析上の閾値により敏感であるため、とくに慎重に解釈する必要がある。
This can be understood as follows.
Under the stricter threshold,
where only highly frequent groups are extracted,
the 90--100\% category is dominated by traffic states in which almost all vehicles
belong to a small set of very frequent groups.
Such conditions are unlikely to occur at the highest densities,
where many rare configurations also appear,
or at the lowest densities,
where leader--follower relations are more fluid.
When the threshold is relaxed to $\lambda^{\mathrm{O}}_{i}\geq 0.1$,
less frequent but still statistically supported groups are included,
which broadens the range of traffic states contributing to the extreme group-proportion classes.
Therefore, the endpoint features in these extreme proportion ranges should be interpreted with particular caution, because they are more sensitive to the analytical threshold than the broader tendencies summarized above.

\paragraph{Connection with entropy: entropy alone is insufficient and richer descriptors may be needed}
% エントロピーとの接続：エントロピー単体では不十分で，より豊かな記述子が必要となり得る
\label{subsubsec:finalDisc_entropy}

The entropy analysis (Section~\ref{subsec:entropy}) indicates that the entropy--speed relationship is not robust across traffic situations.
% エントロピー解析（Section~\ref{subsec:entropy}）は，エントロピーと速度の関係が traffic situation を通じてロバストではないことを示す。
In particular, the state-wise estimates based on all observations with cluster-robust inference show that most effects are not statistically distinguishable from zero, and the sign is not stable across the two occurrence-frequency thresholds (Table~\ref{tab:entropy_cluster_robust}).
% 特に，全観測点に基づくクラスタ・ロバスト推論による traffic situation 別推定では，多くの効果はゼロと統計的に区別できず，また2つの出現頻度閾値の間で符号も安定しない（表\ref{tab:entropy_cluster_robust}）。
Although DJ under $\lambda^{\mathrm{O}}_{i}\geq 0.1$~(/frame) shows a negative association, this pattern does not replicate under the stricter threshold.
% $\lambda^{\mathrm{O}}_{i}\geq 0.1$~(/frame) におけるDJでは負の関連が見られるものの，より厳しい閾値ではこのパターンは再現しない。

Taken together, these results suggest that the internal composition of groups, as summarized by entropy alone, does not provide a robust explanation for speed variations across traffic situations.
% 以上より，エントロピー単体で要約される group の内部構成は，traffic situation を横断して速度変動を頑健に説明するとは言い難い。
This does not rule out the possibility that internal structure matters; rather, it indicates that any effect is likely to depend on additional factors beyond entropy, such as the specific vehicle-type composition, interaction patterns, or traffic situation.
% これは内部構造の重要性を否定するものではなく，効果があるとしても，車種構成の具体形，車両間相互作用のパターン，あるいは traffic situation など，エントロピー以外の要因に依存する可能性を示す。
Accordingly, linking group composition to traffic performance may require measures that describe group structure in greater detail, rather than relying on a single index such as entropy.
% したがって，group の内部構成と交通パフォーマンスを結び付けるには，エントロピーのように単一の指標ではなく，より group の構造を詳細に記述できる指標が必要である可能性がある。
%※例えば車種の比率ベクトル（4種類の割合そのもの）　特定の組合せ（例：HV含有の有無、二輪比率など）　ネットワーク指標（連結度、中心性、密度、階層性 など）group のサイズや持続時間（同じ混合でも規模が違う）

%%%%%%%%%いまここ20250424 method　とconclusionかこう

\section{Conclusion}
\label{sec:concl}
% [UNCHANGED PARAGRAPH]
%無秩序で車線規律の緩いheterogeneous disordered trafficでは、さまざまな車種が相互に作用し、自己組織化的なグループを形成することがある。それがマクロな交通特性に与える影響は十分に解明されていない。本研究は、groupの出現頻度や構成が、heterogeneous disordered trafficにおける流量–密度関係にどのような影響を及ぼすのかを明らかにすることを目的とした。
In heterogeneous disordered traffic, various types of vehicles interact without strict lane discipline, forming self-organized groups that may influence overall traffic performance. While previous studies have shown evidence of such group formation, the impact of these groups on macroscopic traffic characteristics has remained unclear. This study aimed to clarify whether and how the prevalence and composition of vehicle groups affect flow–density relationships in heterogeneous disordered traffic.

% [UNCHANGED PARAGRAPH]
%本研究では、3種類のPCU推定法を用いて交通の異質性を考慮した流量–密度関係を構築し、group割合の変化が交通流に与える影響を分析した。
%なお、結果の頑健性を高めるため、異なる理論背景を持つPCUを3つ選んだ。
% また，本研究では median 解析と scatter 解析を併用し，それぞれが捉える情報（典型構造と分布・極値）を補完的に解釈しつつ，複数設定にまたがる多数決（consensus）により頑健な傾向を抽出する枠組みを採用した。
By employing three different PCU estimation methods, we constructed flow–density diagrams that account for traffic heterogeneity and examined how varying group proportions affect traffic flow.
To enhance the robustness of the analysis, we selected these PCU methods from different theoretical backgrounds.
In addition, to strengthen the interpretability and robustness of the findings, we combined segment-wise median analysis and scatter analysis, which provide complementary views---the former capturing representative (typical) flow--density structures and the latter preserving distributional constraints and extreme cases---and then extracted robust tendencies via a consensus (majority-vote) comparison across settings.

Through this analysis, the study showed that differences in group proportion are systematically associated with traffic flow characteristics depending on the traffic situation, rather than directly establishing a causal or mechanistic model. By contrasting the typical structures captured by medians with the extreme high-flow cases preserved in scatter plots, we also highlighted that these two perspectives do not necessarily coincide, suggesting that traffic may exhibit qualitatively different modes even under similar density and proportion conditions.
%この分析を通じて、本研究は、group proportion の違いが traffic situation に応じて traffic flow characteristics と系統的に関連していることを示したのであり、直接的な因果モデルや機構モデルを確立したものではない。さらに、median が捉える typical structures と scatter plots に保持された extreme high-flow cases とを対比することで、これら2つの視点が必ずしも一致しないことを示し、similar density と proportion の条件下でも traffic が qualitatively different modes を示しうることを示唆した。(3-1)

The main findings are summarized as follows.
% 主な知見を以下にまとめる。

% [UNCHANGED PARAGRAPH]
%要するに「group proportion が中くらいだと、いつも平均的によく流れるわけではない。けれども、うまく流れている“当たりの状態”は、そのあたりの group proportion で現れやすい。だから交通には、普通は効率が低いが、たまに高効率になるモードがあるのではないか。」
%---
% まず，group proportion は，flow--density relationship に対して非線形かつ traffic-situation-dependent な影響を及ぼすことが分かった。
% とくに，中程度の group proportion（30--60\%）は，より高密度な交通条件のもとで，
% median-based analysis において一貫して最も高い代表的 flow を与えるというよりも，
% むしろ peak-flow または high-flow cases とより明確に関連していた。(特徴4, D)
% これに対して，group proportion が 50\% を超えると，
% traffic の分布は低密度または高密度の極端側へ偏る傾向を示し，
% 中密度域での観測は少なくなった。(特徴5, E)
% とくに，scatter analysis は，peak-performance（high-flow）cases が
% 中程度の group proportion で現れやすいことを示唆する一方で，
% median analysis は，中密度域において，group proportion の増加が
% 代表的 flow の低下と結びつく傾向を示している。
% これらを総合すると，traffic には，頻繁に観測される低効率モードと，
% より希少だが高効率なモードとが共存している可能性が示唆される。
First, group proportions were found to exert a nonlinear and traffic-situation-dependent effect on flow–density relationships. 
In particular, moderate group proportions (30--60\%) were more clearly associated with peak-flow or high-flow cases under denser traffic conditions, rather than consistently producing the highest representative flow in the median-based analysis.
%(corresponding to Features 4 and D). 
In contrast, when the group proportion exceeded 50\%, the distribution of traffic tended to shift toward either low- or high-density extremes, with fewer observations in moderate-density ranges. % (Features 5 and E).
Notably, the scatter analysis suggests that peak-performance (high-flow) cases tend to appear at moderate group proportions, whereas the median analysis indicates that, in intermediate-density regimes, increasing group proportion is typically associated with reduced representative flow; taken together, these results suggest the coexistence of a frequently observed lower-efficiency mode and a rarer but higher-efficiency mode.

% [UNCHANGED PARAGRAPH]
%バイク中心group → 台数ベースとPCUベースで差：特徴7
Second, comparisons between group proportions based on vehicle counts and those based on PCUs revealed notable differences, especially when groups were composed predominantly of small-PCU vehicles such as motorcycles.% (Feature 7). 
This emphasizes the importance of selecting an appropriate normalization method when analyzing group effects in heterogeneous traffic environments to improve the flow.

% [UNCHANGED PARAGRAPH]
%自由流ではgroup割合が少ないと良好な流れ：特徴1, A
Third, in free-flow conditions, lower group proportions tended to be associated with higher flow, suggesting that minimal inter-vehicle coordination can be advantageous when traffic is sparse.
% (Features 1 and A). 
In contrast, under more congested conditions, moderate group proportions were more clearly associated with favorable or peak-flow outcomes, especially in the scatter-based results.

% [REVISED PARAGRAPH] (entropy-related bullet updated to match all+cluster-robust results)
Fourth, the entropy-based analysis of group composition did not show a consistent association between entropy and speed across traffic situations when using all observations with cluster-robust inference.
% 第四に，全観測点を用いたクラスタ・ロバスト推論に基づく解析では，エントロピーと速度の関連は traffic situation を通じて一貫して確認されなかった（表\ref{tab:entropy_cluster_robust}）。
Although DJ under $\lambda^{\mathrm{O}}_{i}\geq 0.1$~(/frame) showed a negative association, this pattern did not replicate under the stricter threshold, suggesting that entropy alone is insufficient to robustly explain speed variations.
Future work should therefore consider richer descriptors of internal group structure, including network-based characteristics, differences in density and speed inside and outside groups, and extended interaction models that incorporate weighted and time-varying leader--follower relations, lateral relations, and multilayer-network or hypergraph representations for multi-vehicle interactions.
% ただし $\lambda^{\mathrm{O}}_{i}\geq 0.1$~(/frame) のDJでは負の関連が見られた一方で，より厳しい閾値では再現されず，エントロピー単体では速度変動を頑健に説明するには不十分であることが示唆される。
%したがって今後は、ネットワーク的特徴、重み付きかつ時間変動する leader–follower 関係、ならびに group 内外の密度や速度の差を含む、より豊かな内部構造記述子を検討する必要がある。
%さらに今後の方法論的拡張としては、leader–follower 関係に加えて lateral relations を取り込むこと、weighted かつ time-varying な interaction strength を導入すること、さらに multi-vehicle interaction を扱う multilayer network や hypergraph 表現を検討することが挙げられる。

% [UNCHANGED PARAGRAPH]
% さいごに，median 解析と scatter 解析を対比することで，
% 代表的な flow--density 構造と，極値として現れる高流量事例とが，
% 交通挙動の異なる側面を反映していることが明らかとなった。
% median 解析は，中密度域においてグループ割合の増加が
% 典型的には流量低下と結びつく傾向を捉える一方で，
% scatter 解析は，中程度のグループ割合において出現しやすい，
% 比較的まれではあるが重要な高流量点を保持している。
% この乖離は，同程度の密度およびグループ割合条件の下であっても，
% 交通が，減速に起因するクラスター形成に支配された
% 頻出する低効率状態と，
% より協調的な配置や集団的運動が成立した場合にのみ現れる
% 希少だが高効率な状態との間を取り得ることを示唆している。
Finally, the contrast between the median and scatter analyses reveals that
the representative flow--density structure and the extreme high-flow cases
reflect different aspects of traffic behavior.
While the median analysis captures the typical tendency that increasing group proportion
in intermediate-density regimes is associated with reduced flow,
the scatter analysis preserves rare but important high-flow points
that emerge predominantly at moderate group proportions.
This discrepancy indicates that, even under comparable density and group proportion conditions,
traffic can alternate between a frequently realized, lower-efficiency state
dominated by deceleration-induced clustering,
and a less frequent but higher-efficiency state
in which more cooperative configurations or collective motions are realized.

%全体として、本研究の知見は、自己組織化されたグループの出現とその割合が、異種混合・無秩序交通における巨視的交通ダイナミクスと密接に関係していることを示している。
%さらに重要なのは、本研究が示した group-based perspective が、単なる密度や車種構成の言い換えではないという点である。group proportion を通じて、自己組織化された leader--follower structures の prevalence を明示的に測ることにより、本研究の分析は、従来の集約指標だけでは直接表現されない traffic organization の一側面を捉えている。この relation-based な視点によって、本研究は、group prevalence と flow--density structure のあいだに見られる非線形かつ traffic-situation-dependent な関連、および代表的な median trend と、まれに現れる peak/high-flow cases とのずれを明らかにすることができた。
%同時に、entropy によって捉えられた内部構成の効果は、交通状況をまたいで頑健ではなく、構成に関わる影響を扱うには、単一の entropy 指標を超えるより豊かな構造記述子が必要であることを示唆している。
Overall, the findings demonstrate that the prevalence of self-organized groups and their proportion are closely related to macroscopic traffic dynamics in heterogeneous disordered traffic.
More importantly, the present study shows that a group-based perspective is not merely a restatement of density or vehicle composition. By explicitly measuring the prevalence of self-organized leader--follower structures through group proportion, the analysis captures a dimension of traffic organization that is not directly represented by conventional aggregate indicators alone. This relation-based perspective made it possible to reveal the nonlinear and traffic-situation-dependent association between group prevalence and the flow--density structure, as well as the divergence between representative median trends and rare peak/high-flow cases.
At the same time, effects of internal composition captured by entropy were not robust across traffic situations, indicating that composition-related influences may require richer structural descriptors beyond a single entropy measure.

%さらに、本研究では因果関係を直接的に確立したわけではない。したがって、本研究の知見は、直接的な因果効果の証拠というよりも、記述的な関連および仮説生成的な結果として理解するのが適切である。
%また、group proportion が果たす役割は、車種構成が大きく異なる他地域では変化する可能性がある。したがって、本研究の知見の一般性を評価するには、より広い観測区間と複数地域から収集されたデータが必要である。
%さらに、group-based variables が従来の集約指標に対してどの程度の追加的説明力を持つのかを、より厳密に定量化することは、今後の重要な課題である。
%これらの知見は、とくにインフラ整備や交通規制が限定的な環境において、ボトムアップ型の交通流改善を目指す、group を考慮した交通制御戦略の設計に向けた基盤を与えるものである。
Moreover, the present study did not directly establish causal relationships. The findings are therefore better understood as descriptive associations and hypothesis-generating results, rather than as direct evidence of causal effects.
The role played by group proportion may vary in other regions where vehicle-type composition differs substantially. Assessing the generality of the present findings will therefore require wider observation sections and datasets collected from multiple regions.
A more rigorous quantification of the additional explanatory value of group-based variables relative to conventional aggregate indicators remains an important subject for future work.
In particular, analyzing the life cycle of individual groups, their formation, duration, and dissipation, for example through survival analysis of group lifetimes and its relation to flow, will require a group-identity tracking framework with explicit definitions of formation, continuation, merging, splitting, and dissipation, and is an important direction for future work.

Taken together, these insights provide a foundation for designing group-aware traffic control strategies aimed at enabling bottom-up flow improvement, particularly in environments with limited infrastructure or formal regulation. More specifically, the present findings may suggest condition-dependent directions. In free-flow conditions, such as off-peak urban arterials, one possible direction is for driver-assistance systems and traffic advisories to avoid inducing unnecessarily persistent platooning and to help preserve fluid leader--follower relations. At signalized approaches and sections upstream of bottlenecks in denser mixed traffic with a large share of motorcycles, measures that encourage favorable formations, such as stop-line separation or dedicated waiting areas for two-wheelers~\cite{shiomi2011evaluation} and ITS-based guidance supporting stable following among behaviorally compatible vehicle types, could be worth exploring as bottom-up interventions. Because the present findings are descriptive associations, such strategies should be regarded as hypotheses to be verified through microscopic simulation and field experiments before deployment.

%\begin{acknowledgement}
\section*{Acknowledgements}
We thank Naotaka Ishizawa, Prashant Walawalkar, and other members of NYK Auto Logistics (India) Private Ltd. for their support with field observations in India. We express our gratitude to Akihiro Fujita for his help with field observations in India. We also thank Chuuya Osugi, Ryo Kondo, Zhiyan Xie, Satoshi Yamamoto, Chenkai Zhang, Munkhtushig Munguntsetseg, and Ryo Kawazu for their invaluable support in analyzing the video footage.

\section*{Funding}
This study was supported in part by a research grant for Exploratory Research on Sustainable Humanosphere Science from the Research Institute for Sustainable Humanosphere (RISH) Kyoto University. This work was also supported by JSPS KAKENHI grant numbers JP25287026, JP18H05923, JP19K15246, JP23K13512, and JP26K07940 and partially supported by MEXT under ``Priority Issues and Exploratory Challenges on post-K (Supercomputer Fugaku)'' (Exploratory Challenge 2: Construction of Models for Interaction Among Multiple Socioeconomic Phenomena, Model Development, and its Applications for Enabling Robust and Optimized Social Transportation Systems) (Project ID: hp190163). 

\section*{Declaration of generative AI and AI-assisted technologies in the manuscript preparation process}
During the preparation of this work, the authors used ChatGPT in order to translate part of the text from Japanese and to check grammar/proofread the manuscript. After using this tool/service, the authors reviewed and edited the content as needed and take full responsibility for the content of the published article.

%\end{acknowledgement}

\clearpage % 付録を改ページし参考文献と視覚的に分離（共著者指摘・2026-07-16）
\appendix
\section{Time series of the group proportion}
\label{app:timeseries}
For completeness, this appendix presents the temporal variation of the vehicle-count-based group proportion over the entire observation period. Because the instantaneous proportion fluctuates strongly from frame to frame, a $30$~s moving average is applied to reveal the underlying trend.

\begin{figure}[htbp]
\centering
\includegraphics[width=\textwidth]{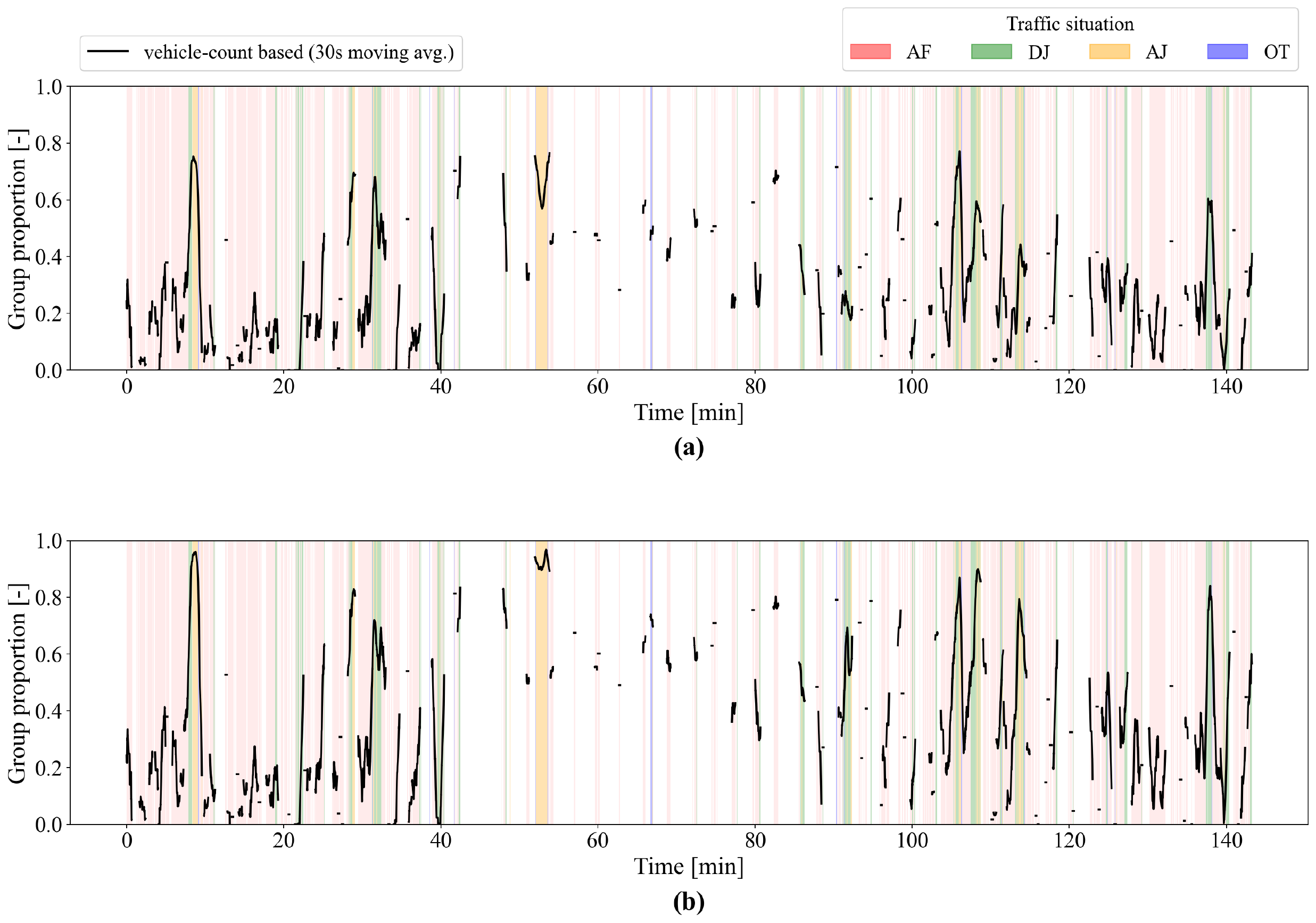}
\caption{Time series of the vehicle-count-based group proportion over the observation period, smoothed with a $30$~s moving average. (a) groups with $\lambda^{\mathrm{O}}_{i}\geq 0.5$~(/frame) are considered; (b) groups with $\lambda^{\mathrm{O}}_{i}\geq 0.1$~(/frame) are considered. The background shading indicates the traffic situation (AF, DJ, AJ, and OT).}
\label{fig:groupPropTimeSeries}
\end{figure}

\bibliographystyle{spmpsci}
\bibliography{mybibfile.bib}

@article{mason1997car,
  title={{Car-following model of multispecies systems of road traffic}},
  author={Mason, Anthony D and Woods, Andrew W},
  journal={Physical Review E},
  volume={55},
  number={3},
  pages={2203},
  year={1997},
  publisher={APS}
}

@phdthesis{lee2007agent,
  title={{An agent-based model to simulate motorcycle behaviour in mixed traffic flow}},
  author={Lee, Tzu-Chang},
  year={2007},
  school={Imperial College London (University of London)}
}

@article{sayer2000effect,
  title={{The effect of lead-vehicle size on driver following behavior}},
  author={Sayer, James R},
  year={2000},
  publisher={University of Michigan, Ann Arbor, Transportation Research Institute}
}

@article{sarvi2013heavy,
  title={{Heavy commercial vehicles-following behavior and interactions with different vehicle classes}},
  author={Sarvi, Majid},
  journal={Journal of advanced transportation},
  volume={47},
  number={6},
  pages={572--580},
  year={2013},
  publisher={Wiley Online Library}
}

@inproceedings{aghabayk2011examining,
  title={{Examining vehicle interactions during a vehicle-following manoeuvre}},
  author={Aghabayk, Kayvan and Young, William and Sarvi, Majid and Wang, Yibing},
  booktitle={Australasian Transport Research Forum (ATRF), 34th, 2011, Adelaide, South Australia, Australia},
  volume={34},
  number={0084},
  year={2011}
}

@article{chen2016car,
  title={{Car-Following and Lane-Changing Behavior Involving Heavy Vehicles}},
  author={Chen, Danjue and Ahn, Soyoung and Bang, Soohyuk and Noyce, David},
  journal={Transportation Research Record: Journal of the Transportation Research Board},
  number={2561},
  pages={89--97},
  year={2016},
  publisher={Transportation Research Board of the National Academies}
}

@article{nagahama2017dependence,
  title={{Dependence of driving characteristics upon follower--leader combination}},
  author={Nagahama, Akihito and Yanagisawa, Daichi and Nishinari, Katsuhiro},
  journal={Physica A: Statistical Mechanics and its Applications},
  volume={483},
  pages={503--516},
  year={2017},
  publisher={Elsevier}
}

@article{nagahama2022certain,
  title={Certain Types of Vehicles in Heterogeneous Traffic in India Tend to Gather},
  author={Nagahama, Akihito and Wada, Takahiro and Yanagisawa, Daichi and Nishinari, Katsuhiro},
  journal={Journal of the Eastern Asia Society for Transportation Studies},
  volume={14},
  pages={1794--1813},
  year={2022},
  publisher={Eastern Asia Society for Transportation Studies}
}

@inproceedings{babenko2009visual,
  title={Visual tracking with online multiple instance learning},
  author={Babenko, Boris and Yang, Ming-Hsuan and Belongie, Serge},
  booktitle={2009 IEEE Conference on computer vision and Pattern Recognition},
  pages={983--990},
  year={2009},
  organization={IEEE}
}

@article{elmansouri2020urban,
  title={{Urban transportation in Libya: An overview}},
  author={Elmansouri, Omar and Almhroog, Abdulmojeb and Badi, Ibrahim},
  journal={Transportation research interdisciplinary perspectives},
  volume={8},
  pages={100161},
  year={2020},
  publisher={Elsevier}
}

@article{fattah2022insights,
  title={{Insights into the socio-economic impacts of traffic congestion in the port and industrial areas of Chittagong city, Bangladesh}},
  author={Fattah, Md Abdul and Morshed, Syed Riad and Kafy, Abdulla-Al},
  journal={Transportation Engineering},
  volume={9},
  pages={100122},
  year={2022},
  publisher={Elsevier}
}

@article{samal2021adverse,
  title={Adverse effect of congestion on economy, health and environment under mixed traffic scenario},
  author={Samal, Satya Ranjan and Mohanty, Malaya and Santhakumar, S Moses},
  journal={Transportation in Developing Economies},
  volume={7},
  number={2},
  pages={15},
  year={2021},
  publisher={Springer}
 }

@article{liu2016modeling,
  title={Modeling and simulation of the car-truck heterogeneous traffic flow based on a nonlinear car-following model},
  author={Liu, Lan and Zhu, Liling and Yang, Da},
  journal={Applied Mathematics and Computation},
  volume={273},
  pages={706--717},
  year={2016},
  publisher={Elsevier}
}

@article{nagahama2021detection,
author = {Nagahama, Akihito and Wada, Takahiro and Yanagisawa, Daichi and Nishinari, Katsuhiro},
year = {2021},
month = {02},
pages = {125789},
title = {Detection of leader-follower combinations frequently observed in mixed traffic with weak lane-discipline},
volume = {570},
journal = {Physica A: Statistical Mechanics and its Applications},
doi = {10.1016/j.physa.2021.125789}
}

@article{nagahama2025grouping,
  title={Grouping mechanisms of vehicles in heterogeneous traffic with weak lane discipline: A single-site observational study focusing on leader--follower relations},
  author={Nagahama, Akihito and Nishinari, Katsuhiro},
  journal={Physica A: Statistical Mechanics and its Applications},
  pages={131032},
  year={2025},
  publisher={Elsevier}
}

@inproceedings{nagahama2022prototype,
  title={Prototype Models for Predicting Vehicle Types Generated in Heterogeneous Traffic Simulation},
  author={Nagahama, Akihito and Wada, Takahiro and Takadama, Keiki and Yanagisawa, Daichi and Nishinari, Katsuhiro and Tanaka, Kenji},
  booktitle={International Conference on Traffic and Granular Flow},
  pages={431--438},
  year={2022},
  organization={Springer}
}

@article{phogat2020study,
  title={Study on effect of mixed traffic in highways},
  author={Phogat, Aman and Gupta, Rakesh and Kumar, Er Neeraj},
  journal={International Research Journal of Engineering and Technology (IRJET)},
  volume={7},
  number={01},
  year={2020}
}

@article{bhardwaj2023understanding,
  title={Understanding sudden traffic jams: From emergence to impact},
  author={Bhardwaj, Ankit and Iyer, Shiva R and Ramesh, Sriram and White, Jerome and Subramanian, Lakshminarayanan},
  journal={Development Engineering},
  volume={8},
  pages={100105},
  year={2023},
  publisher={Elsevier}
}

@article{li2022equilibrium,
  title={Equilibrium modeling of mixed autonomy traffic flow based on game theory},
  author={Li, Jia and Chen, Di and Zhang, Michael},
  journal={Transportation research part B: methodological},
  volume={166},
  pages={110--127},
  year={2022},
  publisher={Elsevier}
}

@misc{chen2025evidence,
      title={Evidence and quantification of cooperation of driving agents in mixed traffic flow}, 
      author={Di Chen and Jia Li and H. Michael Zhang},
      year={2025},
      eprint={2408.07297},
      archivePrefix={arXiv},
      primaryClass={physics.soc-ph},
      url={https://arxiv.org/abs/2408.07297}, 
}

@article{kumar2006self,
  title={Self-organizing traffic at a malfunctioning intersection},
  author={Kumar, Sujai and Mitra, Sugata},
  journal={Journal of Artificial Societies and Social Simulation},
  volume={9},
  number={4},
  year={2006}
}

@article{shiomi2011evaluation,
  title={Evaluation of spatial motorcycle segregation at isolated signalized intersections considering traffic flow conditions},
  author={Shiomi, Yasuhiro and Nishiuchi, Hiroaki},
  journal={Journal of the Eastern Asia Society for Transportation Studies},
  volume={9},
  pages={1644--1659},
  year={2011},
  publisher={Eastern Asia Society for Transportation Studies}
}

@inproceedings{king2015traffic,
  title={Traffic behaviour and compliance with the law in low and middle income countries: are we observing 'pragmatic driving'?},
  author={King, Mark},
  booktitle={Proceedings of the 2015 Australasian Road Safety Conference (ARSC2015)},
  pages={1--11},
  year={2015},
  organization={Australasian College of Road Safety (ACRS)}
}

@article{chandra2017indian,
  title={{Indian highway capacity manual (Indo-HCM)}},
  author={Chandra, Satish and Gangopadhyay, S and Velmurugan, S and Ravinder, Kayitha},
  year={2017}
}

@article{Chandra1995DYNAMICPA,
  title={{Dynamic PCU and estimation of capacity of urban roads}},
  author={Satish Chandra and Kumar and Pradip Kumar Sikdar},
  journal={Indian highways},
  year={1995},
  volume={23},
  url={https://api.semanticscholar.org/CorpusID:110038427}
}

@misc{IDB,
title = {{Urban Road Congestion in Latin America and the Caribbean: Characteristics, Costs, and Mitigation}},
author = {Calatayud, Agustina and Sánchez González, Santiago and Bedoya-Maya, Felipe and Giraldez Zúñiga, Francisca and Márquez, José María},
year = {2021},
doi = {10.18235/0003149},
url = {https://doi.org/10.18235/0003149}
}

@techreport{akbar2023fast,
  title={The fast, the slow, and the congested: Urban transportation in rich and poor countries},
  author={Akbar, Prottoy A and Couture, Victor and Duranton, Gilles and Storeygard, Adam},
  year={2023},
  institution={National Bureau of Economic Research}
}

@article{lee2016agentbased,
  title={Agent-based modeling of motorcycle behavior at signalized intersections},
  author={Lee, Jihun and Park, Byungkyu and Smith, Brian L.},
  journal={Transportation Research Record},
  volume={2560},
  pages={101--109},
  year={2016},
  publisher={SAGE Publications Sage CA: Los Angeles, CA}
}

@article{das2017modelling,
  title={Modelling lateral movement behaviour of motorcycles in non-lane-based heterogeneous traffic using cellular automata},
  author={Das, Sandip and Maurya, A K and Rathi, S},
  journal={Transportmetrica B: Transport Dynamics},
  volume={5},
  number={2},
  pages={213--236},
  year={2017},
  publisher={Taylor \& Francis}
}

@article{sharma2021estimation,
  title={Estimation of passenger car units for heterogeneous traffic using a microscopic simulation model},
  author={Sharma, Aashish and Mathew, Titus K},
  journal={Journal of Traffic and Transportation Engineering (English Edition)},
  volume={8},
  number={5},
  pages={715--727},
  year={2021},
  publisher={Elsevier}
}

@inproceedings{nagahama2020response,
  title={Response Time and Deceleration Affected by Lateral Shift of Leaders in Vehicular Traffic with Weak Lane Discipline},
  author={Nagahama, Akihito and Wada, Takahiro},
  booktitle={Traffic and Granular Flow 2019},
  pages={539--546},
  year={2020},
  organization={Springer}
}

%全編はappendixへ
%\clearpage
%\section*{Appendix}
%\input calcOVM2.tex

\end{document}